# Sublimation Kinetics for Individual Graphite and Graphene Nanoparticles (NPs): NP-to-NP Variations and Evolving Structure-Kinetics and Structure-Emissivity Relationships

Bryan A. Long, Chris Y. Lau, Daniel J. Rodriguez, Susanna An Tang, and Scott L. Anderson*

Department of Chemistry, University of Utah, 315 S. 1400 E., Salt Lake City, Utah 84112, USA

*Corresponding author: anderson@chem.utah.edu

**ABSTRACT:** A single nanoparticle (NP) mass spectrometry method was used to measure sublimation rates as a function of nanoparticle temperature ($T_{NP}$) for sets of individual graphite and graphene NPs. Initially, the NP sublimation rates were ~400 times faster than that for bulk graphite, and there were large NP-to-NP variations. Over time, the rates slowed substantially, though remaining well above the bulk rate. The initial activation energies ($E_a$ values) were correspondingly low and doubled as a few monolayer's worth of material were sublimed from the surfaces. The high initial rates and low $E_a$ values are attributed to large numbers of edge, defect, and other low coordination sites on the NP surfaces, and the changes are attributed to atomic-scale "smoothing" of the surface by preferential sublimation of the less stable sites. The emissivity of the NPs also changed after heating, most frequently increasing. The emissivity and sublimation rates were anti-correlated, leading to the conclusion that high densities of low-coordination sites on the NP surfaces enhances sublimation but suppresses emissivity

## INTRODUCTION

Chemistry on the surface of nanoparticles (NPs) is important in catalysis and in many applications of polycrystalline nanomaterials, yet most measurements of nano surface chemistry average over many NPs or crystallites. Because NPs in any ensemble are expected to have different distributions of surface sites (*e.g.,* facets, vertices, defects, …), and in many cases, large variations in NP size and shape, there is reason to expect that the surface chemistry of individual NPs may deviate substantially from ensemble-averaged chemistry. Understanding these deviations, and associating them with variations in NP structure, is important in developing structure-reactivity relationships for nano surface reactions.

Here we present results from a new method we have developed to measure reaction kinetics and emission spectroscopy for individual NPs at well-defined temperatures. Previous work toward this goal is briefly reviewed in the SI. We apply the technique to sublimation kinetics for graphite and graphene NPs and report on the NP-to-NP variations in the kinetics, the evolution of rates and activation energies ($E_a$ values) that occurs as the NPs sublime, and on correlations between changes in the distribution of surface sites, and the brightness and wavelength dependence of the thermal emission spectra for individual NPs.

There have been many studies over the last 70 years of the equilibrium vapor pressure of graphite at temperatures >2300 K,[1-4] including some with mass analysis of the vapor.[5-7] The vapor consists of $C_n$ clusters, with partial pressures in the following ratio: $C_3 : C_1 : C_2 = \sim77.4 : 20.2 : 2.4$, with much smaller contributions from larger clusters. These data have been used to derive $C_n$ heats of formation, $\Delta_f H(C_n)$, which are: C = 716.7 kJ/mol, $C_2$ = 837.7 kJ/mol, $C_3$ = 820.1 kJ/mol. Unusually, the sticking coefficients for $C_n$ in collisions with hot graphite are well below unity,[4, 7-8] thus equilibrium data do not give the $C_n$ sublimation rates or activation energies ($E_a$ values).

Graphite sublimation in vacuum has also been studied,[9-12] providing total sublimation rates averaged over the $C_n$ species and over the surfaces of the polycrystalline samples. As shown in **Figure S1**, an effective $E_a$ for sublimation can be estimated by fitting the literature data to a rate law of the form $R = A\cdot\exp(-E_a/kT)$, resulting in a value near 862 kJ/mol, substantially in excess of the $C_n$ $\Delta_fH$ values. A laser-heated sublimation mass spectrometry study measured $C_n$ species distributions.[13] As in the equilibrium vapor, $C_3$ was the most abundant, but the $C/C_3$ and $C_2/C_3$ ratios were significantly higher than in the equilibrium vapor.

Both our graphite/graphene NPs, and the bulk graphite materials used in the above experiments, have distributions of exposed surface sites, including perfect basal planes, vacancies and other basal plane defects, exposed basal plane edges, and for polycrystalline materials, grain boundaries. Previous results regarding the energetics of such sites and for site diffusion and annealing, are summarized in the Discussion. Fully-coordinated, basal plane sites are the most stable, hence have the highest $E_a$ for sublimation. Basal plane edges, vacancies, and other defects expose atoms in under-coordinated and/or strained geometries, with lower stabilities, hence lower $E_a$ values for sublimation. The sublimation rates and $E_a$ values for both bulk graphite and for our NPs are weighted averages over the distributions of exposed surface sites. The bulk measurements also averaged over large ensembles of crystallites in the polycrystalline samples.

Our single NP measurements are, in essence, looking at individual crystallites, thus allowing the effects of NP-to-NP heterogeneity on sublimation rates to be observed. In addition, as the NPs sublime, the site distributions evolve, and we are able to observe the effects of this structural evolution on the rates and $E_a$ values. For the bulk measurements, enough material was sublimed to ensure that the rates, averaged over many crystallites, reached steady state. For individual NPs, one question is whether anything like steady state can be reached.

Finally, we previously observed that when graphite and other carbon NPs are heated above ~1900 K, the intensities of their emission spectra change significantly over time, sometimes brightening by up to a factor of two.[14] Here we characterize the changes in NP spectral intensities and wavelength dependence as NPs sublime, and are able to extract a correlation between the emission brightness and the sublimation rate that provides insight into the structural factors that control emissivity.

## EXPERIMENTAL METHODS

The instrument and methods used for trapping a single nanoparticle (NP), and for measuring its charge (Q), mass (M), and temperature ($T_{NP}$), have been discussed previously.[14-16] In brief, electrospray ionization (ESI) is used to get charged NPs into the gas phase, and after passage through hexapole and quadrupole ion guides, the NPs are passed through a split ring-electrode electrodynamic trap (SRET),[17] where single NPs can be trapped. Argon in the 1 mTorr range is added to damp NP motion. A radio frequency voltage of frequency, $F_{RF}$, and amplitude, $V_0$, is applied across the trap, creating an effective potential minimum at the trap center. Any NP that becomes trapped is heated by a laser focused through the trap, and trapping is detected by observing the thermal emission from the hot NP using a Si avalanche photodiode (APD).

A trapped NP undergoes simple harmonic motion about the trap center, with well-defined frequencies associated axial and radial motion. The axial frequency, $F_z$, is given by:

$$F_z = \left(\frac{|Q|}{M}\right)\left(\frac{\sqrt{2}V_0}{4\pi^2 F_{RF} z_0^2}\right), \qquad \textbf{Equation 1}$$

where $z_0$ =2.96 mm. $F_z$ is measured every 30 seconds by a resonance excitation technique, thus giving Q/M. To determine Q, $F_z$ is measured repeatedly as a vacuum ultraviolet lamp is used to induce single electron changes in Q, resulting in $F_z$ steps that are fit to determine Q. $F_z$ measurement precision is ~0.03%, which is sufficient to determine Q exactly for Q ≤ ~70 – typical of the small NPs studied here. Given Q, M can be extracted from the $F_z$ measurements. Spontaneous thermionic emission charge steps, sometimes seen at high $T_{NP}$, are taken into account in calculating M.

To measure kinetics, M is monitored vs. time as the NP is held at constant $T_{NP}$, or at a series of $T_{NP}$ values. The M vs. time data is fit to determine the mass-loss rate at each $T_{NP}$. Rates are expressed as numbers of C atoms lost *per* second, but note that our measurements provide no information about the desorbing $C_n$ species. The NPs were heated by a continuous 532 nm laser, loosely focused through the trap center, such that the beam waist was more significant than the amplitude of the NP motion. We previously showed that there are no significant differences if, instead, a 10.6 μm laser is used for heating.[14] The laser power was measured and controlled in order to stabilize $T_{NP}$.

***Uncertainty in Mass Measurements:*** For NPs with $Q \leq 70$ (most NPs here) we are generally able to determine Q exactly, and in that case, the uncertainty in the M values extracted from equation 1 is primarily due to the uncertainty in $z_0$, which depends on the accuracy with which the trap was machined and assembled. This factor gives uncertainty in the absolute mass calibration of roughly ±1.3%. For a few of the larger NPs where Q was too large to determine exactly, there is additional uncertainty in the absolute mass calibration of roughly ±0.6%. The precision or relative uncertainties in comparing M measurements is much better, ~±0.18%, from a combination of the uncertainties in measuring $V_0$ and $F_z$. Each sublimation rate was determined by fitting at ~20 mass measurements to extract the slope of M vs. time.

Uncertainties in the absolute mass calibration cancel, at least partially, for some quantities reported. In the NP volume ratios used to correct spectral intensities, the absolute uncertainty cancels exactly. To allow rates to be compared for NPs of different sizes, we normalize them to a nominal NP surface area, calculated from M, assuming spherical shape with bulk density (2.265 $g/cm^2$).[18] Thus, the area-normalized rates are proportional to $(\Delta M)/M^{2/3}$, with ±0.4 % uncertainty.

***NP Temperature Determination:*** Emission spectra covering the 600 to 1600 nm range were recorded simultaneous with the mass measurements, using an optical system described previously, along with the calibration methodology, and the uncertainties in $T_{NP}$ values.[14, 16] Briefly, light emitted by the trapped NP was collected and split into $\lambda < 980$ nm and $\lambda > 980$ nm beams, which were dispersed by a pair of spectrographs, equipped with cooled Si CCD and InGaAs photodiode array cameras, respectively. The sensitivity of the optical system vs. wavelength, S(λ), is measured using a $CO_2$ laser-heated micro-thermocouple (type C) as a calibration emitter, allowing NP spectra to be corrected for non-idealities in the optical system. A few example spectra are shown in **Figure S2**.

The NP spectra are fit to a thermal emission model function consisting of Planck's function for ideal blackbodies, multiplied by an emissivity function, ϵ:

$$I(\lambda, T_{NP}) = \frac{2c \cdot \epsilon(\lambda, T_{NP})}{\lambda^4 \cdot \left(\exp\left(\frac{hc}{\lambda k T_{NP}}\right) - 1\right)} \qquad \textbf{Equation 2}$$

The $\epsilon(\lambda, T_{NP})$ function accounts for non-idealities due both to the properties of the emitting material and to optical coupling with NPs much smaller than the emitted wavelengths. According to Mie theory, the emissivity for small spherical particles is: [19]

$$\epsilon(\lambda, T_{NP}) = \frac{8\pi r}{\lambda} \, \mathrm{Im}\left(\frac{(n+ik)^2 - 1}{(n+ik)^2 + 2}\right) \text{ , } \textbf{Equation 3}$$

where r is the NP radius, and n + ik is the complex index of refraction of the material, which depends on both λ and $T_{NP}$. Even for bulk materials, n and k are generally not known over our wavelength range at high temperatures, and for NPs there are additional factors such as quantum confinement and large numbers of surface sites that may affect the optical properties. Therefore, as is commonly done for carbonaceous NPs,[20-24] we use a power-law approximation to the emissivity, $\epsilon(\lambda) \propto \lambda^{-n}$, with n as a fitting parameter, resulting in the thermal emission model function:

$$I(\lambda, T_{NP}) = \frac{K}{\lambda^{(4+n)} \cdot \left(\exp\left(\frac{hc}{\lambda k T_{NP}}\right) - 1\right)} \text{ . } \textbf{Equation 4}$$

Here, K is a normalization constant, and the two parameters that affect the wavelength dependence are $T_{NP}$ and n. The fit quality is generally good, as can been in **figure S2**. We estimated[14] that the extracted $T_{NP}$ values have roughly ±6.2% absolute uncertainty, with ±2% relative uncertainty in comparing extracted $T_{NP}$ values.

***Other Experimental Considerations:*** Details of the NP feedstocks and steps taken to ensure that the NPs are pure carbon are discussed in the SI. The conclusion is that contaminants from ESI or from air exposure are desorbed long before the kinetics experiments start, and that metal contamination should be negligible. **Figure S3** shows an experiment testing possible effects on sublimation rates from background $O_2$. The worst-case effects are estimated to be <5% at our lowest $T_{NP}$ where sublimation is slow, dropping to < 0.1% above ~1950 K.

## RESULTS

Sublimation kinetics were studied in a series of experiments on individual graphite and graphene NPs. We first present a typical experiment and its analysis, and then summarize the results for sets of graphite and graphene NPs, focusing on both NP-to-NP variability, and the evolution of NP properties as the NPs are heated to drive sublimation.

***Single NP Data:*** **Figure 1** shows a typical experiment on a single multilayer graphene platelet NP of initial mass 64.03 MDa. The graphene feedstock has reported average thickness of ~5 nm, corresponding to ~15 layers. The NP temperature ($T_{NP}$) was ramped in 100 K steps from 1800 K to 2100 K, back down to 1800 K, and finally back up to 2100 K. An initial period during which the NP was held at ~1300 K for Q determination is not shown. Gaps in the data indicate periods when $T_{NP}$ was unstable, and these periods were omitted from further analysis. With the exception of a few excursions, $T_{NP}$ was generally stable to ~±10 K (±0.5%) during the constant $T_{NP}$ periods, and even the largest excursions were only ~ 30 K (1.5%).

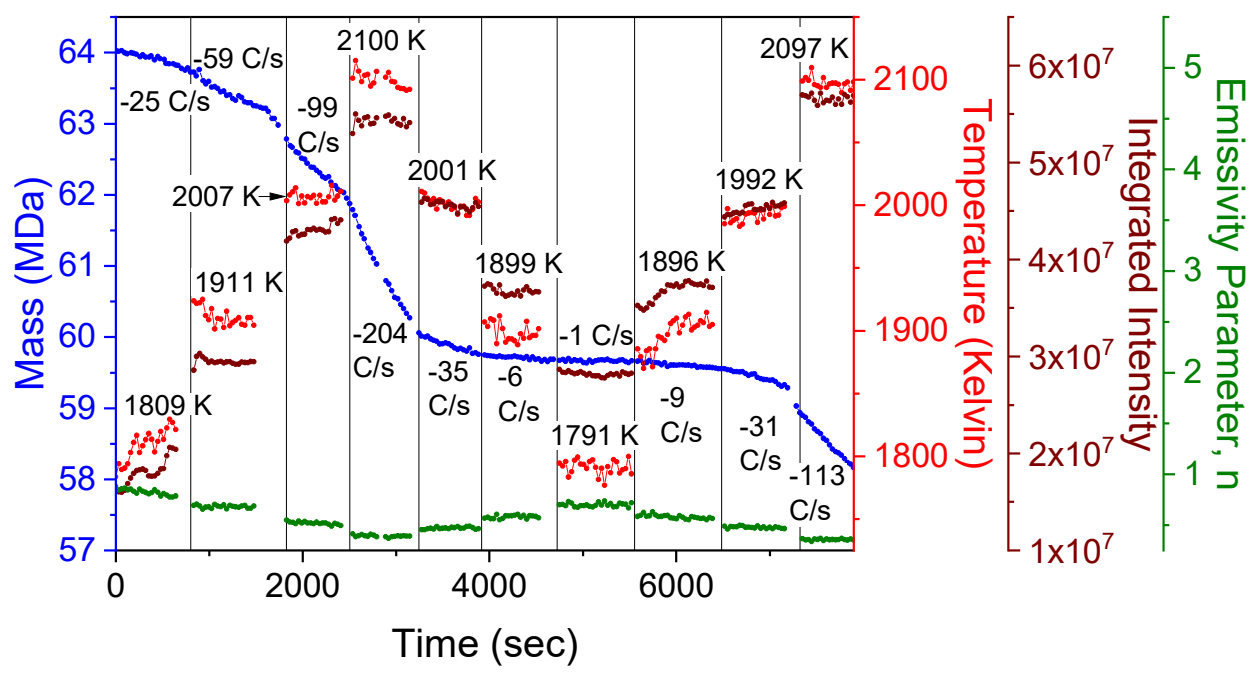

**Figure 1:** Typical experiment in which a single graphene NP of initial mass of 64.03 MDa was heated in step-wise fashion from ~1800 K to ~2100 K, back down to ~1800 K, then back to ~2100 K. The $T_{NP}$ and sublimation rates (given as C atoms/sec) indicated in boxes, are the values averaged over the time intervals defined by the vertical black lines.

M and $T_{NP}$ were measured repeatedly during each constant $T_{NP}$ time interval, which are indicated by solid vertical lines. The $T_{NP}$ values shown are simply the averages over these time intervals. Mass loss rates were determined by linear fits to M vs. time over the same time intervals, thus giving the average rates, which are indicated in terms of C atoms lost/second. As noted, the experiment gives no information about the desorbing $C_n$ product distribution.

In addition to M and $T_{NP}$, the figure shows the integrated emission intensity (emitted photons/second/sr over the 600 – 1600 nm range) and the value of the n parameter in the power law emissivity function. Note that n decreased as $T_{NP}$ increased, and then increased again as the NP cooled, indicating that emissivity is temperature dependent. The integrated emission intensity, I, increased by a factor of ~2.6 as $T_{NP}$ increased by a factor of ~1.17 (~1800 to ~2100 K), corresponding to $I \propto T_{NP}^6$.

The sublimation rates (slopes of M vs. time) increased with increasing $T_{NP}$, as expected, however, note that the rates at similar $T_{NP}$ values measured at the beginning and end of the experiment were significantly different, generally slowing over time. For example, at ~2007 K during the initial heat ramp up, the NP sublimed at a rate of 99 C/sec, but at ~2001 K during the ramp down, the rate was only 35 C/sec. As NPs sublime, some rate slowing is expected due to loss of NP surface area, however, the mass loss over the entire experiment was only ~9%, thus the surface area loss ($\propto \sim M^{2/3}$) would have been on the order of 6%, i.e., far too small to account for the rate slowing.

To allow the kinetic and emission behavior to be observed more easily, it is useful to convert the raw results into more concise and comparable forms, and this process is illustrated in **figure 2** for the NP in **figure 1**. The $T_{NP}$ values, emission intensities, and n parameters are the average values for each time interval in **figure 1**, and the rates are the average slopes in each interval. The uncertainties in mass, rates, $T_{NP}$, etc. are discussed above, and **figure 2** illustrates these using error bars. In this experiment, $T_{NP}$ was ramped up, down, then up again, but for other NPs, different $T_{NP}$ programs were examined. The order and direction of heat ramps applied to each NP are indicated using the following terminology: 1st Ramp: Up, 2nd Ramp: Down, etc.

The emission intensity changes are summarized in **figure 2A**. Ideal blackbody emission intensity scales with emitter surface area, but the emissivity for small NPs (**equation 3**) introduces another factor of r, thus intensity should scale with $r^3$. Therefore, to compensate for the expected effects of decreasing NP size during experiments (and to allow comparisons between different size NPs), the intensities in **figure 2** and all figures below, have been scaled by the NP mass ($\propto r^3$) measured simultaneous with the spectra. In this case, the intensity increased and decreased with $T_{NP}$ as noted, but note that the intensities tended to increase over the course of the experiments.

The horizontal error bars represent the estimated ±2% relative uncertainty in $T_{NP}$ values, because this is what affects trends extracted from comparing data points. Intensity error bars are not shown because they would be smaller than the height of the symbols.

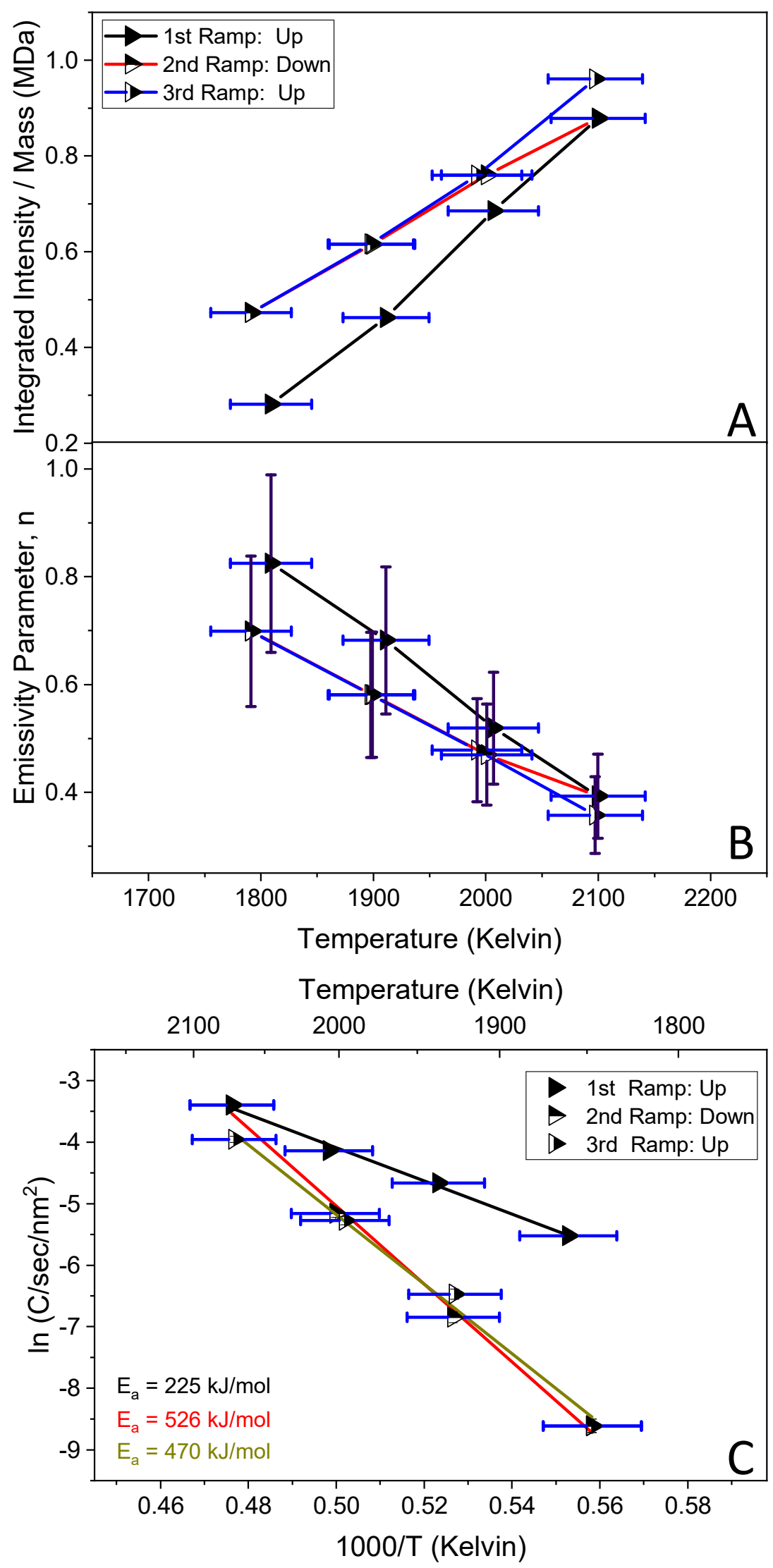

**Figure 2:** Analysis of the graphene experiment in **figure 1.** A: Integrated emission intensity averaged over each constant $T_{NP}$ period. B: Emissivity parameter, n, averaged over each constant $T_{NP}$ interval. C: Area-normalized sublimation rate vs. $T_{NP}$ for each heat ramp, with fits to extract the effective $E_a$ values shown.

**Figure 2B** plots the n parameter, which decreased with increasing $T_{NP}$, then recovered as the $T_{NP}$ dropped, but only to ~80% of its initial value, again indicating an irreversible change during the experiment. The vertical error bars represent the estimated ~±20% uncertainty in the n parameter[14] – large because it takes a large change in n to cause significant change in spectrum fit quality.

**Figure 2C** plots the sublimation rates measured at each $T_{NP}$ during the three heat ramps. Vertical error bars (~±0.4%) are not shown because they would be smaller than height of the symbols. To correct for the ~6% surface area loss during the experiment, and to allow comparisons with other NPs, we would ideally normalize rates to the NP surface areas, but these are unknown. Therefore, we have approximated the area-normalized rates assuming that the NPs are spherical with the bulk density. Obviously, this gives lower limits on the surface areas, hence upper limits on the area-normalized rates, given as $C/sec/nm^2$. The data for all three heat ramps are reasonably linear when plotted as $\ln(C/sec/nm^2)$ vs. 1000/T, thus the fit lines can be used to extract activation energies, $E_a$, and prefactors, A, in rate expressions of the type: $R = A\exp(-E_a/RT_{NP})$.

For this data, the $E_a$ value extracted from 1$^{st}$ ramp up was 225 kJ/mol. The $E_a$ extracted from the 2$^{nd}$ ramp down was 526 kJ/mol, and the 3$^{rd}$ ramp up gave $E_a$ = 470 kJ/mol. We discuss the interpretation of the extracted $E_a$ values, below, after summarizing the results for sets of graphene and graphite NPs.

Because the uncertainties in the mass loss rates are small, the uncertainty in the $E_a$ values comes almost entirely from the ±2% relative uncertainty in $T_{NP}$ measurements. We estimated the $E_a$ uncertainty by simulating the effect of apply random gaussian $T_{NP}$ perturbations to each point in the heat ramps, with 2% standard deviation, refitting the data to extract the $E_a$ values associated with the perturbed $T_{NP}$ values. The simulation was repeated ~10,000 times for each heat ramp to obtain the corresponding spread in $E_a$. The spreads vary somewhat from heat ramp to heat ramp, but average ~±20% of the extracted $E_a$ values, and we take this as our estimate of the uncertainty in $E_a$.

**Figures S4** and **S5** present a similar experiment and analysis for a single graphite NP, with initial mass of 38.93 MDa. As with the graphene NP in **figures 1** and **2**, irreversible changes in the graphite NP's kinetic and spectroscopic properties were observed over the course of the experiment. The volume-weighted emisison intensity increased, the n parameter decreased, and the sublimation kinetics slowed substantially over the course of the experiment, with a substantial increase in the $E_a$ values from the first to second heat ramps.

Before presenting summaries of sublimation experiments on sets of graphite and graphene NPs, we present an example of a simpler kind of experiment, monitoring kinetic and emission behavior over longer time periods at constant $T_{NP}$, as shown in **figures 3** and **S6**, for two graphite NPs held at different $T_{NP}$ values. The NP in **figure 3** had initial mass of 82.88 MDa, and was held at 1900 K for ~9 hours (an initial period for Q determination at low $T_{NP}$ is not shown). This experiment was done using the 532 nm heating laser, augmented by a cw $CO_2$ laser run at nominally constant power. $T_{NP}$ had good long-term stability, but there were short term fluctuations, from power fluctuations of the $CO_2$ laser, which the $T_{NP}$ control program was not able to compensate completely. The emission intensity fluctuated with the $T_{NP}$ fluctuations, as expected from the strong dependence of intensity on $T_{NP}$. The mass loss rate also fluctuated, but much more slowly, and some fluctuations (e.g. between 10,000 and 20,000 seconds), were not obviously correlated with variations in $T_{NP}$. More importantly, there was a gradual slowing of the mass loss rate from an average of 368 C/sec over the first three hours, to an average of 189 C/sec over the final three hours. This ~50% decrease is far larger than can be rationalized by the ~12% loss in total mass (~9% in surface area). More dramatic mass loss rate slowing was observed for a graphite NP at 2000 K, as shown in **figure S6.**

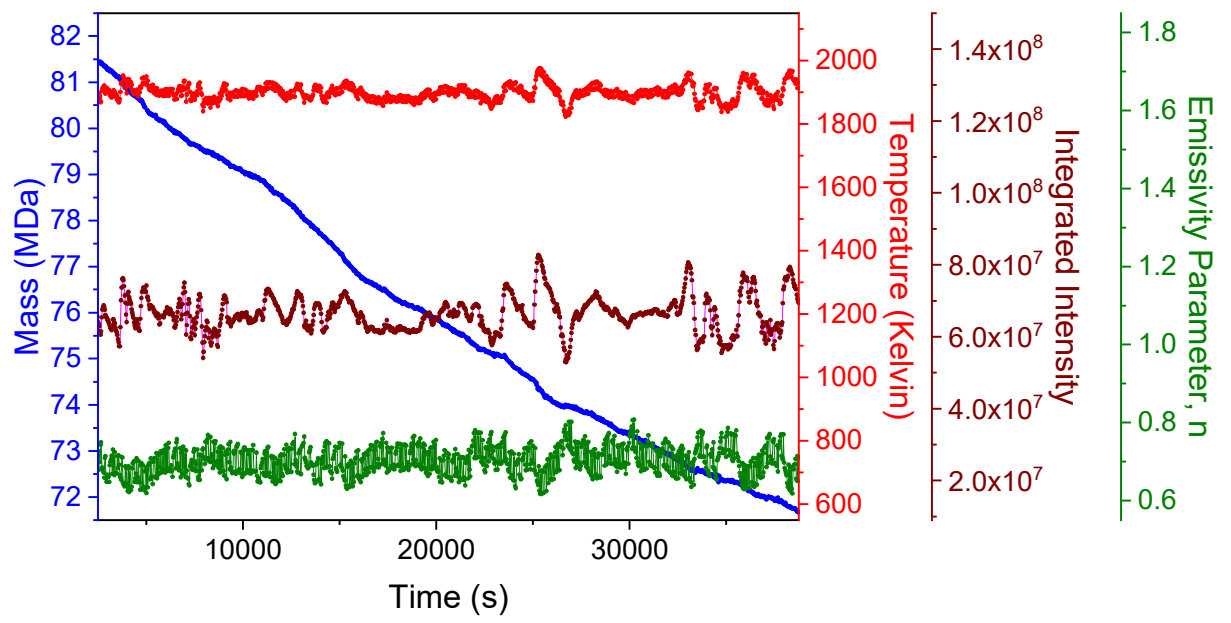

**Figure 3**: Long exposure heating experiment of a graphite NP with initial mass 82.88 MDa. The $T_{NP}$ was held at 1900 K for ~10.8 hours.

***Comparison Data:*** **Figure 4** plots aggregated data for emission properties from 15 graphite NPs with initial masses ranging from ~5 to ~120 MDa (~19 to 56 nm, if spherical), and for 7 graphene NPs with initial masses ranging from ~10 to ~70 MDa. A consistent set of symbols is used in **figures 4-7, S7,** and **S8**; color and shape indicate particular NPs, while symbol filling indicates the 1st, 2nd, or 3rd heat ramps. The arrows after the legend symbols indicate whether each heat ramp was up or down, *i.e.,* to increasing or decreasing $T_{NP}$.

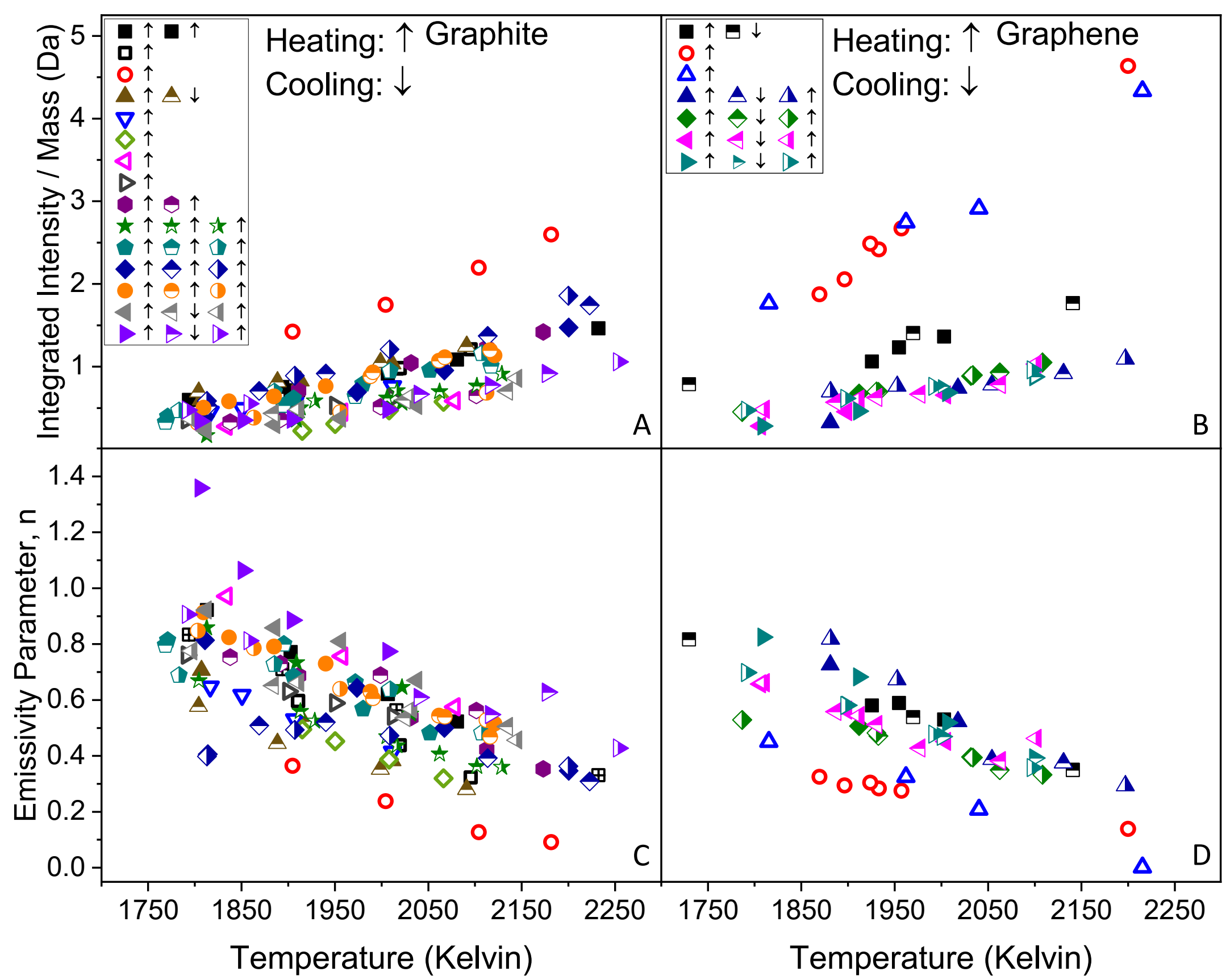

**Figure 4:** A and B: Integrated emission intensity for all graphite and graphene NPs in each heat ramp. Consistent symbol shapes are used for each NP. The symbol fill indicates the heat ramp. The arrows in the legend indicate whether the heat was ramped up or down. C and D: Emissivity parameter n for each NP in all heat ramps.

For some experiments, indicated by open symbols, we were only able to run a single heat ramp (always up) before losing the NP from the trap. The vertical scales are identical for the graphite and graphene NPs, allowing direct comparison of the emission properties.

The top half of the figure shows the integrated emission intensities, scaled by the NP masses measured at the same time as the emission, *i.e.,* approximately corrected for the effects expected from varying NP volume, both from NP-to-NP and for each NP as it sublimed. Error bars are not shown, but would be the same as in **figure 2A** (±2% horizontal, negligible vertical).

The main observations were:

**1.** The volume-corrected emission intensity increased with increasing $T_{NP}$ for all NPs, but with substantial NP-to-NP variation.

**2.** One graphite and two graphene NPs stood out as having substantially higher-than-average volume-corrected emission intensities, even though their masses were similar to those of other NPs (see below).

**3.** The graphite and graphene NPs had similar emission intensities, *i.e.,* while there were large NP-to-NP variations within each data set, the intensity ranges for graphite and graphene NPs overlapped.

**4.** Examples were seen where the emission brightness increased, decreased, or did not change significantly between the heat ramps.

The bottom half of **figure 4** shows data for the n parameter, which controls the shape of the thermal emission model function used in spectral fitting. Low (high) n corresponds to flatter (sharper) λ dependence. Error bars are not shown, but would be the same as in **figure 2B** (±2% horizontal, ±20% vertical). The main observations were:

**1.** The n parameter tended to decrease with increasing $T_{NP}$ but with NP-to-NP variation of more than a factor of 2.

**2.** The one graphite and two graphene NPs that had unusually high emission intensity also had unusually low n parameters, *i.e.,* less λ-dependent spectra.

**3.** The graphite and graphene NPs had n parameters in overlapping ranges, but graphite had slightly higher n parameters on average.

**Figure 5** summarizes the sublimation rates, R, plotted as ln(R) vs. 1000/T, such that the slopes are proportional to the activation energies ($E_a$). Data points are the experimental rates for each NP, and lines are single exponential fits to the data. As in **figures 2C** and **S5C**, the raw sublimation rates (C atoms/sec) have been normalized to nominal NP surface areas, calculated assuming spherical shape with the bulk density. Error bars are not shown, but would be the same as those in **figure 2C** (±2% horizontal, negligible vertical).

Rates are reported separately for the 1$^{st}$ heat ramp (always up) and for the 2$^{nd}$ and 3$^{rd}$ heat ramps, the direction of which can be seen in the legend for **figure 4**. The masses of the NPs in each heat ramp can be seen by comparison to **figure 6**, and the charges can be seen by comparison to **figure S8**. Also plotted for comparison in each frame of **figure 5** is an estimate of the sublimation rate for bulk graphite from measurements at temperatures above 2300 K (see **figure S1)**.[9-12]

Most of the data are reasonably well fit by single exponential functions, but during the 1$^{st}$ heat ramps, the fits for some NPs deviated by up to a factor of ~5 at the lowest or highest $T_{NP}$ points. In those cases, the fit lines shown are the best single exponential fits to the remaining points. In the 2$^{nd}$ and 3$^{rd}$ ramps, the data are generally well fit by single exponential rate laws.

To facilitate comparisons between different heat ramps and with the rates for bulk graphite, we also show the ensemble-averaged rates as solid red dots with red fit lines. Calculating the ensemble averages is complicated by the fact that the measured $T_{NP}$ values varied from NP-to-NP. To obtain the average rate, we used the single exponential fits to each NP data set to estimate what the rates for each NP would have been at 1800, 1900, 2000, and 2100 K. These estimated rates were ensemble-averaged to obtain the red dots, which were then fit to single exponentials (red lines).

There are several important points to note:

**1.** During the 1$^{st}$ heat ramp, the average area-normalized rate for graphite NPs at 1800 K was ~400 times faster than the bulk graphite rate, and that for graphene was ~360 times faster. Our assumption of spherical shape in calculating the NP surface area gives an upper limit on the rate *per* $nm^2$, however, the actual NP surface areas are certainly not underestimated by anything like a factor of 400. Thus, we conclude that the NPs had substantially higher inherent sublimation rates *per* unit area than bulk graphite.

**2.** The NP sublimation rates increased more slowly with increasing $T_{NP}$ than that for bulk graphite, such that for $T_{NP}$ = 2100 K, the averaged NP rates were only a factor of ~three higher than that for bulk graphite.

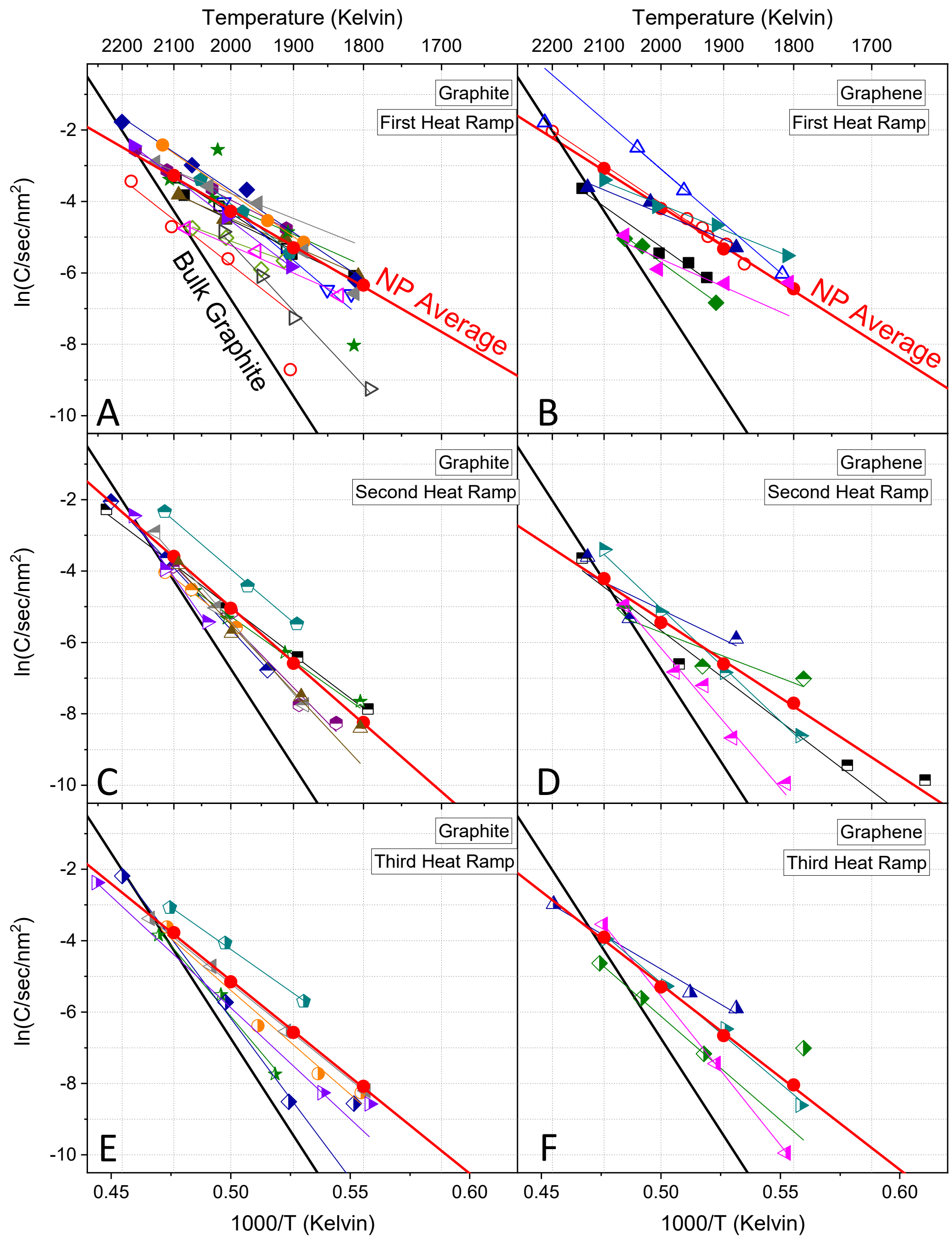

**Figure 5:** Top: Area-normalized rates vs. $T_{NP}$ for graphite (right) and graphene (left) NPs. Points indicate measured rates using the same symbols as in **figure 4**, and lines through each set of points are single exponential fits. The solid red circles show the average rates, and the red lines are fits to the average rates. The solid black line is the extrapolated rate for bulk graphite. Middle and bottom rows: Analogous plots for the 2$^{nd}$ and 3$^{rd}$ heat ramps.

**3.** The anomalously bright graphite NP (red circles) had the slowest sublimation rates and the steepest slope of any of the graphite NPs. The two anomalously bright graphene NPs (red circles, blue triangles) also had somewhat steeper-than-average slopes, but had higher-than-average sublimation rates.

**4.** The rates at low $T_{NP}$ in the 2$^{nd}$ and 3$^{rd}$ heat ramps were substantially lower than those in the 1$^{st}$ heat ramp,

although still substantially higher than the bulk graphite rate. For example, at 1800 K the averaged rate for graphite NPs in the 2nd heat ramp was ~6.6 times slower than in the 1st heat ramp, but still ~60 times higher than that for bulk graphite.

**5.** The rates in the 2nd and 3rd heat ramps increased substantially faster with $T_{NP}$ compared to the 1st heat ramp so that for $T_{NP}$ > 2100 K, the rates for all three heat ramps were similar and also similar to that for bulk graphite.

**6.** The area-normalized rates at low $T_{NP}$ in the different heat ramps varied from NP to NP by factors of up to 10 for the graphite NPs, and by factors of up to 20 for the graphene NPs. The variability of the rates decreased at higher $T_{NP}$ but still exceeded a factor of 4.

**7.** The NP-to-NP variability in the sublimation rate was much larger than the differences between the ensemble-averaged rates for graphite and graphene, *i.e.,* there was no significant difference between graphite and graphene NPs.

Finally, the activation energies ($E_a$ values) extracted from the single exponential fits in **figure 5** are plotted in **figure 6** against the NP mass, measured at the beginning of each heat ramp. To maintain readability, error bars are not shown, however, the uncertainties in mass are negligible, and as discussed above, the uncertainties in the $E_a$ values are roughly ±20%. **Figure S7** in the SI gives the prefactors (A) using the same format. For NPs where only a single heat ramp was completed, a single point is shown (open symbols). For NPs where multiple ramps were completed, filled symbols connected with lines are used. For a few of the more massive NPs, arrows are superimposed on the lines to show the order of the heat ramps (1 → 2 → 3). For other NPs, the $E_a$ values can be associated with the 1st, 2nd, and 3rd heat ramps by recognizing that the NP mass always decreased from 1st to 2nd to 3rd heat ramp. The NP-ensemble-averaged $E_a$ values, and the value for bulk graphite are shown as horizontal dashed lines.

The important points for this figure are:

**1.** There were no obvious trends in $E_a$ with NP mass.

**2.** In 1st heat ramp for each NP, the $E_a$ values varied significantly from NP-to-NP, and many were unexpectedly low. For graphite NPs, the $E_a$ values ranged from 202 to 597 kJ/mol, with an ensemble-averaged $E_a$ of 322 kJ/mol. For graphene, the $E_a$ values ranged from 224 to 482 kJ/mol with average of 352 kJ/mol. (The ensemble-averaged values given in the figure include only NPs for which at least two heat ramps are available).

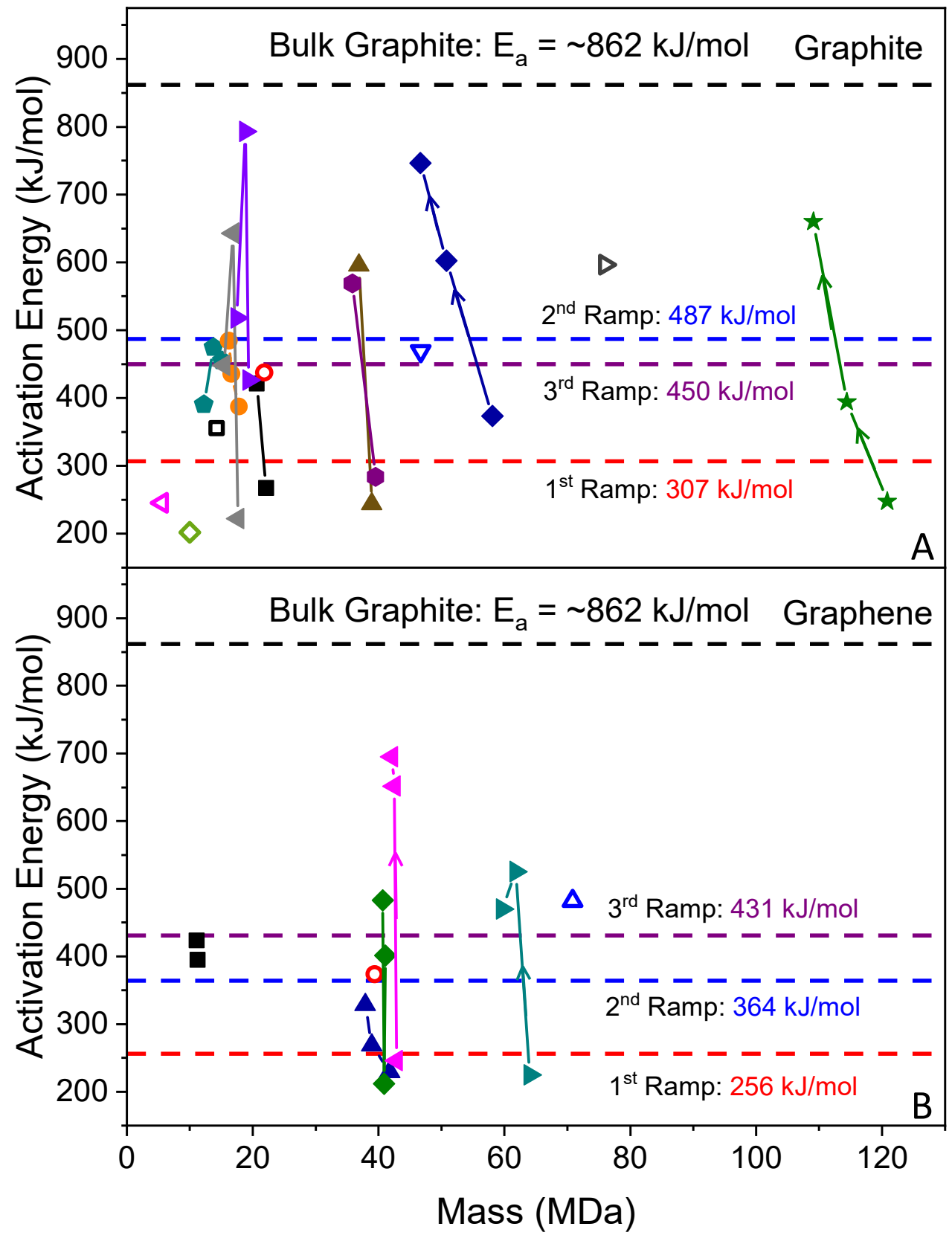

**Figure 6**: $E_a$ values extracted from fits to the rates in **figure 5** for graphite (top) and graphene (bottom). Pre factor values are found in **figure** S6. Also shown in all figures are the average values for the NPs and the values obtained by fitting bulk graphite data.

**3.** These spreads in $E_a$ values are not uncertainties, but rather indicate the inherent variations between different NPs. Measuring this NP-to-NP variability is one of the motivations for studying single NP kinetics.

**4.** In the 2nd heat ramp, the $E_a$ values increased for all NPs, but there were still substantial NP-to-NP variations (394 to 792 kJ/mol for graphite, 212 to 652 kJ/mol for graphene). The ensemble-averaged $E_a$ values increased to 487 kJ/mol and 364 kJ/mol for graphite and graphene, respectively.

**5.** In the 3rd heat ramp, the $E_a$ values for a few of the graphite NPs approached the bulk graphite value, but most remained well below, and for a few NPs, the $E_a$ values decreased, in one case substantially. As a result, the average $E_a$ decreased slightly for graphite.

**6.** On average, the $E_a$ values for graphene NPs were lower than those for graphite NPs, but the differences were much smaller than the NP-to-NP variations within each data set.

## DISCUSSION

The results presented here show that there are substantial NP-to-NP variations in volume-scaled emission intensities, n parameters, area-normalized sublimation rates, and activation energies, and that all these properties evolve significantly as the NPs sublime. It is not so obvious what factors influence the emission properties, however, the sublimation kinetics clearly must depend on the distribution of sites present on each NP's surface. Sites with different structures have different energetics, hence different $E_a$ values and rates for sublimation, and the measured rates for each NP are averages over the site distributions present at the time of measurement. The substantial NP-to-NP variation in sublimation behavior is attributed to the heterogeneity of the graphene and graphite feedstocks, which have distributions of size, shape, and presumably of atomic-scale surface structure. The mass losses during each experiment correspond to removing several monolayer's worth of material, thus it is reasonable to expect that the surface structures should evolve, thereby accounting for the observed evolution of the kinetics. Before discussing the structure-kinetics relationship, we need to consider two other properties which varied significantly from NP-to-NP, and thus could conceivably have affected sublimation: the NP size/mass and the charge.

***Effects of NP Mass on Sublimation Rates:*** **Figure 5** shows that the surface-area-normalized rates for the NPs in each heat ramp varied substantially and that they changed from one heat ramp to the next. The mass of each NP at the start of each heat ramp can be identified by comparison to **figure 6**, where the same symbols are used. Close examination shows no correlation between mass and rates in the 1$^{st}$ heat ramp – some NPs with similar masses had very different rates, and some NPs with very different masses had similar rates. In the 2$^{nd}$ and 3$^{rd}$ heat ramps, the NP-to-NP rate variations were smaller, and one apparent anomaly emerged. Most of the graphite NPs had rates within a factor of ~two of the average rate, and these showed no correlation to the NP mass. One graphite NP (turquoise pentagons) had distinctly higher rates, and it happened to be the lowest mass graphite NP (initial mass ~17.5 MDa) for which multiple heat ramps were completed. Note, however, that there were four other graphite NPs with an initial mass just a few percent higher, and these had rates at or below the average. Therefore, we discount the possibility that the high rates for this NP resulted from an inherent effect of mass on sublimation.

**Figure 6** plots $E_a$ vs. NP mass in each heat ramp on each NP. $E_a$ varied substantially for the set of NPs, but there is no obvious correlation with NP mass. It is true that the $E_a$ values generally increased during the three heat ramps, while M decreased, but the mass changes are small compared to the differences in masses between different NPs. Therefore, the changes in $E_a$ as each NP sublimed are attributed to evolution of the NP surface site distributions, rather than to any inherent effect of the small changes in mass.

***Effects of NP Charge on Sublimation Rates:*** In these experiments, the NP charge, Q, varied from +30e to +110e, generally increasing with NP mass. We previously showed that changing Q has no effect on emission spectra or $T_{NP}$,[14] and here the question is whether Q might have effects on sublimation kinetics. In fact, such effects would be quite surprising because Q/M is so low – between 7 x 10$^{-7}$ and 6 x 10$^{-6}$ e/Da, corresponding to between 8 x 10$^{-6}$ and 7 x 10$^{-5}$ e/carbon atom. **Figure S8** shows the measured $E_a$ values for the graphite NPs plotted against the Q values measured at the beginning of each heat ramp. A comparison of **figures 5** and **S8** allows the rates for each graphite NP to be associated with its charge. As discussed in the SI, we conclude that there are no direct effects of Q on either sublimation rates or extracted $E_a$ values.

***Effects of NP Surface Site Distributions on Sublimation Rates:*** For graphitic NPs, the two major classes of sites are perfect basal planes where all atoms are fully coordinated with unstrained bonds, and less stable sites such as basal plane defects or edges. Because such sites are less stable, they have lower heats of sublimation, and because they have fewer or strained C-C bonds, the transition states for sublimation should also be lower in energy (*i.e.,* lower $E_a$ values), compared to those for loss of $C_n$ from perfect basal planes.

For both graphite and graphene NPs, the rates measured at low $T_{NP}$ during the 1$^{st}$ heat ramps averaged about 400 times faster than the rate for bulk graphite, indicating that the NPs initially had large numbers of defective or under-coordinated surface sites. At low $T_{NP}$ during the 2$^{nd}$ heat ramps, the rates were much slower, indicating that the numbers of such sites decreased when the NPs were heated, driving sublimation. For most, but not all NPs, the rates in the 3$^{rd}$ heat ramps were slower yet, indicating that sublimation tends to drive the NPs toward more stable surface structures.

This evolution of the surface site distributions explains why the $E_a$ values extracted from the 1st heat ramp data were so low – below 300 kJ/mol for many NPs. Such low $E_a$ values are unphysical because any sites with such low $E_a$ values would have sublimed away long before the start of the kinetics experiments, during the >20 minutes that each NP was held at ~1300 K for Q determination. For example, assuming a 1st order desorption rate constant of the form, $k_1 = \nu_1 \cdot \exp(-E_a/RT)$, and $\nu_1 \approx 10^{14}$ sec$^{-1}$, in the range typically seen for surface desorption,[25-26] we can estimate the lifetime ($1/k_1$) for a site with $E_a$ = 300 kJ/mol to be ~10 msec at 1300 K.

Instead, the low 1st heat ramp $E_a$ values reflect the *evolution* of the site distribution during the heat ramp. Initially, the NPs had a large number of sites contributing to sublimation, resulting in high initial rates. As the NPs were heated, the coverage of such sites dropped, resulting in rates that increased more slowly with $T_{NP}$ than would have been the case for a constant site distribution. Evolution, thus, artificially decreased the slopes in **figure 5**, giving artificially low $E_a$ values.

The measured rates can be used to roughly estimate the average $E_a$ for sites contributing to sublimation during the 1st heat ramp. At 1800 K, the average mass loss rate was ~0.002 C/sec/nm$^2$, corresponding to an average surface atom lifetime of ~50 seconds. Using the $k_1$ expression given above, this would imply $E_a \approx 540$ kJ/mol, averaged over the sites contributing to sublimation at that $T_{NP}$. By the end of the 1st heat ramp, at ~2100 K, the average rate was ~0.04 C/sec/nm$^2$, implying an average surface atom lifetime of ~2.5 seconds and $E_a \approx 580$ kJ/mol, consistent with the conclusion that the surface site distributions evolved to higher average site stability.

It is useful to compare these values to what is known about the energetics of different graphitic sites and diffusion processes. The net energy to remove an atom from a perfect basal plane, creating a mono-vacancy, is reported to be in the 705 to 725 kJ/mol range,[27-34] and removing $C_2$ to create a di-vacancy has net formation energy estimated to be in the 700 - 790 kJ/mol range.[33-35] It is likely that the $E_a$ values for these processes are higher yet, thus, under our conditions, sublimation of C or $C_2$ from perfect basal plane sites should make only minor contributions to the total rate. We were unable to find energetics for loss of larger $C_n$ from basal planes, but given the numbers of bonds that would have to be broken, these are also likely to be minor processes under our conditions.

The fast NP sublimation rates are, instead, attributed to the presence of low-coordination, high formation energy sites on the NP surfaces. Some types of low-coordination sites probably are not important, however. For example, adatoms on top of graphitic basal planes are bound by only 145 to 200 kJ/mol,[33-34, 36-37] and have $E_a$ values for lateral diffusion in the 34 to 51 kJ/mol range.[38] Therefore, any adatoms initially present on the NPs would have desorbed or diffused to defects or edges[38] long before the start of our experiments.

One type of site that must contribute strongly to sublimation of NPs are basal plane edges. Several symmetric edge structures are possible, including "armchair" with formation energy estimated to be in the range of 110 - 250 kJ/mol *per* angstrom of edge length, "zigzag" (116 - 300 kJ/mol/angstrom), and Klein ( 172 kJ/mol/angstrom).[27, 38-39] Armchair edges expose pairs of doubly coordinated atoms, zigzag edges expose individual doubly coordinated atoms, and the Klein edge exposes a row of singly coordinated atoms. In addition, there are lower symmetry edge structures that can form by reconstruction from the symmetric edge structures with low $E_a$ values (< 200 kJ/mol).[38, 40] Thus, interconversion between different edge structures should be rapid on our time scale.

Another important type of site is basal plane defects, which are likely to be present to some extent in the reactant NPs. The $E_a$ values for sublimation of strained or under-coordinated atoms at defects are similar to those for edge sites, and as material sublimes around the defects, holes or pits form that expose sites similar to those around the basal plane edges.[41-46]

An obvious question is the extent to which annealing is important during sublimation, *i.e.,* is annealing fast enough to significantly modify the surface site distribution, as it evolves during sublimation? The $E_a$ for lateral diffusion of C atoms within perfect basal planes is estimated to be in the ~675 to 780 kJ/mol range,[28] and annealing of dislocation loops by a vacancy creation/migration mechanism has $E_a$ ~800 kJ/mol,[47] both substantially higher than the $E_a$ values for NP sublimation. As already noted, interconversion between different edge structures should be fast, however, that simply converts between different types of under-coordinated sites with similar energies, rather than reducing their numbers. If the NPs have existing mono-vacancies, their $E_a$ values for lateral diffusion are only *ca.* 115 to 135 kJ/mol,[33-34, 48] which would give diffusion rates fast compared to sublimation. Diffusion of a mono-vacancy to a basal plane edge is energetically favorable (and for NPs, the distances are small), and since this effectively heals the vacancy, we expect mono-vacancies to be short-lived. Note, however, that merging two mono-vacancies to form a di-vacancy is also energetically favorable, and the energy for lateral diffusion of di-vacancies is estimated to be 580 - 840 kJ/mol,[33-34, 48] *i.e.,* once formed, di-vacancies are much more stable with respect to diffusion than mono-vacancies. This is presumably also true of larger vacancies. We conclude that atomic-scale transformations (mono-vacancy diffusion, edge interconversion) should be fast on the sublimation time scale, but that large scale restructuring of the NPs in ways that would significantly change the number of under-coordinated surface sites, is probably not. Therefore, sublimation – specifically preferential sublimation from defective/under-coordinated sites – is the dominant process driving the evolution of the surface site distributions during sublimation.

As shown in **figure 6**, in the 3rd heat ramps, the $E_a$ values for a few of the NPs approached that for bulk graphite, but the average remained substantially lower, raising the question of how best to compare our single NP measurements with the bulk sublimation experiments described above.[9-12] The bulk experiments, on polycrystalline graphite, would also have had contributions from a range of sites, including basal planes, exposed basal plane edges or defects, and grain boundaries. Furthermore, the individual crystallites in the bulk graphite undoubtedly evolved with time, similar to the evolution shown here for single NPs. The bulk rates average over many crystallites, however, and because macroscopic amounts of material were sublimed, the area-averaged site distributions and associated kinetics would have reached steady state.

The obvious question is whether steady state can ever be reached during sublimation of individual graphite or graphene NPs. Consideration of graphitic NP structures, and of experiments like those in **figures 3, 5**, **6**, and **S6,** suggest that the answer may be "only partially". These figures all show a tendency of rates to slow and $E_a$ values to increase as sublimation proceeds, corresponding to the evolution of the NP surface sites toward greater stability, however, this evolution is not monotonic. Rates were observed to fluctuate with time in the constant $T_{NP}$ experiments (**figures 3** and **S6**), and there were a few NPs where the rates accelerated and $E_a$ values decreased in the 3rd heat ramp (**figures 5 and 6)**. This behavior suggests that there is a propensity for NP structures to evolve in ways that generally decreases the number of defective or low coordination surface sites, but that it is possible for there to be temporary increases in the number of such sites.

This behavior is not surprising for graphitic NPs that are initially rough, with irregular edges, defects, and surface asperities. Such NPs should sublime quickly at first, but the combination of preferential sublimation at low-coordination sites, and annealing should tend to reduce the numbers of such sites. On the other hand, there are processes, such as sublimation around basal defects, or sublimation that exposes underlying defective reactions, which increase the numbers of low-coordination sites. Therefore, as sublimation and annealing proceed, the rate should approach a near-steady-state, punctuated by occasional rate accelerations as is observed.

The motivation for studying both graphite and graphene NPs was to examine the effects of changing the ratio of basal plane area to basal plane edge length. The SI discusses the expected numbers and fractions of basal plane vs. edge sites for idealized graphite and graphene NPs, for a mass (52 MDa) typical of the NPs studied here. For idealized cube-shaped graphite, the edge length would be ~33.5 nm, and given the unit cell dimensions, just over 50% of the exposed surface atoms would be in edge sites. For graphene with the ~5 nm average thickness reported for our NP feedstock, and idealized square NP would have much larger surface area, but with only ~6% of the exposed surface atoms in edge sites. More importantly, the total *number* of exposed edge atoms in the graphene NP would be ~1/3rd that for the graphite NP. Therefore, if sublimation occurs primarily at exposed edge sites, we might expect graphene NPs to sublime with about 1/3rd the rate of equal mass graphite NPs.

In fact, however, the difference between the average sublimation rates for graphite and graphene is much smaller than the variations within the sets of graphite and graphene NPs. This is not surprising, because the NPs are not idealized structures with regular-shaped, defect-free basal planes. Both irregularities of the basal plane edges, and vacancies and other defects in the basal planes increase the number of low coordination sites, thus increasing the sublimation rates.

***Correlations between Emission Intensity and Sublimation Rates:*** **Figures 1**, **2**, **S4**, and **S5** show examples in which the integrated emission intensity, I, increased after heating, and in those examples, the sublimation rate, R, also slowed significantly. This raises the question of whether emissivity might be correlated with sublimation rate, and if so, what the correlation reveals about the factors that control graphitic NP emissivity. One way to extract such correlations would be to compare I and R for all the NPs before and after periods at high $T_{NP}$ where significant mass loss occurred. The difficulty is that our control over $T_{NP}$ was not adequate to hit exactly the same $T_{NP}$ values in different heat ramps and for different NPs. Therefore, the following approach was used. The rates (**figure 5**) and emission intensities (**figure 4**) measured in each heat ramp for each NP, were fit to determine how they varied with $T_{NP}$, using $I \propto T_{NP}^{s}$ and $R = A\cdot\exp(-E_a/kT_{NP})$, with s, A, and $E_a$ as fitting parameters. The fits were then used to interpolate between the I and R measurements, to estimate the values for $T_{NP}$ = 2000 K.

Because of the large NP-to-NP variation in both I and R, it is difficult to deduce much about correlations by comparing the I and R values directly, as shown in **figure S9**. The correlation is clearer if we compare *changes* in I and R for each NP after heating, because this cancels out the large NP-to-NP variations in absolute intensity and rate. **Figure 7** plots the percent change in I vs. the percent change in R for each NP, comparing the interpolated 2000 K I and R values measured in the 2$^{nd}$ and 3$^{rd}$ heat ramps relative to those measured in the 1$^{st}$. Filled symbols compare I and R values measured in the 2$^{nd}$ and 1$^{st}$ heat ramps, and open symbols compare the net change between the 3$^{rd}$ and 1$^{st}$ heat ramps, with circles and diamonds for graphite and graphene, respectively. Arrows connecting the symbols show how particular NPs evolved. There are a few unconnected points for NPs where only two heat ramps were completed.

It can be seen that most of the points lie in the upper left quadrant, *i.e.,* there is a reasonably strong anti-correlation between changes in sublimation rate and changes in emission intensity, the most common behavior being for intensities to increase while rates decrease. The tendency of the sublimation rates to decrease over time is attributed to evolution of the NP surface site distributions, i.e., to loss of low-coordination/defective sites, as discussed above. Therefore, **figure 7** suggests that loss of such sites increases the emission brightness of graphitic NPs. It is not surprising that the NP surface structure should affect emission properties, because defects and low coordination surface sites must affect the NP electronic and vibronic (phonon) state distributions. Understanding the structure-emissivity relationship quantitatively is not so

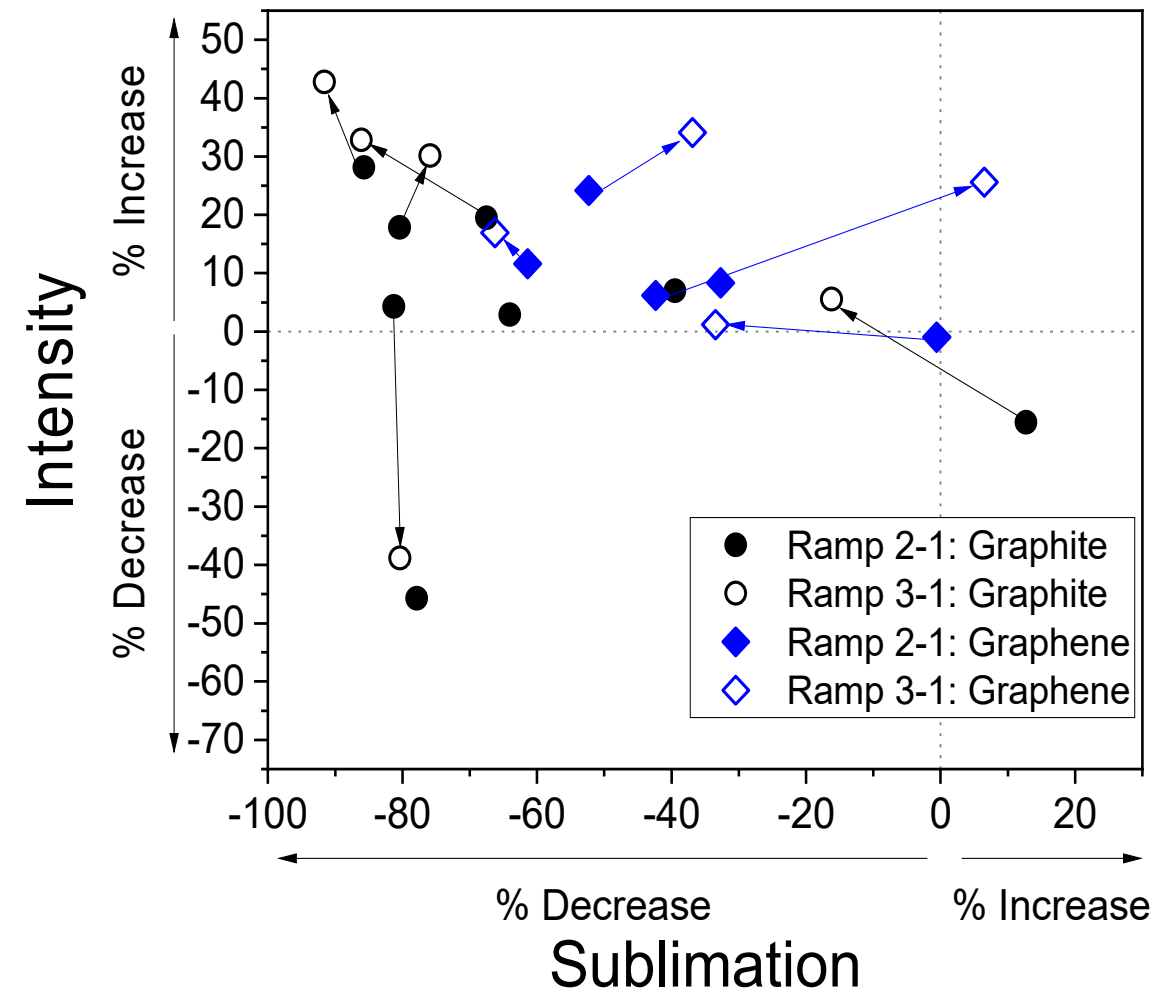

**Figure 7:** Correlations between heat-induced changes in emission intensity and sublimation rate at 2000 K, where each point represents a different NP. 'Ramp 2-1" means the percent change in intensity and rate between the 2$^{nd}$ and 1$^{st}$ heat ramps.

straightforward, and would appear to be an interesting topic for theoretical investigation.

## CONCLUSIONS

We have, for the first time, reported sublimation rates as a function of temperature for individual graphite and graphene NPs. The average rates *per* unit area were initially several orders of magnitude faster than that for bulk graphite but decreased significantly as sublimation was driven at high $T_{NP}$ over the course of several hours. There were, however, large NP-to-NP variations in the rates – much larger than the difference between the graphite and graphene rates, for example.

No significant correlations were observed of sublimation rates with NP mass or NP charge, both of which varied significantly as the NPs were repeatedly heated. Instead, the NP-to-NP variations in sublimation kinetics are attributed to the heterogeneity of the NPs and the resulting variations in the numbers of defective/under-coordinated surface sites. The evolution of the kinetics as the NPs were heated is, therefore, attributed to the evolution of the surface site distributions driven by preferential sublimation from less stable sites, and annealing.

The activation energies extracted from the rates increased substantially in repeated heat ramps, reflecting the evolution of the NP surface structure toward greater stability, however, the evolution during both repeated temperature cycling and in long term heating experi-ments was non-monotonic. The NPs approached a partial steady state in which their surfaces had relatively stable fractions of low coordination and defect sites, but individual NPs never reached a true steady state.

The thermal emission intensity underwent only partially reversible changes as the NP temperature was repeatedly cycled. Most commonly, the emission brightened after heating, while the sublimation rates decreased, suggesting that emission intensity is depressed by high coverage of defects and low coordination surface sites.

## ASSOCIATED CONTENT

**Supporting Information.**

The Supporting Information is available free of charge via the Internet at http://pubs.acs.org.

The following files are available free of charge:

**Figures S1-S9** show differences in equilibrium and no-equilibrium rates vs temperature, single NP emission spectra, oxidation as a function of time, example of a graphite NP experiment, long term heating experiment, prefactor (A) information, NP charge state vs. activation energy, and how normalized intensity varies for different NPs vs mass loss rate.

Data underlying **Figures 1-7** are provided in tabular form (XLSX).

## AUTHOR INFORMATION

**Corresponding Author**

*Scott L. Anderson, anderson@chem.utah.edu, (801)585-7289

**Author Contributions**

The manuscript was written through contributions of all authors. All authors have given approval to the final version of the manuscript.

**Funding Sources**

U.S. Department of Energy, Office of Science, Office of Basic Energy Sciences, under Award Number DE-SC- 0018049.

## ACKNOWLEDGMENT

The nIR spectrograph was purchased with funds from the Albaugh Scientific Equipment Endowment of the College of Science, University of Utah. This material is based upon work supported by the U.S. Department of Energy, Office of Science, Office of Basic Energy Sciences, under Award Number DE-SC- 0018049. We thank Dr. Bruce Kay (PNNL) and Prof. Timothy Minton (Montana State University) for helpful discussions.

## ABBREVIATIONS

NP, nanoparticle; $T_{NP}$, nanoparticle temperature;

## REFERENCES

1. Brewer, L.; Gilles, P. W.; Jenkins, F. A., The Vapor Pressure and Heat of Sublimation of Graphite. *J. Chem. Phys.* **1948,** *16*, 797-807.
2. Honig, R. E.; Kramer, D. A., *Vapor Pressure Data for the Solid and Liquid Elements*. RCA Laboratories, David Sarnoff Research Center: 1969.
3. Zavitsanos, P. D.; Carlson, G. A., Experimental study of the sublimation of graphite at high temperatures. *J. Chem. Phys.* **1973,** *59* (6), 2966-2973.
4. Chupka, W. A.; Berkowitz, J.; Meschi, D. J.; Tasman, H. A., Mass Spectrometric Studies of High Temperature Systems. In *Advances in Mass Spectrometry*, Elliott, R. M., Ed. Pergamon: 1963; pp 99-109.
5. Drowart, J.; Burns, R. P.; DeMaria, G.; Inghram, M. G., Mass Spectrometric Study of Carbon Vapor. *J. Chem. Phys.* **1959,** *31* (4), 1131-1132.
6. Chupka, W. A.; Inghram, M. G., Molecular Species Evaporating from a Carbon Surface. *J. Chem. Phys.* **1953,** *21* (7), 1313-1313.
7. Steele, W. C.; Bourgelas, F. N. *Studies of Graphite Vaporization Using a Modulated Beam Mass Spectrometer, Technical report AFML-TR-72-222, AVSD-0048-73-CR*; Air Force Materials Laboratory, Air Force Systems Command: Wright-Patterson Air Force Base, Ohio, 1972.
8. Doehaerd, T.; Goldfinger, P.; Waelbroeck, F., Direct Determination of the Sublimation Energy of Carbon. *J. Chem. Phys.* **1952,** *20*, 757.
9. Clarke, J. T.; Fox, B. R., Rate and Heat of Vaporization of Graphite above 3000°K. *J. Chem. Phys.* **1969,** *51* (8), 3231-3240.
10. Marshall, A. L.; Norton, F. J., Carbon Vapor Pressure and Heat of Vaporization. *J. Am. Chem. Soc.* **1950,** *72* (5), 2166-2171.
11. Tsai, C. C.; Gabriel, T. A.; Haines, J. R.; Rasmussen, D. A., Graphite Sublimation Tests For Target Development For The Muon Collider/Neutrino Factory *IEEE Symp. Fusion Eng.* **2006,** *21*, 377-379.
12. Haines, J. R.; Tsai, C. C. *Graphite Sublimation Tests for the Muon Collider/Neutrino Factory Target Development Program*; Oak Ridge, 2002; p 6 pages.
13. Pflieger, R.; Sheindlin, M.; Colle, J.-Y., Advances in the mass spectrometric study of the laser vaporization of graphite. *J. Appl. Phys.* **2008,** *104* (5), 054902.
14. Long, B. A.; Rodriguez, D. J.; Lau, C. Y.; Schultz, M.; Anderson, S. L., Thermal Emission Spectroscopy of Single, Isolated Carbon Nanoparticles: Effects of Particle Size, Material, Charge, Excitation Wavelength, and Thermal History. *J. Phys. Chem. C* **2020,** *124*, 1704-1716.
15. Howder, C. R.; Long, B. A.; Alley, R. N.; Gerlich, D.; Anderson, S. L., Single Nanoparticle Mass Spectrometry as a High Temperature Kinetics Tool: Emission Spectra, Sublimation, and Oxidation of Hot Carbon Nanoparticles. *J. Phys. Chem. A* **2015,** *119*, 12538–12550.
16. Long, B. A.; Rodriguez, D. J.; Lau, C. Y.; Anderson, S. L., Thermal emission spectroscopy for single nanoparticle temperature measurement: optical system design and calibration. *Appl. Opt.* **2019,** *58* (3), 642-649.
17. Gerlich, D.; Decker, S., Trapping Ions at High Temperatures: Thermal Decay of $C_{60}^{+}$. *Appl. Phys. B: Lasers Opt.* **2014,** *114*, 257-266.
18. Rossman, R. P.; Smith, W. R., Density of Carbon Black by Helium Displacement. *Ind. Eng. Chem. Res.* **1943,** *35* (9), 972-976.
19. Bohren, C. F.; Huffman, D. R., *Absorption and Scattering of Light by Small Particles*. Wiley: New York, 1983.
20. Siddall, R. G.; McGrath, I. A., The emissivity of luminous flames. *Symp. (Int.) Combust., [Proc.]* **1963,** *9* (1), 102-110.
21. Köylü, Ü. Ö.; Faeth, G. M., Optical Properties of Overfire Soot in Buoyant Turbulent Diffusion Flames at Long Residence Times. *J. Heat Transfer* **1994,** *116* (1), 152-159.
22. Landstrom, L.; Elihn, K.; Boman, M.; Granqvist, C. G.; Heszler, P., Analysis of ThermalRradiation fromLlaser-Heated Nanoparticles Formed by Laser-Induced Decomposition of Ferrocene. *Appl. Phys. A* **2005,** *81*, 827-833.
23. Landstrom, L.; Heszler, P., Analysis of Blackbody-Like Radiation from Laser-Heated Gas-Phase Tungsten Nanoparticles. *J. Phys. Chem. B* **2004,** *108*, 6216-6221.

24. Michelsen, H. A., Understanding and predicting the temporal response of laser-induced incandescence from carbonaceous particles. *J. Chem. Phys.* **2003,** *118*, 2012-2045.
25. Kaden, W. E.; Kunkel, W. A.; Roberts, F. S.; Kane, M.; Anderson, S. L., CO Adsorption and Desorption on Size-selected $Pd_n/TiO_2(110)$ Model Catalysts: Size Dependence of Binding Sites and Energies, and Support-mediated Adsorption. *J. Chem. Phys.* **2012,** *136*, 204705/1-204705/12.
26. Kolasinski, K. W., *Surface Science. Foundations of Catalysis and Nanoscience*. 3rd ed.; Wiley: Chichester, 2012.
27. Zobelli, A.; Ivanovskaya, V.; Wagner, P.; Suarez-Martinez, I.; Yaya, A.; Ewels, C. P., A comparative study of density functional and density functional tight binding calculations of defects in graphene. *Phys. Stat. Sol. B* **2012,** *249* (2), 276-282.
28. Thrower, P. A.; Mayer, R. M., Point defects and self-diffusion in graphite. *Phys. Stat. Sol. A* **1978,** *47*, 11-37.
29. Kaxiras, E.; Pandey, K. C., Energetics of defects and diffusion mechanisms in graphite. *Phys. Rev. Letts.* **1988,** *61* (23), 2693-2696.
30. El-Barbary, A. A.; Telling, R. H.; Ewels, C. P.; Heggie, M. I.; Briddon, P. R., Structure and energetics of the vacancy in graphite. *Phys. Rev. B* **2003,** *68* (14), 144107.
31. Li, L.; Reich, S.; Robertson, J., Defect energies of graphite: Density-functional calculations. *Physical Review B* **2005,** *72* (18), 184109.
32. Hahn, J. R.; Kang, H., Vacancy and interstitial defects at graphite surfaces: Scanning tunneling microscopic study of the structure, electronic property, and yield for ion-induced defect creation. *Phys. Rev. B* **1999,** *60* (8), 6007-6017.
33. Banhart, F.; Kotakoski, J.; Krasheninnikov, A. V., Structural Defects in Graphene. *ACS Nano* **2011,** *5* (1), 26-41.
34. Ivanovskii, A. L., Graphene-based and graphene-like materials. *Russ. Chem. Rev.* **2012,** *81* (7), 571-605.
35. Yamashita, K.; Saito, M.; Oda, T., Atomic Geometry and Stability of Mono-, Di-, and Trivacancies in Graphene. *Jpn. J. Appl. Phys.* **2006,** *45* (8A), 6534-6536.
36. Lehtinen, P. O.; Foster, A. S.; Ayuela, A.; Krasheninnikov, A.; Nordlund, K.; Nieminen, R. M., Magnetic Properties and Diffusion of Adatoms on a Graphene Sheet. *Phys. Rev. Letts.* **2003,** *91* (1), 017202.
37. Krasheninnikov, A. V.; Lehtinen, P. O.; Foster, A. S.; Nieminen, R. M., Bending the rules: Contrasting vacancy energetics and migration in graphite and carbon nanotubes. *Chem. Phys. Lett.* **2006,** *418*, 132-136.
38. Skowron, S. T.; Lebedeva, I. V.; Popov, A. M.; Bichoutskaia, E., Energetics of atomic scale structure changes in graphene. *Chem. Soc. Rev.* **2015,** *44*, 3143-3176.
39. Suarez-Martinez, I.; Savini, G.; Haffenden, G.; Campanera, J.-M.; Heggie, M. I., Dislocations of Burgers vector c /2 in graphite. *Phys. Stat. Sol. C* **2007,** *4* (8), 2958-2962.
40. Li, J.; Li, Z.; Zhou, G.; Liu, Z.; Wu, J.; Gu, B.-L.; Ihm, J.; Duan, W., Spontaneous edge-defect formation and defect-induced conductance suppression in graphene nanoribbons. *Phys. Rev. B* **2010,** *82* (11), 115410.
41. Chang, H.; Bard, A. J., Formation of monolayer pits of controlled nanometer size on highly oriented pyrolytic graphite by gasification reactions as studied by scanning tunneling microscopy. *J. Am. Chem. Soc.* **1990,** *112* (11), 4598-9.
42. McCarroll, B.; McKee, D. W., Reactivity of graphite surfaces with atoms and molecules of hydrogen, oxygen, and nitrogen. *Carbon* **1971,** *9* (3), 301-311.
43. Myers, G. E.; Gordon, M. D., Anomalous hexagonal etch pits produced in the graphite-water vapor reaction. *Carbon (Oxford)* **1968,** *6* (3), 422-3.
44. Patrick, D. L.; Cee, V. J.; Beebe, T. P., Jr., "Molecule corrals" for studies of monolayer organic films. *Science (Washington, D. C.)* **1994,** *265* (5169), 231-4.
45. Hahn, J. R.; Kang, H., Conversion efficiency of graphite atomic-scale defects to etched pits in thermal oxidation reaction. *J. Vac. Sci. Technol., A* **1999,** *17* (4, Pt. 1), 1606-1609.
46. Hahn, J. R.; Kang, H.; Lee, S. M.; Lee, Y. H., Mechanistic Study of Defect-Induced Oxidation of Graphite. *J. Phys. Chem. B* **1999,** *103* (45), 9944-9951.
47. Turnbull, J. A.; Stagg, M. S., Isothermal annealing studies on vacancy and interstitial loops in single crystal graphite. *Philos. Mag.* **1966,** *14* (131), 1049-66.
48. Zhang, H.; Zhao, M.; Yang, X.; Xia, H.; Liu, X.; Xia, Y., Diffusion and coalescence of vacancies and interstitials in graphite: A first-principles study. *Diamond Relat. Mater.* **2010,** *19* (10), 1240-1244.

TOC figure

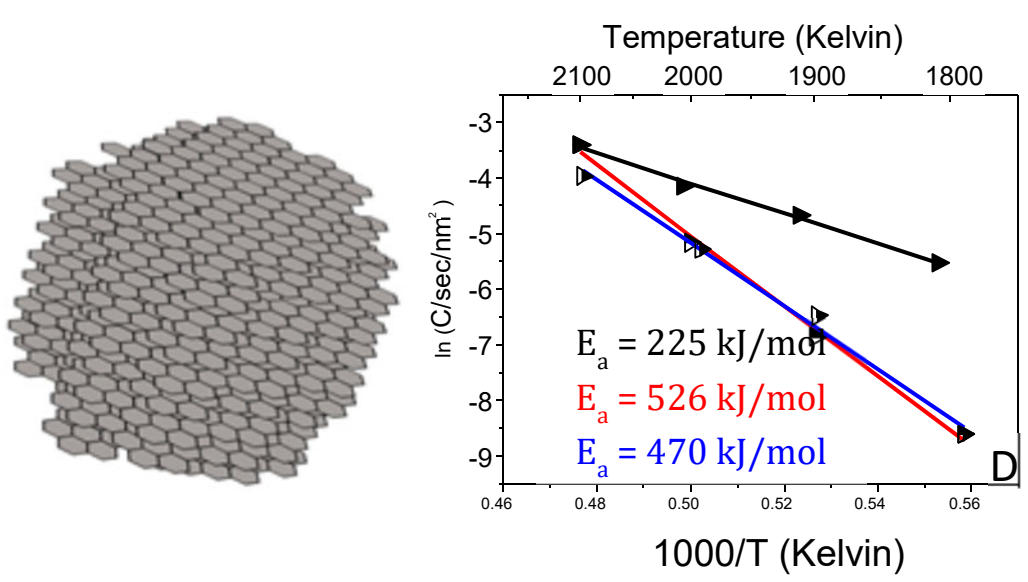
Temperature (Kelvin)
2100
2000
1900
1800
-3
-4
-5
-6
-7
-8
-9
ln (C/sec/nm²)
Ea = 225 kJ/mol
Ea = 526 kJ/mol
Ea = 470 kJ/mol
D
0.46
0.48
0.50
0.52
0.54
0.56
1000/T (Kelvin)

# Sublimation Kinetics for Individual Graphite and Graphene Nanoparticles (NPs): NP-to-NP Variations and Evolving Structure-Kinetics and Structure-Emissivity Relationships

Bryan A. Long, Chris Y. Lau, Daniel J. Rodriguez, Susanna An Tang, and Scott L. Anderson*

**Department of Chemistry, University of Utah, 315 S. 1400 E., Salt Lake City, Utah 84112, USA**
***Corresponding author: anderson@chem.utah.edu**

## Supporting Information

### Previous work toward the goal of single NP reaction kinetics:

Our single NP kinetics technique is based on earlier NP mass spectrometry and fluorescence spectroscopy work by several groups,[1-6] which laid a foundation of methods for measuring NP mass and charge using optical detection of the NP secular (motional) frequencies. Some experiments have also examined the kinetics for mass gain or loss, although until now, none could directly measure the NP temperature or vary it quantitatively. For example, in 2004, Schlemmer et al. measured the kinetics for mass gain by a ~500 nm diameter silica NP, as molecules from the vacuum system adsorbed on the NP surface at near room temperature. Recently Esser *et al.*[9] reported a new instrument and measurements of argon adsorption/desorption for silica NPs at temperatures below 40 K, but without direct temperature measurement.

Our group developed an electrospray ionization (ESI) NP source to introduce CdSe/ZnS core/shell NPs to a trap, detecting them by laser-induced luminescence, and used this instrument to study changes in the luminescence behavior as the NP lost mass by slow sublimation.[12-13] We also previously examined sublimation of hot carbon NPs,[15] however, in that experiment, as with all previous single NP kinetics work, the NP temperature was unknown, so that rates could not be associated with $T_{NP}$ to extract energetic information. Temperature determination required development of an optical system and calibration method capable of measuring visible and near IR spectra for single NPs with quantitative intensity information, and we previously reported the optical system design, as well as the steps necessary to convert raw spectra to temperatures.[16-17] The present report is the first in which single NP kinetic measurements have been done at well-defined and variable temperature.

**Bulk graphite sublimation:**

**Figure S1** presents measured graphite sublimation rates and a fit to an Arrhenius rate law, $R=A\exp(-E_a/kT)$, used to extract an $E_a$ of 862 kJ/mol, and to extrapolate the rate to the $T_{NP}$ range of our experiments. For materials where the sticking coefficients (S) are unity, sublimation rates can be calculated from the equilibrium vapor pressures, and the figure also shows the rates that would be obtained if this assumption is erroneously used in conjunction with measured equilibrium vapor pressures.[14] In fact, however, graphite has S values for various $C_n$

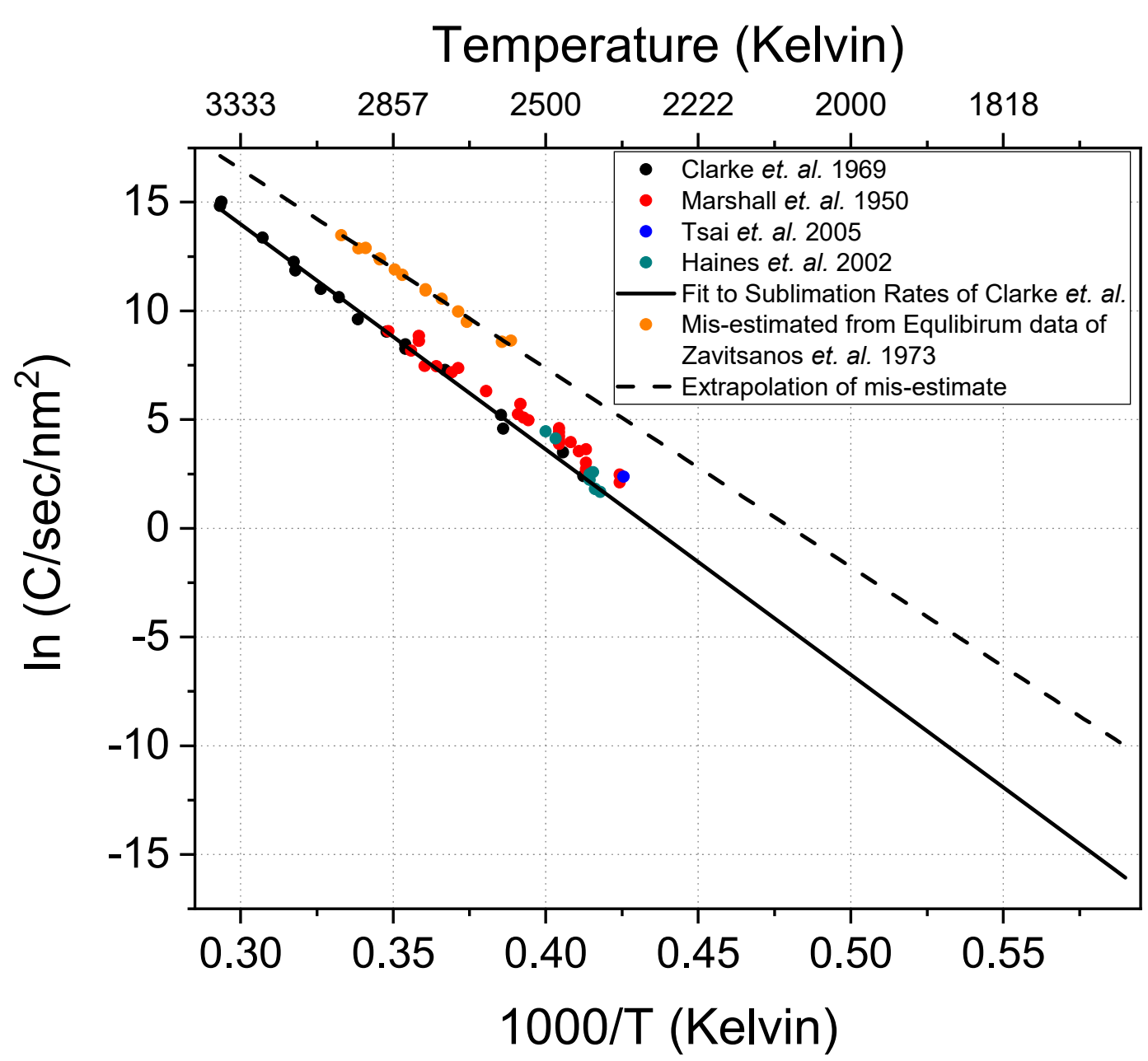

**Figure S1:** Measurements of the sublimation rates for graphite in vacuum by Clarke *et al.*[7], Marshall *et al.*,[8] Tsai *et al.*,[10] and Haines *et al.*,[11] The solid line is an Arrhenius fit to the Clarke data. The orange points and dashed fit line show the sublimation rate that would *incorrectly* be estimated if equilibrium vapor pressure data (in this case, by Zavitsanos *et al.*[14]) are analyzed using the common, but in this case, erroneous assumption that the $C_n$ sticking coefficient is unity with hot graphite surfaces. Note, Zavitsanos *et al.* did not make this error – the mis-estimate of the sublimation rate is included only to make the point that graphite sublimation is quite different from that typical for metals.

vapor species that temperature dependent and well below unity.[18-20] For example, S(C) was found to vary from 0.7 ± 0.15 at $T_{surface}$ = 800 K, to 0.4 ± 0.2 at 2300 K,[18] and S($C_3$) was found to be zero within experimental uncertainty, for surface temperatures above 1000 K.[18, 20]

**Typical Carbon NP spectra:**

A few typical emission spectra are shown for a single laser-heated graphite NP (black points), along with fits (colored curves) using the model thermal emission function given in equation 4 in the main text. The $T_{NP}$ and n parameters extracted from the fits are given in the figure legend.

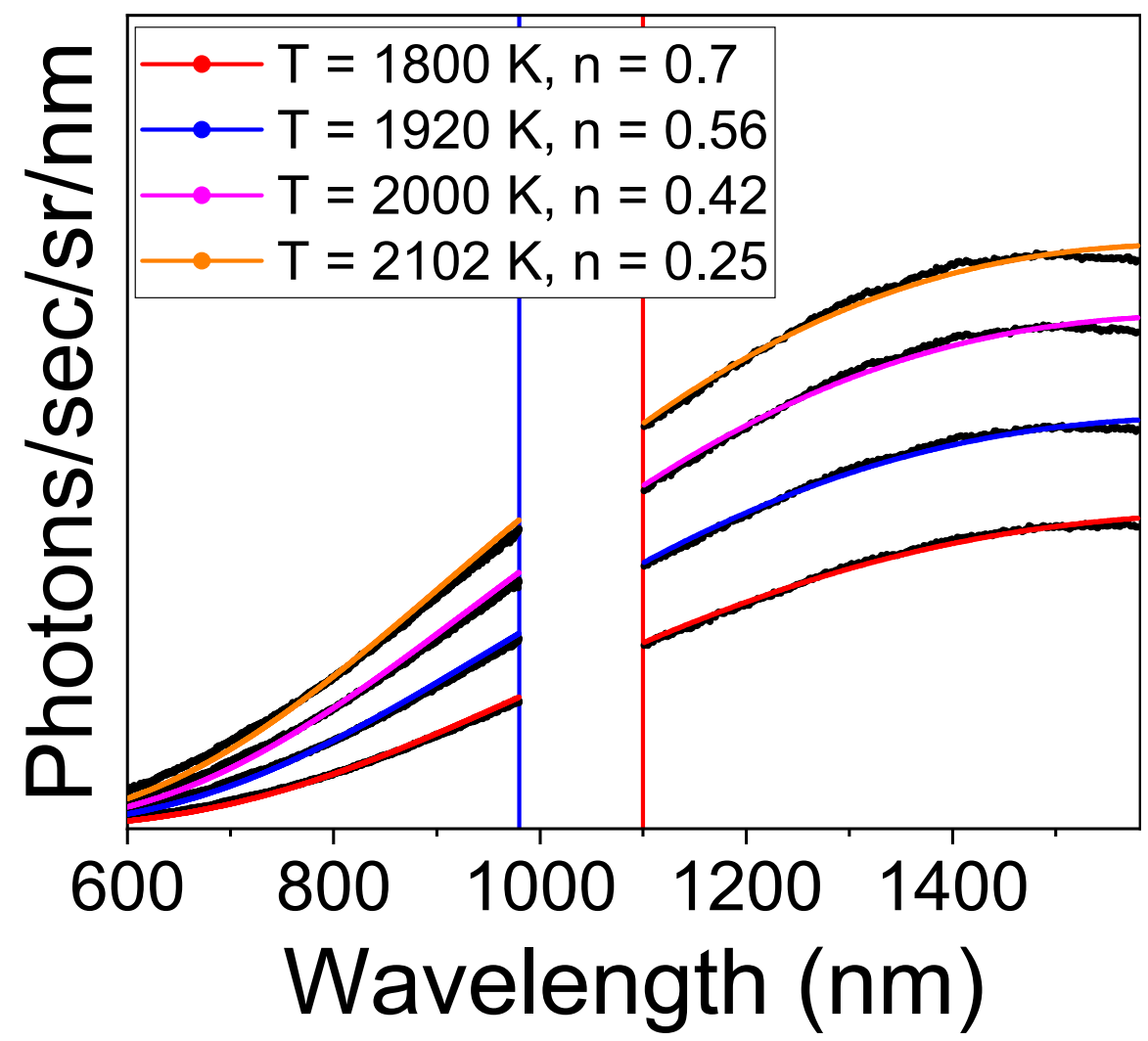

**Figure S2:** Emission spectra for a single graphite NP. $T_{NP}$ are extracted by fitting the raw emission intensity using a Planckian fitting function. The emissivity parameter n is also shown for each $T_{NP}$ value.

**Graphite and Graphene NP purity in the experiments:**

The NPMS method only monitors the total NP mass, thus it is important that we work with materials with the highest possible purity, with particular concern regarding any contaminants (*e.g.,* metals) that might survive at high temperatures, and catalyze sublimation or other reactions. The entire ESI source is carefully cleaned whenever the NP material is changed, with parts exposed to the ESI solutions extensively cleaned or replaced. The graphite NPs examined here have stated purity of 99.9999% (metals basis, Alfa Aesar), and the graphene platelet material (US Research Nanomaterials, Inc.) has stated composition of 99.7% C, <0.3% O. As received, the NPs presumably have oxidized surface groups and adventitious adsorbates from air exposure. Electrospray ionization was done from 2 mM ammonium acetate solutions in methanol, thus the electrosprayed NPs likely also have excess ammonium acetate and methanol adsorbed. Most stable molecules, like water or methanol, desorb from graphite surfaces at low temperatures, and ammonium acetate was chosen specifically because it decomposes to volatile products well below the $T_{NP}$ range of interest here.[21] Similarly, oxidized surface groups have been shown to desorb as CO or $CO_2$ below ~1400 K when heated in vacuum,[22-25] and adventitious organic contaminants would be expected to pyrolyze to carbon plus gaseous products. Trapped NPs are initially held at $T_{NP} \leq 1300$ K during charge stepping (20 – 60 minutes), then heated over the course of a few minutes to 1800 K for the start of the sublimation experiments. We expect, therefore, both contaminants and heteroatom functional groups on the NP surfaces should have desorbed, i.e., the NPs should be quite pure.

**Possible effects of background oxygen on the sublimation rates:**

Another consideration is that the presence of $O_2$ or other oxidizers in the trap chamber background could lead to NP mass change from reactions, interfering with measuring sublimation rates. UHP Ar (99.999%) at pressures of ~1 mTorr is used as a buffer gas to damp NP motion during normal operations, with UHV-compatible gas lines and fittings. The trap chamber is also mostly UHV-compatible but because it is exposed to methanol from the ESI source during every NP injection, its base pressure is typically ~7 x $10^{-7}$ Torr. To test for the effects of small additions of $O_2$, experiments were done in which a separate flow controller was used to add $O_2$ to the Argon buffer gas at pressures ~10,000 times higher than the worst-case $O_2$ background pressure (i.e., assuming the background is air).

One such experiment is summarized in **Figure S3**, which tests of the effects of deliberately adding $O_2$ on the mass-loss rates. The mass-loss rates were measured without and with 0.18 mTorr of $O_2$ added to the argon buffer gas in the trap. From the difference in mass loss rates, the rate of oxidative mass loss can be determined. As has been observed previously[26-27] the rate for graphite oxidation by $O_2$ decreases with increasing surface temperature in our $T_{NP}$ range. As a result, in the worst case, where the $O_2$ pressure would have been ~$10^{-4}$ of that used in **figure S3**, background oxidation would have had < 5% effect on the sublimation rates at our lowest temperatures, decreasing with increasing $T_{NP}$ to < 0.1% above ~1950 K.

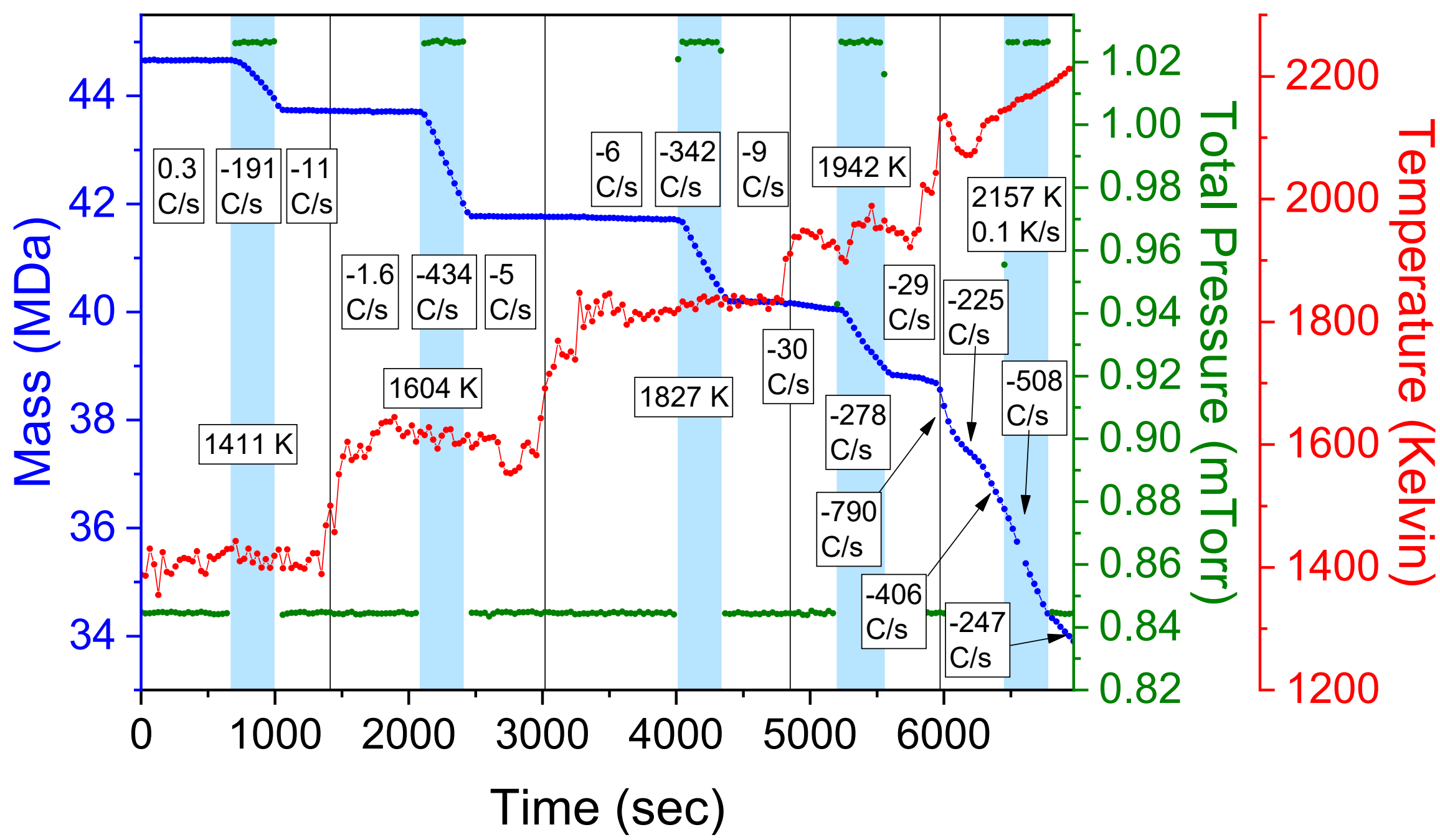

**Figure S3:** A 44.66 MDa graphite NP was heated to ~5 different temperatures, measuring the mass loss rate, with and without added $O_2$.

**Typical experiment for a single graphite NP:**

**Figure S4** shows an experiment analogous to that shown in **Figure 1** of the main text. A graphite NP with initial mass 38.9 MDa (~38 nm diameter, if spherical), was heated in a series of steps, from ~1800 K to ~2100 K. The initial period, during which the NP was held at ~1300 K for Q determination is not shown. Gaps in the data indicate periods when $T_{NP}$ was unstable, and these periods were omitted from further analysis. $T_{NP}$ values and mass loss rates, averaged over the roughly constant $T_{NP}$ periods defined by the solid vertical lines, are given in boxes. In addition to M and $T_{NP}$, the figure shows the integrated emission intensity (emitted photons/second/sr over the 600 – 1600 nm range) and the value of the n parameter in the power law emissivity function. Note that n decreased as $T_{NP}$ increased, and then increased again as the NP cooled, indicating that emissivity is temperature dependent. Similarly, the integrated emission intensity, I, increased by a factor of ~2 as $T_{NP}$ increased by a factor of ~1.16 (~1800 to ~2100 K), corresponding to $I \propto T_{NP}^{\sim 5}$.

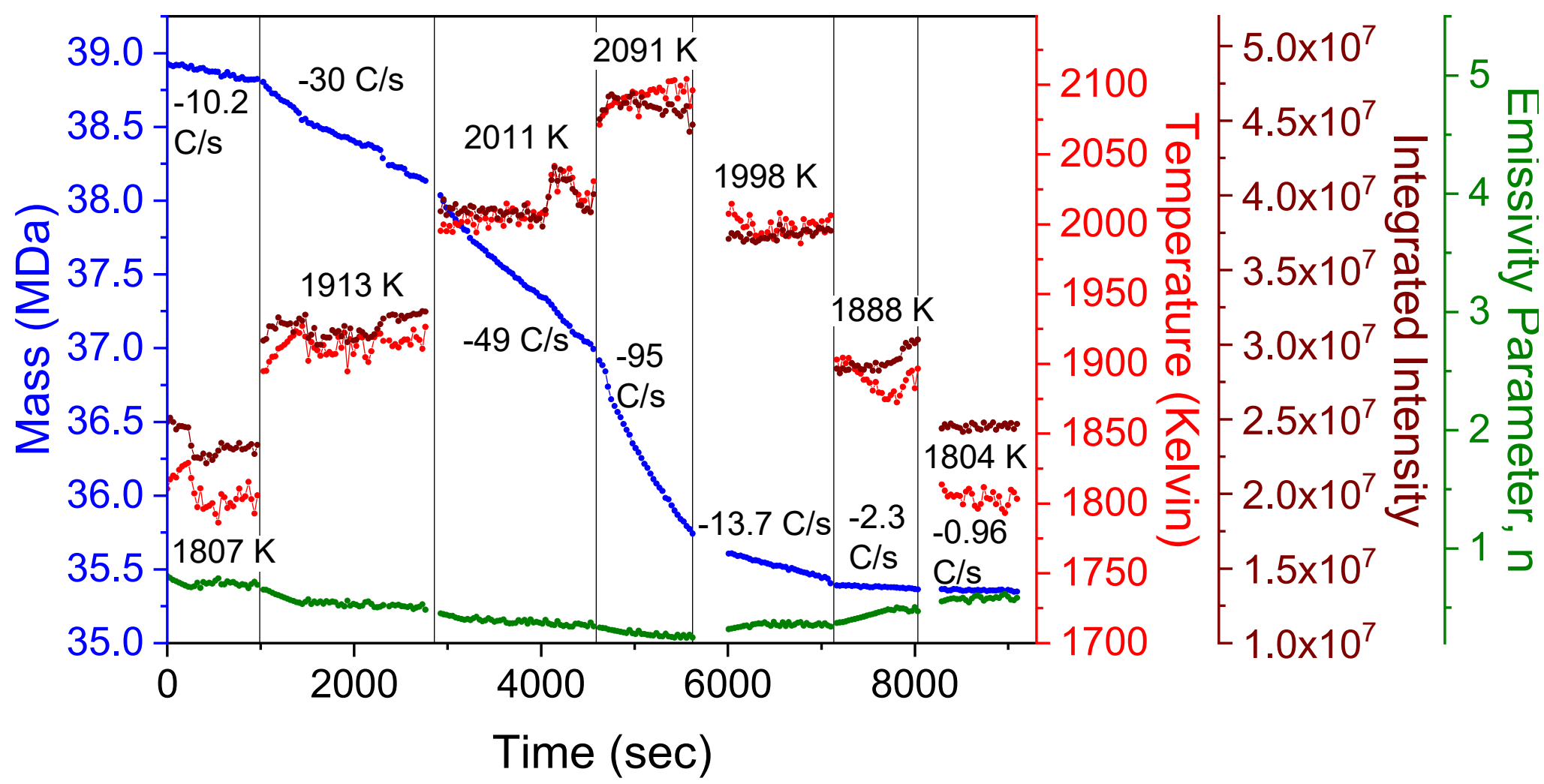

**Figure S4:** Typical sublimation kinetics experiment, in which a single graphite NP of initial mass of 38.93 MDa was heated in step-wise fashion from ~1800 K to ~2100 K, then back down to 1800 K. The temperatures and sublimation rates (given as C atoms/sec) indicated in boxes are the values averaged over the time intervals defined by the vertical black lines.

The sublimation rate increased with increasing $T_{NP}$, as expected, however, the rates at similar $T_{NP}$ values measured during the heating and cooling phases of the experiment were significantly different. For example, at ~1807 K at the beginning of the experiment, the NP sublimed at a rate of 10.2 C/sec, while during the period with $T_{NP} \approx 1804$ K at the end of the experiment, the rate was only 0.96 C/sec. Decreases in sublimation rate were also observed for other temperatures, although in those cases the $T_{NP}$ values didn't match as well during the heating and cooling phases, complicating the comparison. Note also that during the period at ~2091 K, there was a slow increase in $T_{NP}$. Nonetheless, the sublimation rate slowed significantly. As in the example in **figure 1**, As NPs sublime, some rate slowing is expected due to reduction in NP surface area, however, the mass loss over the entire experiment was only ~9%, thus the surface area loss ($\propto \sim M^{2/3}$) would have been on the order of 6 %. Clearly, this small surface area loss cannot account for rates slowing by up to an order of magnitude. **Figure S5** summarizes the results, as in **figure 2** of the main text. **Figure S5A** shows the variation of the integrated emission intensity with $T_{NP}$. Ideal blackbody emission intensity scales with emitter surface area and the emissivity for small NPs (equation 3) introduces another factor of r, thus intensity should scale with $r^3$. Therefore, to compensate for the expected effects of decreasing NP mass during the experiment (and to allow comparisons to NPs with different masses), the intensities have been scaled by the NP mass ($\propto r^3$) measured during the spectral measurements. Note that the intensity increased and decreased with $T_{NP}$, but the final intensity at 1800 K was ~15% higher than the intensity measured at the same $T_{NP}$ prior to the heat ramps. Horizontal Error bars based on the ±2% uncertainty in $T_{NP}$ are plotted with each data point, as explained in the main text. The uncertainties in relative emission intensity are negligible.

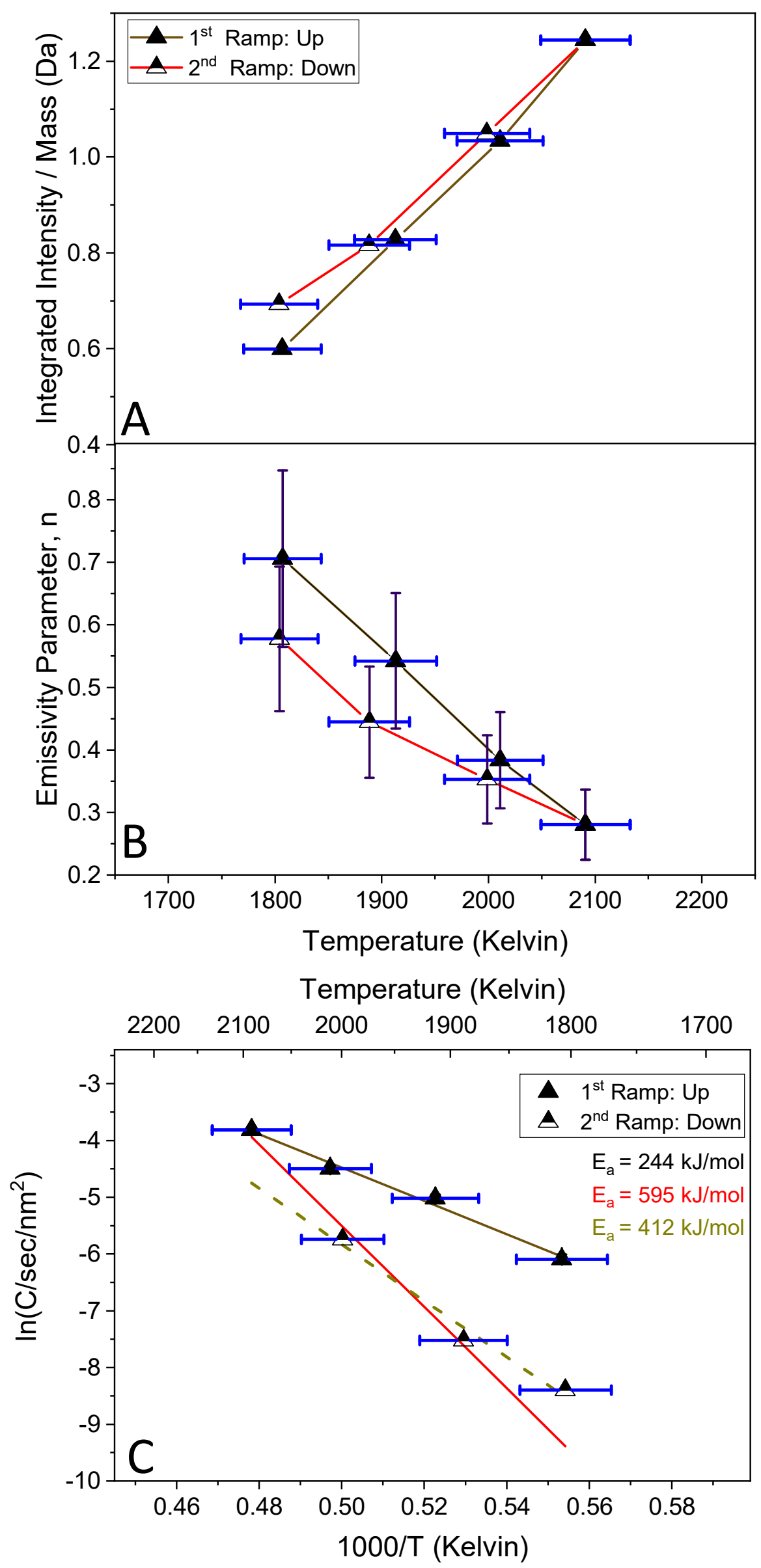

**Figure S5:** Analysis of the experiment shown in **figure 1**. A: Integrated emission intensity averaged over each constant $T_{NP}$ period along with fits used to extract $T_{NP}$ (colored curves).. B: Emissivity parameter, n, averaged over each constant $T_{NP}$ interval. C: Area-normalized sublimation rate vs. $T_{NP}$, along with fits to extract the effective $E_a$ values shown. For the 2nd ramp, fits are shown to the highest and lowest $T_{NP}$ points.

**Figure S5B** plots the n parameter, which decreased with increasing $T_{NP}$, then recovered as the $T_{NP}$ dropped, but only to ~80% of its initial value. These changes in intensity and n parameter indicate that in addition to driving sublimation, heating caused irreversible changes in the NP emissivity. The estimated ±20% uncertainties in n are shown as vertical error bars

**Figure S2C** plots the sublimation rates measured at each $T_{NP}$ as the NP was heated then cooled. The rates and $T_{NP}$ values are simply the rates and temperatures averaged over the constant $T_{NP}$ intervals indicated by the vertical lines in **figure S1**. To correct for the ~6% surface area loss during the experiment, and to allow comparisons with other NPs, we would ideally normalize rates to the NP surface areas, but these are unknown. We have approximately area-normalized the rates assuming that the NPs are spherical with the bulk density, thus obtaining upper limits on the area-normalized rates, given as C/sec/nm$^2$. The data for both the 1$^{st}$ ramp up and 2$^{nd}$ ramp down are reasonably linear when plotted as ln(C/sec/nm$^2$) vs. 1000/T, thus the fit lines can be used to extract activation energies, $E_a$, and prefactors, A, in rate expressions of the type: $R = A \exp(-E_a/RT_{NP})$. For this data, the $E_a$ value extracted from 1$^{st}$ ramp up was 244 kJ/mol. During the 2$^{nd}$ ramp down, the rates are not fit well by a single $E_a$ value. The three highest $T_{NP}$ points are reasonably well fit (red line) to give $E_a$ = 595 kJ/mol, and the three lowest $T_{NP}$ points (dashed line) give 412 kJ/mol. In either case, compared to the 1$^{st}$ ramp up, the $E_a$ values are substantially higher, and the rates at low $T_{NP}$ substantially lower, during the 2$^{nd}$ ramp down. Error bars indicate the ±2% uncertainty in $T_{NP}$. The ~±1.3% uncertainty in the mass loss rate is small compared to the size of the symbols.

**Another constant $T_{NP}$ sublimation experiment:**

**Figure S6** shows an experiment in which a graphite NP (initial mass 33.7 MDa, ~36.1 nm if spherical) was heated, initially to 1900 K for ~2 hours, then increased to 2000 K to drive faster mass loss. In this experiment, heating was initially done with just the 532 nm laser, but when $T_{NP}$ was increased to 2000 K, a cw $CO_2$ laser was used to help heat the NP. The $CO_2$ laser is less stable, and decreased the stability of $T_{NP}$, which, nonetheless, only fluctuated by ±40 K, (±2%) with standard deviation of 17 K (<1%). The total mass loss was ~23%.

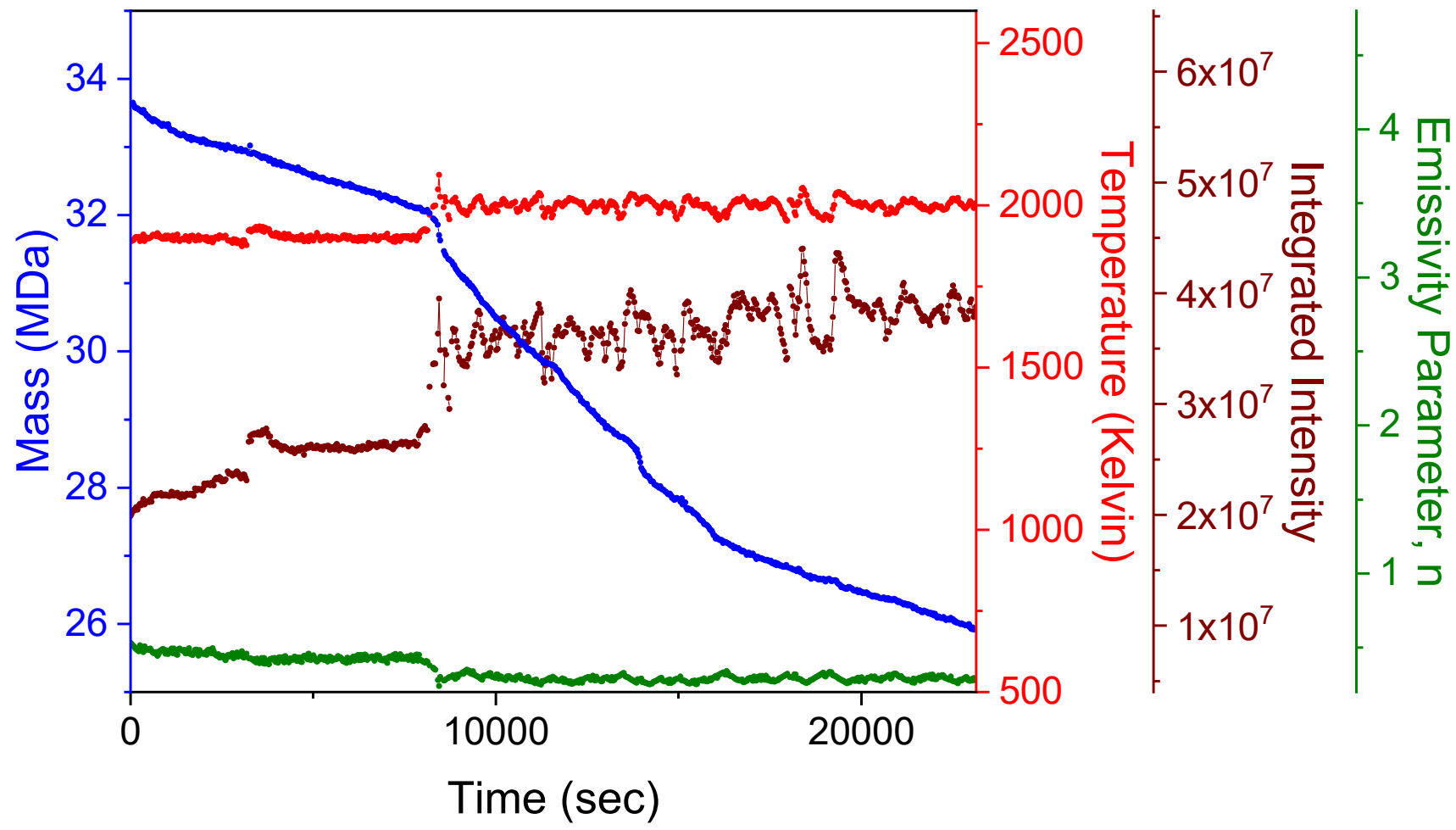

**Figure S6:** Long exposure heating experiments. A graphite NP with initial mass 33.7 MDa initially heated at ~1900 K ~2 hours, then at 2000 K for an additional ~4.4 hours.

During the 2000 K period, the sublimation rate slowed by ~67%, from an average of 51 C/sec over the first 10 minutes to 17 C/sec over the final 10 minutes, whereas the surface area (assuming spherical shape) would have decreased by only ~13%. Note also that there were significant fluctuations in the sublimation rate (*e.g.,* at ~14,000 sec) that are not correlated with fluctuations in either $T_{NP}$ or the emission intensity. The emission intensity generally tracked the fluctuations in $T_{NP}$, as expected, but the intensity also slowly increased during each constant $T_{NP}$ period. For this NP, the n parameter decreased with time at 1900 K but was roughly constant at 2000 K.

**A factors extracted from fitting rates in Figure 5 of the main text.**

The A factors extracted by fitting the rates shown in **figure 5** of the main text are shown in **figure S7.** The logs of the A factors are correlated with the $E_a$ values, as can be seen by comparing to **figure 6** of the main text. The A factors tended to increase substantially from heat ramp to heat ramp, but remained well below the A factor value extracted by fitting the bulk graphite sublimation rate data shown in **figure S1.**

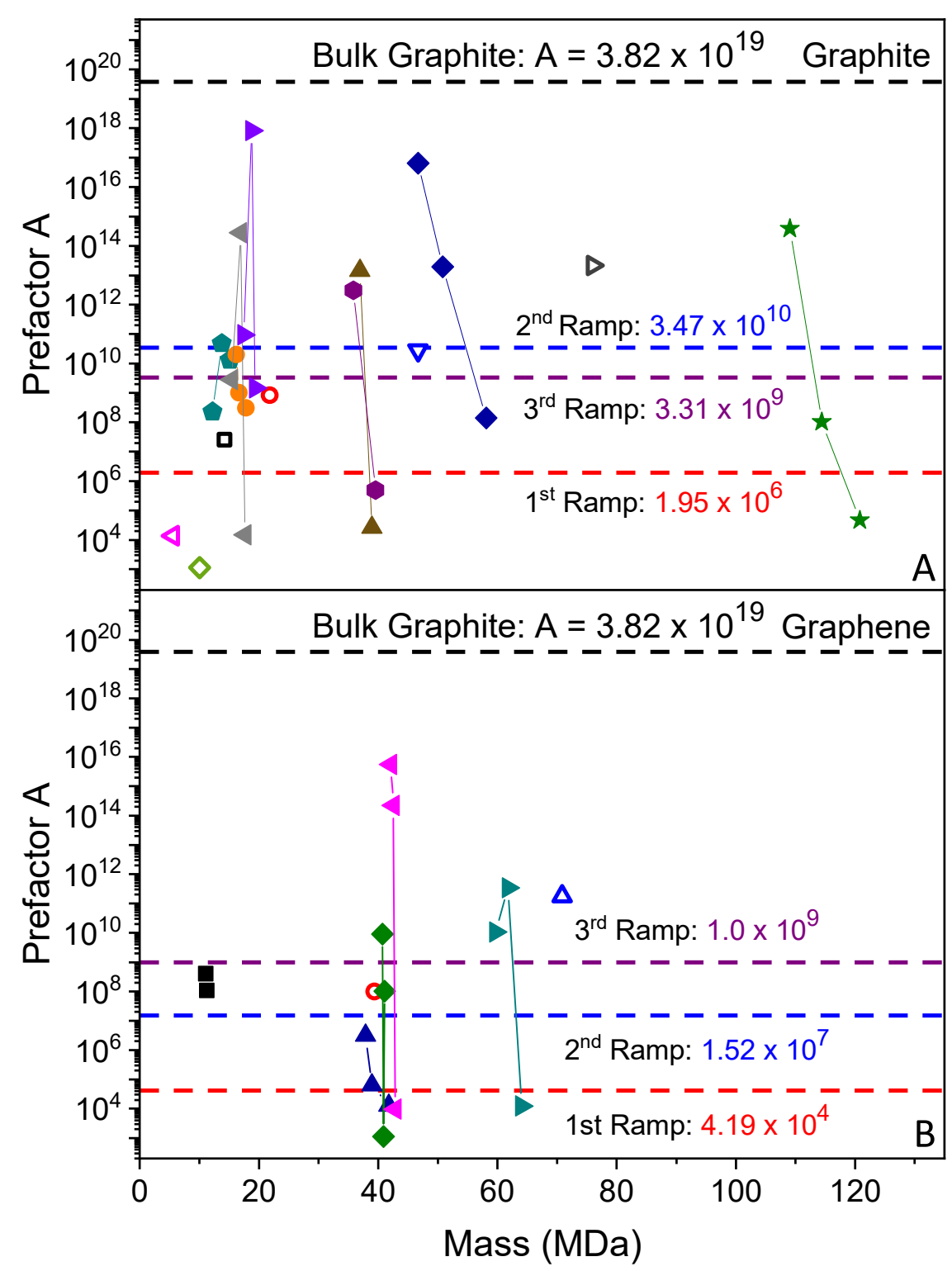

**Figure S7:** Prefactor, A, values for the graphite and graphene NP fits found in **figure 6** in. Also shown in all figures are the average values for the NPs and the values obtained by fitting bulk graphite data.

**Relation between NP charge (Q) and kinetics.**

**Figure S8** shows the $E_a$ values for all the graphite NPs plotted vs. the Q value at the beginning of each heat ramp. The Q values generally increased, particularly between the 1st and 2nd heat ramps, due to thermionic emission at high $T_{NP}$. The inherent effects of Q can be judged by comparing the $E_a$ values measured for each of the heat ramps across the ensemble of NPs. As with the situation for NP mass, the large variations in Q across the ensemble are clearly not correlated to the kinetics. For individual NPs, Q tends to increase in successive heat ramps, and $E_a$ also increases, however, we believe these are simply the consequences of two unrelated processes that occur at high $T_{NP}$ – thermionic emission that increases Q, and sublimation that changes the surface site distribution, thus changing the sublimation kinetics. The effects of Q on the actual rates can be observed by comparing **figure 5** and **figure S8**, which allows the Q value for each NP in **figure 5** to be identified, because the symbol colors and shapes are consistent. Again, there is no correlation between Q and the rates.

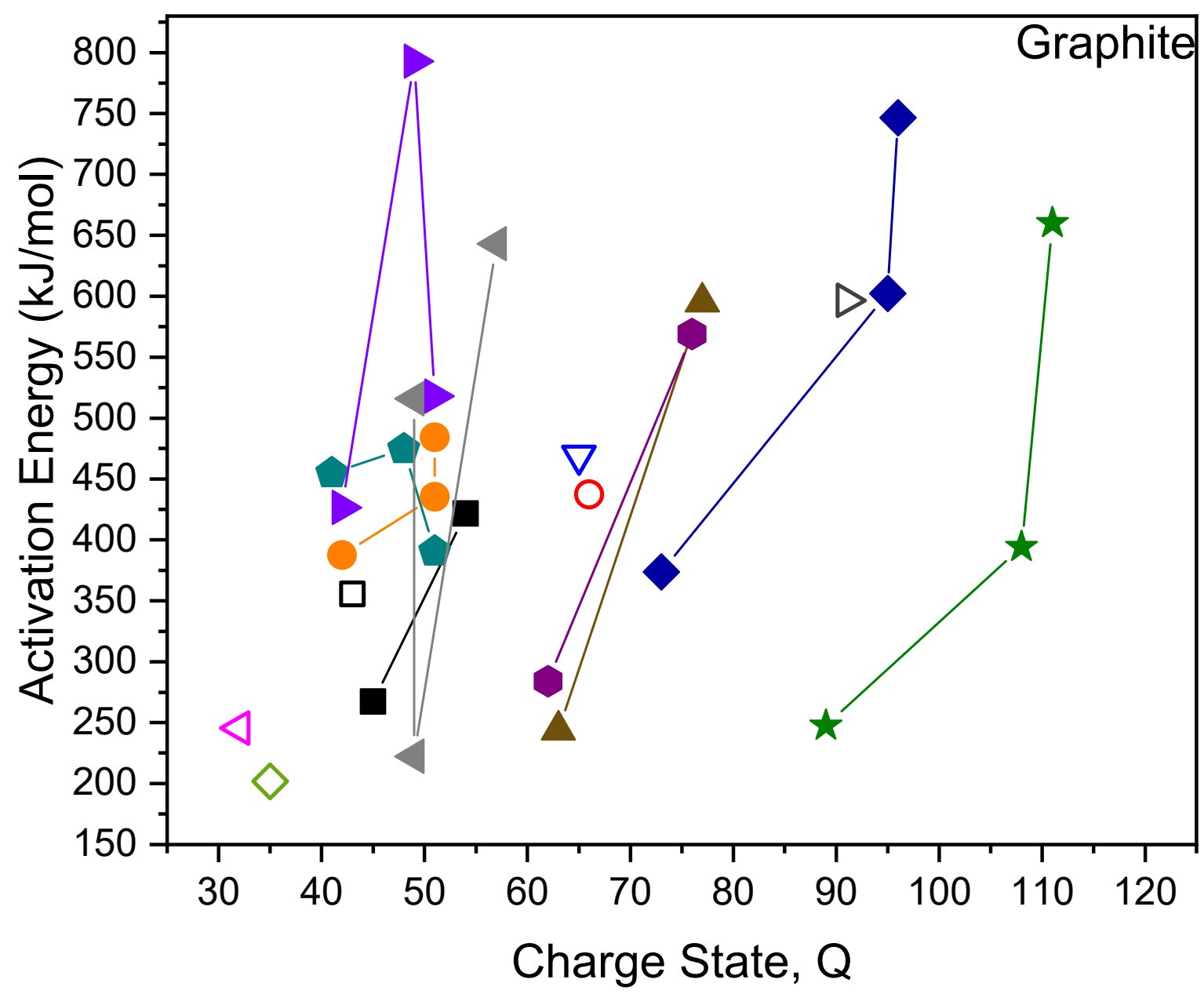

**Figure S8:** Showing Ea vs Q. In this case it seems that there no noticeable changes due to Q state changes. In the figure the initial Q state was use for each Ea that was determined for each heat ramp.

**Comparison of the fraction of under-coordinated edge sites for graphite and graphene NPs:**

The sublimation rate for an NP is expected to be strongly influenced by the number of defective or under-coordinated sites exposed on the NP surface because such sites have lower stability, and therefore should have lower activation energy ($E_a$) for sublimation. For NPs, the class of such sites that clearly must be present in large numbers is edge sites. An idealized graphite NP might consist of a cubical stack of graphene layers. For example, given graphite's 0.3353 nm interlayer spacing and 0.2461 nm basal plane unit cell edge length (2 C/unit cell),[28] a stack with thickness of 100 layers would be ~33.53 nm thick, and if we assume square layers with 33.53 nm edges, then the resulting cubical NP would have mass of ~52 MDa – in the mass range for our NPs.

This cubical NP would have ~8.7 x $10^4$ fully coordinated C atoms in the exposed top and bottom basal planes, and ~9.4 x $10^4$ under-coordinated atoms in the edges of the stacked basal planes. Thus, for idealized graphite NPs with aspect ratios near unity, roughly 50% of the NP surface consists of under-coordinated atoms in basal plane edge sites. This ratio is independent of NP size but is strongly affected by the NP shape. For example, for the same NP thickness and basal plane area (i.e., the same mass), if the basal planes are elongated or irregularly shaped such that the perimeter/area ratio increases, the fraction of under-coordinated edge sites would also increase. On the other hand, if the NP is thinner, with fewer but larger basal planes (to keep the same mass), the fraction of edge sites decreases.

Consider graphene NPs. Our graphene NPs feedstock is stated to have 2 to 8 nm thickness, i.e., the average thickness is ~5 nm. A 5.03 nm thick NP would have 15 layers. If these layers were square, they would need to have a lateral dimension of 86.5 nm to give the same 52 MDa mass as the cubical NP discussed above. In such a particle, there would be ~3.6 x $10^4$ under-coordinated atoms in the edges of the 15 stacked layers, with ~5.8 x $10^5$ fully coordinated atoms exposed in the top and bottom basal planes. Thus the idealized graphene NP would have only ~6.3% of its surface atoms in under-coordinated edge sites. More importantly, the total number of under-coordinated edge atoms would be about one third the number in an idealized graphite NP with an aspect ratio of 1.0.

Of course, real graphite and graphene NPs would have additional under-coordinated sites due to irregular shapes and to defects in the basal planes. Nonetheless, we expect that graphene NPs should have a significantly smaller fraction of under-coordinated surface atoms than graphite NPs, which should result in slower sublimation kinetics.

Another factor that must be taken into account in comparing the sublimation rates for graphite and graphene NPs is that the reported rates for each NP have been normalized to the nominal surface area, calculated as the surface area of a spherical particle with the same mass, assuming the bulk density. This sets a lower limit on the surface area, hence giving area-normalized rates that are upper limits. For the cubical NP discussed above, the actual surface area would be 1.23 times greater than the nominal area, and for the 5 nm graphene NP, the actual area would 1.68 times greater than the nominal area. Roughness or other irregularities would further increase the actual surface areas.

**Comparison of integrated intensities and sublimation rates in the three heat ramps:**

**Figure S9** plots the 2000 K integrated intensities vs. sublimation rates for all the NP. Because of the large NP-to-NP variations, it is difficult to see much correlation, other than a general tendency for sublimation rates to decrease in subsequent heat ramps.

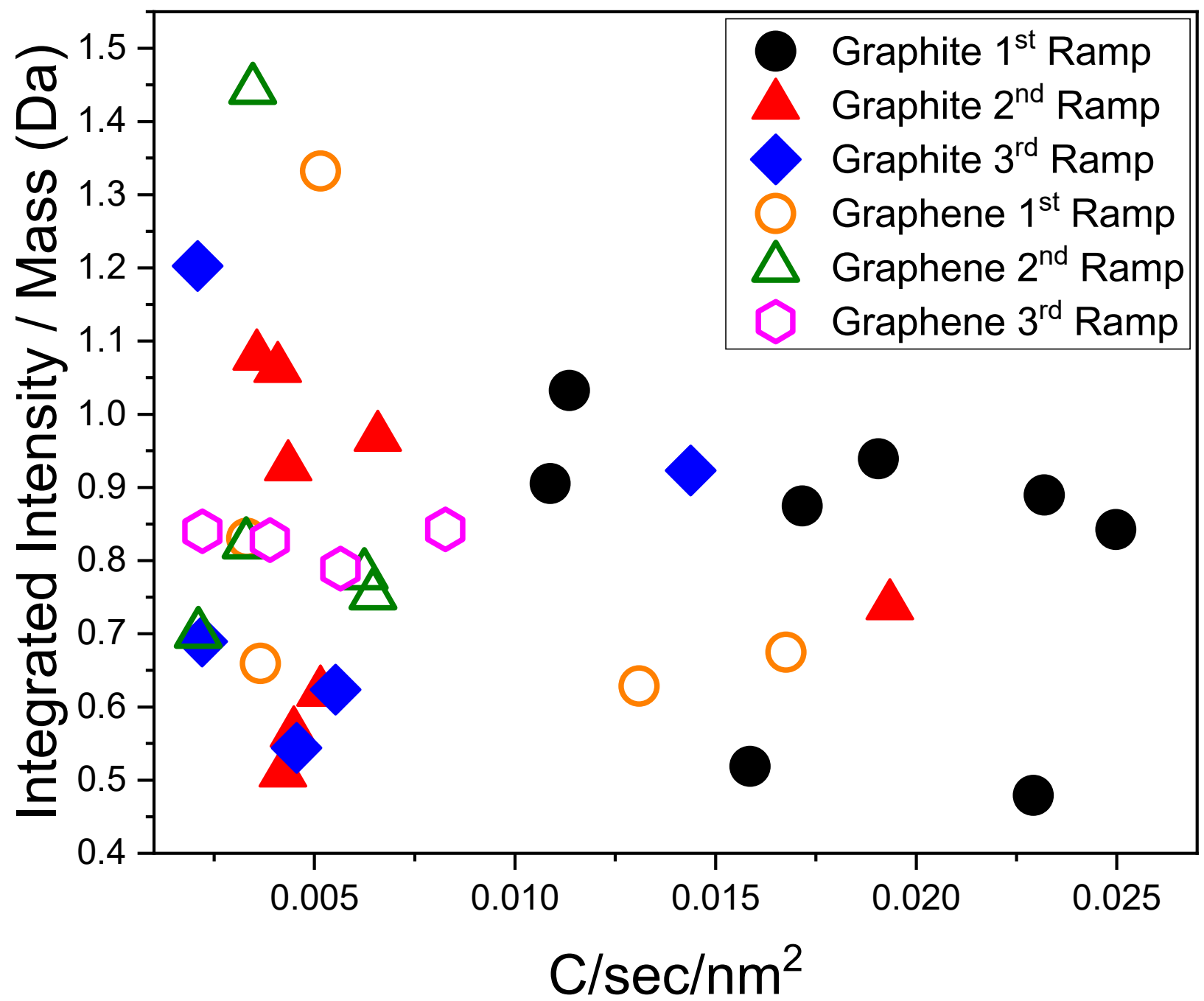

**Figure S9**: Interpolated integrated intensities plotted vs. area-normalized sublimation rates, both at 2000 K, for all NPs where more than one temperature ramp was completed. On average, the rates decreased substantially from the 1st to 2nd heat ramps, and the NP-to-NP spread in emission brightness increased. Comparing the 2nd and 3rd ramps, the effects on rates and intensities were much smaller, making generalizations difficult.

## References

1. Schlemmer, S.; Illemann, J.; Wellert, S.; Gerlich, D., Nondestructive High-Resolution And Absolute Mass Determination Of Single Charged Particles In A Three-Dimensional Quadrupole Trap. *J. Appl. Phys.* **2001,** *90* (10), 5410-5418.

2. Cai, Y.; Peng, W. P.; Kuo, S. J.; Lee, Y. T.; Chang, H. C., Single-Particle Mass Spectrometry of Polystyrene Microspheres and Diamond Nanocrystals. *Anal. Chem.* **2002,** *74* (1), 232-238.

3. Gerlich, D., Molecular Ions and Nanoparticles in RF and AC Traps. *Hyperfine Interact.* **2003,** *146/147* (1-4/1-4), 293-306.

4. Trevitt, A. J.; Wearne, P. J.; Bieske, E. J.; Schuder, M. D., Observation of Nondegenerate Cavity Modes for a Distorted Polystyrene Microsphere. *Opt. Lett.* **2006,** *31* (14), 2211-2213.

5. Graf, C.; Lewinski, R.; Dembski, S.; Langer, B.; Ruehl, E., Experiments on single levitated particles: a novel approach for investigations of electronic properties of structured II-VI-semiconductor nanoparticles in selected environments. *Phys. Status Solidi C* **2007,** *4* (9), 3244-3259.

6. Nagornykh, P.; Coppock, J. E.; Kane, B. E., Cooling of levitated graphene nanoplatelets in high vacuum. *Appl. Phys. Lett.* **2015,** *106* (24), 244102.

7. Clarke, J. T.; Fox, B. R., Rate and Heat of Vaporization of Graphite above 3000°K. *J. Chem. Phys.* **1969,** *51* (8), 3231-3240.

8. Marshall, A. L.; Norton, F. J., Carbon Vapor Pressure and Heat of Vaporization. *J. Am. Chem. Soc.* **1950,** *72* (5), 2166-2171.

9. Esser, T. K.; Hoffmann, B.; Anderson, S. L.; Asmis, K. R., A cryogenic single nanoparticle action spectrometer. *Review of Scientific Instruments* **2019,** *90* (12), 125110.

10. Tsai, C. C.; Gabriel, T. A.; Haines, J. R.; Rasmussen, D. A., Graphite Sublimation Tests For Target Development For The Muon Collider/Neutrino Factory *IEEE Symp. Fusion Eng.* **2006,** *21*, 377-379.

11. Haines, J. R.; Tsai, C. C. *Graphite Sublimation Tests for the Muon Collider/Neutrino Factory Target Development Program*; Oak Ridge, 2002; p 6 pages.

12. Bell, D. M.; Howder, C. R.; Johnson, R. C.; Anderson, S. L., Single CdSe/ZnS Nanocrystals in an Ion Trap: Charge and Mass Determination and Photophysics Evolution with Changing Mass, Charge, and Temperature. *ACS Nano* **2014,** *8*, 2387–2398.

13. Howder, C. R.; Bell, D. M.; Anderson, S. L., Optically Detected, Single Nanoparticle Mass Spectrometer With Pre-Filtered Electrospray Nanoparticle Source. *Rev. Sci. Instrum.* **2014,** *85*, 014104 -014110.

14. Zavitsanos, P. D.; Carlson, G. A., Experimental study of the sublimation of graphite at high temperatures. *J. Chem. Phys.* **1973,** *59* (6), 2966-2973.

15. Howder, C. R.; Long, B. A.; Alley, R. N.; Gerlich, D.; Anderson, S. L., Single Nanoparticle Mass Spectrometry as a High Temperature Kinetics Tool: Emission Spectra, Sublimation, and Oxidation of Hot Carbon Nanoparticles. *J. Phys. Chem. A* **2015,** *119*, 12538–12550.
16. Long, B. A.; Rodriguez, D. J.; Lau, C. Y.; Anderson, S. L., Thermal emission spectroscopy for single nanoparticle temperature measurement: optical system design and calibration. *Appl. Opt.* **2019,** *58* (3), 642-649.

17. Long, B. A.; Rodriguez, D. J.; Lau, C. Y.; Schultz, M.; Anderson, S. L., Thermal Emission Spectroscopy of Single, Isolated Carbon Nanoparticles: Effects of Particle Size, Material, Charge, Excitation Wavelength, and Thermal History. *J. Phys. Chem. C* **2020,** *124*, 1704-1716.

18. Chupka, W. A.; Berkowitz, J.; Meschi, D. J.; Tasman, H. A., Mass Spectrometric Studies of High Temperature Systems. In *Advances in Mass Spectrometry*, Elliott, R. M., Ed. Pergamon: 1963; pp 99-109.

19. Doehaerd, T.; Goldfinger, P.; Waelbroeck, F., Direct Determination of the Sublimation Energy of Carbon. *J. Chem. Phys.* **1952,** *20*, 757.

20. Steele, W. C.; Bourgelas, F. N. *Studies of Graphite Vaporization Using a Modulated Beam Mass Spectrometer, Technical report AFML-TR-72-222, AVSD-0048-73-CR*; Air Force Materials Laboratory, Air Force Systems Command: Wright-Patterson Air Force Base, Ohio, 1972.

21. Olszak-Humienik, M., On the thermal stability of some ammonium salts. *Thermochim. Acta* **2001,** *378* (1), 107-112.

22. Tremblay, G.; Vastola, F. J.; Walker, P. L., Jr., Characterization of Graphite Surface Complexes by Temperature-Programmed Desorption. *Extended Abstracts - Bienneial Conf. on Carbon* **1975,** *12*, 177-178.

23. Marchon, B.; Carrazza, J.; Heinemann, H.; Somorjai, G. A., TPD and XPS studies of O2, CO2, and H2O adsorption on clean polycrystalline graphite. *Carbon* **1988,** *26* (4), 507-514.

24. Nowakowski, M. J.; Vohs, J. M.; Bonnell, D. A., Surface reactivity of oxygen on cleaved and sputtered graphite. *Surf. Sci.* **1992,** *271* (3), L351-L356.

25. Pan, Z.; Yang, R. T., Strongly Bonded Oxygen in Graphite: Detection by High-Temperature TPD and Characterization *Ind. Eng. Chem. Res.* **1992,** *31*, 2675-2680.

26. Olander, D. R.; Siekhaus, W.; Jones, R.; Schwarz, J. A., Reactions of Modulated Molecular Beams with Pyrolytic Graphite. I. Oxidation of the Basal Plane. *J. Chem. Phys.* **1972,** *57* (1), 408-420.

27. Murray, V. J.; Smoll, E. J.; Minton, T. K., Dynamics of Graphite Oxidation at High Temperature. *J. Phys. Chem. C* **2018,** *122* (12), 6602-6617.

28. Bacon, G. E., The interlayer spacing of graphite. *Acta Crystallogr. A* **1951,** *4* (6), 558-561.

Figure 1 Graphene Experiment

| Time (sec) | M (Mda) | Time (sec) | Integration | n | T |
|---|---|---|---|---|---|
| 0 | 64.03095 | -1 | 1.61E+07 | 0.88158 | 1783.647 |
| 31 | 64.01698 | 32 | 1.60E+07 | 0.8435 | 1794.097 |
| 63 | 64.02718 | 64 | 1.60E+07 | 0.85637 | 1789.552 |
| 95 | 63.99884 | 96 | 1.62E+07 | 0.86252 | 1790.65 |
| 122 | 63.98377 | 128 | 1.69E+07 | 0.84647 | 1793.929 |
| 165 | 63.99905 | 160 | 1.75E+07 | 0.85355 | 1802.868 |
| 192 | 63.98872 | 192 | 1.79E+07 | 0.83089 | 1810.943 |
| 219 | 63.97677 | 224 | 1.82E+07 | 0.81364 | 1816.617 |
| 261 | 63.9404 | 256 | 1.83E+07 | 0.85311 | 1802.99 |
| 289 | 63.97053 | 288 | 1.84E+07 | 0.8322 | 1808.295 |
| 320 | 63.93606 | 321 | 1.82E+07 | 0.82052 | 1813.213 |
| 358 | 63.93691 | 353 | 1.77E+07 | 0.80522 | 1818.126 |
| 379 | 63.95389 | 385 | 1.76E+07 | 0.81945 | 1813.529 |
| 416 | 63.8977 | 417 | 1.77E+07 | 0.84643 | 1803.605 |
| 449 | 63.94471 | 449 | 1.79E+07 | 0.82281 | 1811.945 |
| 475 | 63.91391 | 481 | 1.84E+07 | 0.78552 | 1819.799 |
| 518 | 63.90223 | 513 | 1.84E+07 | 0.81008 | 1813.514 |
| 539 | 63.84826 | 545 | 1.98E+07 | 0.79365 | 1822.623 |
| 572 | 63.84357 | 578 | 2.06E+07 | 0.77625 | 1829.492 |
| 615 | 63.84325 | 610 | 2.05E+07 | 0.77441 | 1827.048 |
| 641 | 63.82505 | 642 | 2.04E+07 | 0.78625 | 1821.399 |
| 668 | 63.82035 | 674 | | | |
| 706 | 63.77752 | 706 | | | |
| 743 | 63.7502 | 738 | | | |
| 770 | 63.78432 | 770 | | | |
| 802 | 63.72711 | 802 | | | |
| 835 | 63.72515 | 835 | 2.86E+07 | 0.69425 | 1923.923 |
| 861 | 63.66925 | 867 | 3.00E+07 | 0.70444 | 1921.671 |
| 894 | 63.76052 | 899 | 3.03E+07 | 0.68823 | 1922.264 |
| 935 | 63.6043 | 931 | 2.99E+07 | 0.65605 | 1924.964 |
| 964 | 63.558 | 963 | 2.97E+07 | 0.67692 | 1912.164 |
| 996 | 63.59515 | 995 | 2.94E+07 | 0.6801 | 1908.744 |
| 1026 | 63.51324 | 1027 | 2.95E+07 | 0.66276 | 1917.622 |
| 1060 | 63.51948 | 1059 | 2.94E+07 | 0.7035 | 1901.511 |
| 1091 | 63.49712 | 1092 | 2.94E+07 | 0.68749 | 1909.779 |
| 1123 | 63.44715 | 1124 | 2.95E+07 | 0.68132 | 1909.167 |
| 1156 | 63.45754 | 1156 | 2.93E+07 | 0.69803 | 1902.692 |
| 1187 | 63.39956 | 1188 | 2.94E+07 | 0.65941 | 1915.53 |
| 1220 | 63.4053 | 1220 | 2.95E+07 | 0.69456 | 1904.033 |
| 1252 | 63.35105 | 1252 | 2.93E+07 | 0.68373 | 1906.672 |
| 1284 | 63.3312 | 1284 | 2.92E+07 | 0.6788 | 1908.297 |
| 1317 | 63.39275 | 1317 | 2.92E+07 | 0.67514 | 1910.375 |
| 1348 | 63.36893 | 1349 | 2.94E+07 | 0.687 | 1905.333 |
| 1380 | 63.3074 | 1381 | 2.94E+07 | 0.69241 | 1904.396 |
| 1413 | 63.29213 | 1413 | 2.93E+07 | 0.6643 | 1910.369 |
| 1445 | 63.29467 | 1445 | 2.95E+07 | 0.66523 | 1910.597 |
| 1476 | 63.25211 | 1477 | 2.95E+07 | 0.68842 | 1904.525 |
| 1504 | 63.24926 | 1509 | | | |
| 1546 | 63.2294 | 1541 | | | |
| 1573 | 63.2148 | 1573 | | | |
| 1605 | 63.21553 | 1605 | | | |
| 1631 | 63.17339 | 1637 | | | |
| 1674 | 63.07437 | 1670 | | | |
| 1697 | 63.07411 | 1702 | | | |
| 1738 | 62.9803 | 1734 | | | |
| 1766 | | 1766 | | | |
| 1795 | | 1798 | | | |
| 1828 | 62.78631 | 1830 | 4.19E+07 | 0.54787 | 2003.595 |
| 1867 | 62.68979 | 1862 | 4.22E+07 | 0.52985 | 2008.22 |
| 1889 | 62.66752 | 1894 | 4.29E+07 | 0.52552 | 2010.825 |
| 1931 | 62.60146 | 1926 | 4.30E+07 | 0.51438 | 2013.813 |
| 1958 | 62.58499 | 1958 | 4.24E+07 | 0.53752 | 2001.343 |
| 1989 | 62.51765 | 1990 | 4.26E+07 | 0.52199 | 2007.306 |
| 2022 | 62.50095 | 2022 | 4.26E+07 | 0.53514 | 2001.792 |
| 2048 | 62.44316 | 2054 | 4.31E+07 | 0.5192 | 2006.838 |
| 2092 | 62.39154 | 2087 | 4.31E+07 | 0.52188 | 2005.904 |
| 2119 | 62.38311 | 2119 | 4.32E+07 | 0.51223 | 2007.898 |
| 2150 | 62.345 | 2151 | 4.30E+07 | 0.5287 | 2001.641 |
| 2183 | 62.31073 | 2183 | 4.31E+07 | 0.51235 | 2007.955 |
| 2214 | 62.25511 | 2215 | 4.30E+07 | 0.5214 | 2002.735 |
| 2247 | 62.2345 | 2247 | 4.28E+07 | 0.51925 | 2002.722 |
| 2279 | 62.25354 | 2279 | 4.29E+07 | 0.50863 | 2006.732 |
| 2305 | 62.17728 | 2311 | 4.41E+07 | 0.4899 | 2015.867 |
| 2349 | 62.12605 | 2344 | 4.42E+07 | 0.51639 | 2004.182 |
| 2375 | 62.08019 | 2376 | 4.37E+07 | 0.50774 | 2007.355 |
| 2408 | 62.05093 | 2408 | 4.41E+07 | 0.49525 | 2011.096 |
| 2440 | 62.03061 | 2440 | | | |
| 2471 | 61.95349 | 2472 | | | |
| 2504 | 61.87553 | 2504 | | | |
| 2536 | 61.79144 | 2536 | 5.30E+07 | 0.41298 | 2101.201 |
| 2562 | 61.70892 | 2568 | 5.50E+07 | 0.38591 | 2115.308 |
| 2605 | 61.55329 | 2600 | 5.43E+07 | 0.39441 | 2106.902 |
| 2633 | 61.45257 | 2632 | 5.38E+07 | 0.40669 | 2098.571 |
| 2659 | 61.38159 | 2665 | 5.46E+07 | 0.39509 | 2103.501 |
| 2701 | 61.25246 | 2697 | 5.47E+07 | 0.38933 | 2104.122 |
| 2729 | 61.18941 | 2729 | 5.40E+07 | 0.40787 | 2095.046 |
| 2760 | 61.10189 | 2761 | 5.40E+07 | 0.40131 | 2097.423 |
| 2793 | 61.02709 | 2793 | 5.46E+07 | 0.38495 | 2106.652 |
| 2825 | | 2825 | | | |
| 2856 | | 2857 | | | |
| 2892 | 60.78997 | 2889 | 5.48E+07 | 0.38462 | 2102.089 |
| 2923 | 60.74515 | 2921 | 5.49E+07 | 0.37024 | 2105.818 |
| 2953 | 60.65148 | 2953 | 5.43E+07 | 0.38442 | 2098.844 |
| 2985 | 60.55817 | 2986 | 5.45E+07 | 0.3884 | 2098.322 |
| 3018 | 60.53887 | 3018 | 5.40E+07 | 0.39487 | 2093.799 |
| 3049 | 60.45837 | 3050 | 5.39E+07 | 0.39026 | 2093.171 |
| 3082 | 60.39976 | 3082 | 5.45E+07 | 0.39277 | 2092.884 |
| 3114 | 60.33373 | 3114 | 5.37E+07 | 0.39223 | 2091.472 |
| 3146 | 60.26894 | 3146 | 5.41E+07 | 0.39125 | 2092.288 |
| 3178 | | 3178 | | | |
| 3209 | | 3210 | | | |
| 3245 | 60.0586 | 3242 | | | |
| 3270 | 60.01174 | 3274 | 4.59E+07 | 0.45265 | 2010.921 |
| 3312 | 60.00127 | 3306 | 4.63E+07 | 0.4525 | 2009.783 |
| 3338 | 60.01456 | 3338 | 4.62E+07 | 0.46343 | 2005.49 |
| 3370 | 59.97458 | 3371 | 4.57E+07 | 0.46809 | 2003.355 |
| 3402 | 59.95672 | 3403 | 4.57E+07 | 0.45973 | 2005.307 |
| 3434 | 59.94235 | 3435 | 4.60E+07 | 0.48711 | 1996.933 |
| 3466 | 59.9044 | 3467 | 4.59E+07 | 0.46438 | 2004.303 |
| 3499 | 59.90518 | 3499 | 4.60E+07 | 0.46799 | 2002.625 |
| 3531 | 59.9186 | 3531 | 4.54E+07 | 0.47129 | 1999.825 |
| 3563 | 59.90563 | 3563 | 4.57E+07 | 0.46367 | 2002.595 |
| 3595 | 59.87819 | 3595 | 4.57E+07 | 0.47808 | 1996.804 |
| 3627 | 59.84872 | 3627 | 4.55E+07 | 0.47423 | 1998.048 |
| 3659 | 59.82662 | 3659 | 4.48E+07 | 0.48404 | 1992.833 |
| 3692 | 59.84193 | 3692 | 4.54E+07 | 0.46827 | 1999.231 |
| 3723 | 59.81958 | 3724 | 4.54E+07 | 0.46928 | 1999.28 |
| 3755 | 59.80091 | 3756 | 4.52E+07 | 0.48582 | 1991.874 |
| 3788 | 59.84651 | 3788 | 4.49E+07 | 0.4879 | 1991.838 |
| 3818 | 59.77022 | 3820 | 4.53E+07 | 0.47418 | 1996.657 |
| 3852 | 59.78083 | 3852 | 4.59E+07 | 0.45681 | 2005.406 |
| 3884 | 59.75885 | 3884 | 4.55E+07 | 0.46282 | 2002.648 |
| 3916 | | 3916 | | | |
| 3948 | 59.74964 | 3948 | 3.73E+07 | 0.56761 | 1907.056 |
| 3980 | 59.73987 | 3981 | 3.75E+07 | 0.57815 | 1903.39 |
| 4013 | 59.73569 | 4013 | 3.74E+07 | 0.56368 | 1909.655 |
| 4044 | 59.72923 | 4045 | 3.69E+07 | 0.56356 | 1908.345 |
| 4077 | 59.72244 | 4077 | 3.65E+07 | 0.60425 | 1889.85 |
| 4109 | 59.73472 | 4109 | 3.74E+07 | 0.55182 | 1911.949 |
| 4140 | 59.71927 | 4141 | 3.64E+07 | 0.56541 | 1901.842 |
| 4173 | 59.7182 | 4173 | 3.62E+07 | 0.5917 | 1891.045 |
| 4205 | 59.72163 | 4205 | 3.63E+07 | 0.59032 | 1893.58 |
| 4237 | 59.72067 | 4237 | 3.65E+07 | 0.59193 | 1893.388 |
| 4269 | 59.73438 | 4269 | 3.63E+07 | 0.59651 | 1892.068 |
| 4301 | 59.72008 | 4302 | 3.67E+07 | 0.6057 | 1888.893 |
| 4333 | 59.71497 | 4334 | 3.62E+07 | 0.57547 | 1897.527 |
| 4365 | 59.70106 | 4366 | 3.69E+07 | 0.59625 | 1894.125 |
| 4398 | 59.72507 | 4398 | 3.72E+07 | 0.56015 | 1907.388 |
| 4429 | 59.68347 | 4430 | 3.68E+07 | 0.59166 | 1894.545 |
| 4462 | 59.72222 | 4462 | 3.66E+07 | 0.58933 | 1896.268 |
| 4493 | 59.68614 | 4494 | 3.67E+07 | 0.58494 | 1896.295 |
| 4526 | 59.69527 | 4526 | 3.66E+07 | 0.57185 | 1901.683 |
| 4558 | 59.66192 | 4558 | | | |
| 4591 | 59.68379 | 4591 | | | |
| 4622 | 59.655 | 4623 | | | |
| 4655 | 59.68696 | 4655 | | | |
| 4686 | 59.68667 | 4687 | | | |
| 4725 | | 4719 | | | |
| 4750 | 59.67103 | 4751 | 2.87E+07 | 0.69961 | 1794.504 |
| 4789 | 59.68508 | 4783 | 2.84E+07 | 0.70201 | 1792.185 |
| 4814 | 59.65858 | 4815 | 2.84E+07 | 0.69048 | 1795.926 |
| 4847 | 59.69945 | 4847 | 2.85E+07 | 0.73012 | 1783.513 |
| 4878 | 59.66195 | 4879 | 2.84E+07 | 0.71549 | 1787.72 |
| 4912 | 59.68198 | 4911 | 2.82E+07 | 0.68026 | 1796.793 |
| 4943 | 59.66885 | 4944 | 2.83E+07 | 0.675 | 1799.189 |
| 4975 | 59.67679 | 4976 | 2.84E+07 | 0.69116 | 1793.871 |
| 5007 | 59.64229 | 5008 | 2.83E+07 | 0.68861 | 1793.439 |
| 5040 | 59.63728 | 5040 | 2.81E+07 | 0.69764 | 1790.344 |
| 5072 | 59.67505 | 5072 | 2.82E+07 | 0.68438 | 1795.722 |
| 5103 | 59.64604 | 5104 | 2.80E+07 | 0.67787 | 1797.068 |
| 5136 | 59.66615 | 5136 | 2.80E+07 | 0.69506 | 1791.113 |
| 5168 | 59.67583 | 5168 | 2.80E+07 | 0.72722 | 1781.593 |
| 5200 | 59.68072 | 5200 | 2.78E+07 | 0.69662 | 1790.504 |
| 5232 | 59.6773 | 5233 | 2.78E+07 | 0.7365 | 1777.023 |
| 5264 | 59.68709 | 5265 | 2.80E+07 | 0.6937 | 1791.68 |
| 5296 | 59.66811 | 5297 | 2.82E+07 | 0.71133 | 1787.044 |
| 5329 | 59.65836 | 5329 | 2.81E+07 | 0.67897 | 1797.416 |
| 5361 | 59.65762 | 5361 | 2.82E+07 | 0.68329 | 1795.588 |
| 5393 | 59.67842 | 5393 | 2.80E+07 | 0.71403 | 1786.638 |
| 5425 | 59.68 | 5425 | 2.84E+07 | 0.70724 | 1789.203 |
| 5457 | 59.6562 | 5457 | 2.83E+07 | 0.70416 | 1790.2 |
| 5489 | 59.6403 | 5489 | 2.82E+07 | 0.66839 | 1799.962 |
| 5522 | 59.66641 | 5521 | 2.83E+07 | 0.71855 | 1786.046 |
| 5551 | | 5554 | | | |
| 5585 | 59.66074 | 5586 | 3.53E+07 | 0.58422 | 1885.461 |
| 5615 | 59.67859 | 5618 | 3.50E+07 | 0.59688 | 1879.409 |
| 5649 | 59.63097 | 5650 | 3.48E+07 | 0.62429 | 1870.146 |
| 5682 | 59.64157 | 5682 | 3.50E+07 | 0.586 | 1883.98 |
| 5714 | 59.66973 | 5714 | 3.55E+07 | 0.59168 | 1885.611 |
| 5745 | 59.63879 | 5746 | 3.54E+07 | 0.63109 | 1871.348 |
| 5778 | 59.63917 | 5778 | 3.61E+07 | 0.59972 | 1885.654 |
| 5810 | 59.63588 | 5810 | 3.63E+07 | 0.58294 | 1895.606 |
| 5842 | 59.61007 | 5842 | 3.67E+07 | 0.58536 | 1894.386 |
| 5875 | 59.63893 | 5875 | 3.67E+07 | 0.59325 | 1892.308 |
| 5906 | 59.5848 | 5907 | 3.74E+07 | 0.59429 | 1897.099 |
| 5939 | 59.60672 | 5939 | 3.73E+07 | 0.5716 | 1905.788 |
| 5971 | 59.60679 | 5971 | 3.73E+07 | 0.56719 | 1907.95 |
| 6003 | 59.60473 | 6003 | 3.75E+07 | 0.56147 | 1909.983 |
| 6035 | 59.60809 | 6035 | 3.71E+07 | 0.59222 | 1898.253 |
| 6067 | 59.60754 | 6067 | 3.73E+07 | 0.57618 | 1904.393 |
| 6099 | 59.604 | 6099 | 3.77E+07 | 0.55735 | 1910.919 |
| 6131 | 59.60594 | 6132 | 3.78E+07 | 0.55157 | 1913.136 |
| 6163 | 59.59731 | 6164 | 3.76E+07 | 0.58265 | 1900.734 |
| 6196 | 59.60313 | 6196 | 3.74E+07 | 0.57871 | 1902.911 |
| 6227 | 59.57467 | 6228 | 3.74E+07 | 0.57149 | 1904.502 |
| 6260 | 59.58173 | 6260 | 3.75E+07 | 0.56393 | 1908.045 |
| 6292 | 59.57874 | 6292 | 3.73E+07 | 0.56732 | 1906.53 |
| 6324 | 59.5762 | 6324 | 3.78E+07 | 0.55085 | 1914.555 |
| 6356 | 59.57442 | 6356 | 3.72E+07 | 0.57342 | 1903.548 |
| 6388 | 59.5715 | 6388 | 3.71E+07 | 0.56439 | 1904.937 |
| 6420 | 59.5627 | 6421 | | | |
| 6452 | 59.55663 | 6453 | | | |
| 6484 | 59.55287 | 6485 | | | |
| 6516 | 59.55459 | 6517 | 4.44E+07 | 0.4884 | 1985.364 |
| 6548 | 59.54266 | 6549 | 4.46E+07 | 0.45909 | 1997.161 |
| 6581 | 59.53505 | 6581 | 4.46E+07 | 0.48623 | 1986.069 |
| 6613 | 59.51345 | 6613 | 4.48E+07 | 0.48123 | 1988.061 |
| 6645 | 59.51523 | 6645 | 4.49E+07 | 0.4832 | 1989.386 |
| 6677 | 59.51199 | 6677 | 4.47E+07 | 0.49707 | 1983.307 |
| 6709 | 59.51669 | 6709 | 4.49E+07 | 0.49485 | 1984.976 |
| 6741 | 59.48614 | 6742 | 4.56E+07 | 0.47826 | 1993.052 |
| 6773 | 59.46047 | 6774 | 4.57E+07 | 0.47941 | 1993.815 |
| 6806 | 59.47832 | 6806 | 4.57E+07 | 0.47611 | 1994.208 |
| 6838 | 59.48351 | 6838 | 4.52E+07 | 0.47159 | 1993.666 |
| 6869 | 59.43204 | 6870 | 4.51E+07 | 0.49097 | 1986.463 |
| 6902 | 59.41359 | 6902 | 4.52E+07 | 0.46843 | 1996 |
| 6934 | 59.40436 | 6934 | 4.51E+07 | 0.48292 | 1989.144 |
| 6966 | 59.40264 | 6966 | 4.56E+07 | 0.46319 | 2000.093 |
| 6998 | 59.39271 | 6999 | 4.53E+07 | 0.47871 | 1992.64 |
| 7030 | 59.39619 | 7031 | 4.59E+07 | 0.4634 | 1999.732 |
| 7062 | 59.37447 | 7063 | 4.55E+07 | 0.48358 | 1991.598 |
| 7094 | 59.33555 | 7095 | 4.55E+07 | 0.47681 | 1992.157 |
| 7127 | 59.30594 | 7127 | 4.57E+07 | 0.46619 | 1998.027 |
| 7159 | 59.33103 | 7159 | 4.58E+07 | 0.46474 | 1998.927 |
| 7190 | 59.28857 | 7191 | | | |
| 7223 | | 7223 | | | |
| 7248 | | 7255 | | | |
| 7285 | 59.03108 | 7287 | | | |
| 7324 | 58.93498 | 7319 | | | |
| 7352 | 58.89805 | 7351 | 5.70E+07 | 0.36573 | 2099.096 |
| 7383 | 58.87045 | 7384 | 5.69E+07 | 0.35351 | 2101.71 |
| 7415 | 58.80675 | 7416 | 5.67E+07 | 0.35724 | 2099.6 |
| 7448 | 58.7753 | 7448 | 5.72E+07 | 0.33614 | 2109.519 |
| 7479 | 58.70892 | 7480 | 5.64E+07 | 0.36347 | 2096.2 |
| 7512 | 58.671 | 7512 | 5.59E+07 | 0.35901 | 2095.173 |
| 7544 | 58.64068 | 7544 | 5.72E+07 | 0.36778 | 2093.829 |
| 7575 | 58.5758 | 7576 | 5.62E+07 | 0.34666 | 2101.578 |
| 7608 | 58.54636 | 7608 | 5.64E+07 | 0.36313 | 2093.896 |
| 7640 | 58.48357 | 7640 | 5.64E+07 | 0.35446 | 2096.786 |
| 7672 | 58.42762 | 7672 | 5.67E+07 | 0.35844 | 2097.293 |
| 7704 | 58.3899 | 7705 | 5.60E+07 | 0.35475 | 2097.352 |
| 7737 | 58.37112 | 7737 | 5.66E+07 | 0.35736 | 2096.432 |
| 7768 | 58.33925 | 7769 | 5.60E+07 | 0.36993 | 2089.825 |
| 7800 | 58.29131 | 7801 | 5.70E+07 | 0.35606 | 2096.706 |
| 7833 | 58.23771 | 7833 | 5.68E+07 | 0.3513 | 2098.118 |
| 7865 | 58.21165 | 7865 | 5.62E+07 | 0.36328 | 2091.535 |
| 7896 | 58.16551 | 7898 | 5.69E+07 | 0.35356 | 2098.134 |

Figure 2A

| T avg | Iavg/Mi |
|---|---|
| 1808.947 | 0.28177 |
| 1911.173 | 0.46236 |
| 2006.727 | 0.68556 |
| 2099.856 | 0.87808 |
| 2000.788 | 0.7601 |
| 1898.889 | 0.6152 |
| 1791.371 | 0.47266 |
| 1897.584 | 0.61588 |
| 1992.088 | 0.75949 |
| 2097.377 | 0.96054 |

Figure 2B

| T avg | n |
|---|---|
| 1808.947 | 0.8245 |
| 1911.173 | 0.682 |
| 2006.727 | 0.51922 |
| 2099.856 | 0.39264 |
| 2000.788 | 0.46961 |
| 1898.889 | 0.58107 |
| 1791.371 | 0.69871 |
| 1897.584 | 0.58077 |
| 1992.088 | 0.47783 |
| 2097.377 | 0.35732 |

| Figure 2C | 1st Ramp: UP | | 2nd Ramp: Down | | | 3rd Ramp: Up | | |
|---|---|---|---|---|---|---|---|---|
| 1000/T | Fit data | Exp Data | 1000/T | Fit data | Exp data | 1000/T | Fit data | Exp Data |
| 0.55281 | -5.5152 | -5.51999 | 0.47622 | -3.53839 | -3.40E+00 | 0.55823 | -8.46645 | -8.61109 |
| 0.52324 | -4.71678 | -4.66442 | 0.4998 | -5.02902 | -5.16E+00 | 0.52699 | -6.70072 | -6.47127 |
| 0.49832 | -4.04401 | -4.13881 | 0.52662 | -6.7245 | -6.85E+00 | 0.50199 | -5.28795 | -5.27756 |
| 0.47622 | -3.44725 | -3.40003 | 0.55823 | -8.72262 | -8.61E+00 | 0.47679 | -3.86389 | -3.9591 |

Figure 3

| Time (sec) | M(Mda) | Time (sec) | Integration | n | T |
|---|---|---|---|---|---|
| 64 | 82.8823 | 1 | 6.33E+07 | 0.68703 | 1904.586 |
| 97 | 82.8405 | 33 | 6.86E+07 | 0.67902 | 1923.848 |
| 129 | 82.8185 | 65 | 6.87E+07 | 0.67095 | 1928.852 |
| 158 | 82.9673 | 97 | 6.34E+07 | 0.6715 | 1908.618 |
| 191 | 82.6584 | 129 | 6.34E+07 | 0.71542 | 1894.623 |
| 231 | 82.6561 | 161 | 6.98E+07 | 0.71903 | 1917.587 |
| 257 | 82.5882 | 193 | 6.76E+07 | 0.6652 | 1922.113 |
| 290 | 82.5934 | 226 | 6.67E+07 | 0.654 | 1923.515 |
| 321 | 82.5495 | 258 | 7.29E+07 | 0.65078 | 1945.863 |
| 354 | 82.5322 | 290 | 6.94E+07 | 0.71493 | 1914.812 |
| 386 | 82.5011 | 322 | 6.56E+07 | 0.71483 | 1899.784 |
| 418 | 82.4649 | 354 | 6.97E+07 | 0.65535 | 1931.729 |
| 451 | 82.4772 | 386 | 6.81E+07 | 0.65388 | 1927.426 |
| 482 | 82.4407 | 419 | 6.68E+07 | 0.71185 | 1904.793 |
| 515 | 82.4088 | 451 | 6.48E+07 | 0.72467 | 1893.694 |
| 547 | 82.418 | 483 | 6.26E+07 | 0.72672 | 1887.051 |
| 579 | | 515 | 6.34E+07 | 0.66043 | 1907.139 |
| 611 | 82.3883 | 547 | 6.25E+07 | 0.66073 | 1904.983 |
| 643 | 82.3626 | 579 | 6.28E+07 | 0.73172 | 1886.124 |
| 675 | 82.3364 | 611 | 6.47E+07 | 0.66581 | 1911.201 |
| 708 | 82.3291 | 644 | 6.48E+07 | 0.66753 | 1910.86 |
| 740 | 82.3168 | 676 | 6.44E+07 | 0.71111 | 1898.964 |
| 772 | 82.3225 | 708 | 6.33E+07 | 0.65301 | 1909.458 |
| 804 | 82.2838 | 740 | 6.11E+07 | 0.67042 | 1895.042 |
| 832 | 82.2547 | 772 | 5.95E+07 | 0.65984 | 1892.873 |
| 873 | 82.2213 | 804 | 5.95E+07 | 0.706 | 1878.666 |
| 901 | 82.2122 | 837 | 6.34E+07 | 0.71408 | 1892.13 |
| 933 | 82.1727 | 869 | 6.65E+07 | 0.64664 | 1922.345 |
| 965 | 82.1757 | 901 | 6.75E+07 | 0.64774 | 1924.127 |
| 997 | 82.1694 | 933 | 6.74E+07 | 0.64111 | 1927.112 |
| 1025 | 82.1222 | 966 | 6.68E+07 | 0.64383 | 1923.196 |
| 1066 | 82.0847 | 998 | 6.59E+07 | 0.64348 | 1919.625 |
| 1094 | 82.0944 | 1030 | 6.41E+07 | 0.67679 | 1902.118 |
| 1126 | 82.0707 | 1062 | 6.28E+07 | 0.68312 | 1895.048 |
| 1158 | 82.0547 | 1094 | 6.14E+07 | 0.69055 | 1889.587 |
| 1190 | 82.0502 | 1126 | 6.03E+07 | 0.67162 | 1889.543 |
| 1223 | 82.0453 | 1159 | 5.94E+07 | 0.70677 | 1875.417 |
| 1255 | 82.0343 | 1191 | 5.94E+07 | 0.72006 | 1872.352 |
| 1287 | 82.009 | 1223 | 6.01E+07 | 0.72914 | 1872.922 |
| 1319 | 81.9769 | 1255 | 6.02E+07 | 0.70112 | 1878.561 |
| 1352 | | 1287 | 6.03E+07 | 0.74103 | 1869.484 |
| 1383 | 81.9947 | 1319 | 6.11E+07 | 0.68364 | 1888.676 |
| 1416 | 81.9673 | 1352 | 6.14E+07 | 0.67925 | 1891.845 |
| 1448 | 81.9401 | 1384 | 6.11E+07 | 0.74218 | 1873.532 |
| 1480 | 81.9469 | 1416 | 6.08E+07 | 0.74491 | 1868.602 |
| 1512 | 81.9406 | 1448 | 6.04E+07 | 0.70661 | 1877.004 |
| 1544 | 81.9269 | 1480 | 6.02E+07 | 0.74688 | 1867.181 |
| 1576 | 81.9348 | 1512 | 6.17E+07 | 0.68777 | 1888.48 |
| 1608 | 81.9214 | 1544 | 6.20E+07 | 0.68939 | 1888.909 |
| 1641 | | 1576 | 6.20E+07 | 0.75087 | 1870.213 |
| 1674 | 81.9326 | 1609 | 6.19E+07 | 0.69596 | 1887.163 |
| 1705 | 81.8917 | 1641 | 6.34E+07 | 0.68807 | 1894.648 |
| 1738 | 81.8722 | 1673 | 6.37E+07 | 0.73753 | 1883.056 |
| 1770 | 81.838 | 1706 | 6.32E+07 | 0.68166 | 1896.049 |
| 1802 | 81.8163 | 1738 | 6.26E+07 | 0.68041 | 1893.58 |
| 1834 | 81.7986 | 1770 | 6.33E+07 | 0.66725 | 1901.551 |
| 1866 | 81.7857 | 1802 | 6.31E+07 | 0.72553 | 1884.957 |
| 1899 | 81.8105 | 1834 | 6.32E+07 | 0.67392 | 1898.823 |
| 1930 | 81.7708 | 1867 | 6.30E+07 | 0.72004 | 1883.658 |
| 1963 | 81.7039 | 1899 | 6.31E+07 | 0.65974 | 1901.733 |
| 1995 | 81.715 | 1931 | 6.37E+07 | 0.65483 | 1905.659 |
| 2027 | 81.7356 | 1963 | 6.45E+07 | 0.65247 | 1910.048 |
| 2059 | 81.6769 | 1995 | 6.66E+07 | 0.64161 | 1920.657 |
| 2092 | 81.6794 | 2027 | 6.82E+07 | 0.69903 | 1908.721 |
| 2123 | 81.6571 | 2059 | 7.04E+07 | 0.62846 | 1939.275 |
| 2156 | 81.6363 | 2092 | 7.19E+07 | 0.68804 | 1926.016 |
| 2188 | 81.6353 | 2124 | 7.29E+07 | 0.62883 | 1946.507 |
| 2220 | 81.6041 | 2156 | 7.31E+07 | 0.62433 | 1948.487 |
| 2252 | 81.5797 | 2188 | 7.31E+07 | 0.62105 | 1948.738 |
| 2284 | 81.5837 | 2220 | 7.28E+07 | 0.67005 | 1933.443 |
| 2316 | 81.5341 | 2252 | 7.28E+07 | 0.61541 | 1950.508 |
| 2349 | 81.5294 | 2284 | 7.25E+07 | 0.66353 | 1935.682 |
| 2381 | 81.5116 | 2317 | 7.16E+07 | 0.67082 | 1929.671 |
| 2413 | 81.4963 | 2349 | 7.06E+07 | 0.68058 | 1921.99 |
| 2445 | 81.4868 | 2381 | 6.92E+07 | 0.69308 | 1916.557 |
| 2477 | 81.4739 | 2413 | 6.84E+07 | 0.64393 | 1925.223 |
| 2509 | 81.4286 | 2445 | 6.68E+07 | 0.70391 | 1901.773 |
| 2541 | 81.4159 | 2477 | 6.55E+07 | 0.71239 | 1894.375 |
| 2573 | 81.4208 | 2509 | 6.41E+07 | 0.72112 | 1886.404 |
| 2605 | 81.4236 | 2541 | 6.35E+07 | 0.67864 | 1896.321 |
| 2638 | 81.4438 | 2574 | 6.29E+07 | 0.7266 | 1880.269 |
| 2669 | 81.3981 | 2606 | 6.15E+07 | 0.69894 | 1884.061 |
| 2702 | 81.4017 | 2638 | 6.06E+07 | 0.74291 | 1868.221 |
| 2734 | 81.407 | 2670 | 6.15E+07 | 0.74932 | 1870.443 |
| 2766 | 81.3782 | 2702 | 6.53E+07 | 0.68332 | 1903.434 |
| 2799 | 81.3613 | 2734 | 6.75E+07 | 0.68421 | 1910.869 |
| 2831 | 81.3435 | 2767 | 6.82E+07 | 0.67615 | 1915.464 |
| 2863 | 81.3524 | 2799 | 6.87E+07 | 0.67183 | 1918.127 |
| 2895 | 81.3495 | 2831 | 6.83E+07 | 0.65832 | 1921.248 |
| 2927 | 81.3342 | 2863 | 6.75E+07 | 0.7085 | 1905.309 |
| 2959 | 81.2954 | 2895 | 6.60E+07 | 0.71379 | 1895.843 |
| 2991 | 81.2712 | 2927 | 6.55E+07 | 0.65642 | 1911.311 |
| 3023 | 81.2976 | 2959 | 6.53E+07 | 0.64768 | 1914.093 |
| 3055 | 81.2815 | 2992 | 6.47E+07 | 0.66612 | 1905.287 |
| 3088 | 81.2828 | 3024 | 6.47E+07 | 0.66394 | 1905.604 |
| 3120 | 81.2337 | 3056 | 6.40E+07 | 0.66095 | 1903.886 |
| 3152 | 81.2532 | 3088 | 6.36E+07 | 0.64771 | 1907.3 |
| 3184 | 81.234 | 3120 | 6.31E+07 | 0.69014 | 1892.165 |
| 3216 | 81.2452 | 3152 | 6.24E+07 | 0.69906 | 1889.319 |
| 3248 | 81.2262 | 3185 | 6.22E+07 | 0.70598 | 1885.875 |
| 3281 | 81.1981 | 3217 | 6.18E+07 | 0.71481 | 1883.867 |
| 3313 | 81.2054 | 3249 | 6.29E+07 | 0.66868 | 1899.109 |
| 3345 | 81.1619 | 3281 | 6.48E+07 | 0.73052 | 1888.301 |
| 3377 | 81.1805 | 3313 | 6.48E+07 | 0.66862 | 1905.819 |
| 3409 | 81.1642 | 3345 | 6.40E+07 | 0.66528 | 1903.436 |
| 3441 | 81.1304 | 3377 | 6.36E+07 | 0.72406 | 1886.698 |
| 3474 | 81.1382 | 3409 | 6.25E+07 | 0.66977 | 1896.451 |
| 3505 | 81.1148 | 3442 | 6.18E+07 | 0.67127 | 1893.564 |
| 3538 | | 3474 | 6.13E+07 | 0.66984 | 1891.534 |
| 3570 | 81.1197 | 3506 | 6.04E+07 | 0.71405 | 1876.218 |
| 3602 | 81.1102 | 3538 | 5.93E+07 | 0.68294 | 1880.473 |
| 3634 | 81.0595 | 3570 | 5.84E+07 | 0.71565 | 1868.244 |
| 3667 | 81.0784 | 3602 | 5.73E+07 | 0.69142 | 1869.538 |
| 3698 | 81.0857 | 3634 | 5.73E+07 | 0.66596 | 1878.635 |
| 3730 | 81.0608 | 3667 | 6.26E+07 | 0.6593 | 1900.945 |
| 3763 | 81.0384 | 3699 | 7.01E+07 | 0.65372 | 1927.455 |
| 3795 | 81.0092 | 3731 | 7.36E+07 | 0.71107 | 1926.021 |
| 3827 | 80.9995 | 3763 | 7.60E+07 | 0.71011 | 1932.17 |
| 3854 | 80.9339 | 3795 | 7.63E+07 | 0.64603 | 1952.199 |
| 3896 | 80.8816 | 3827 | 7.56E+07 | 0.64225 | 1949.467 |
| 3924 | 80.8849 | 3859 | 7.41E+07 | 0.63901 | 1943.986 |
| 3955 | 80.8777 | 3892 | 7.15E+07 | 0.63621 | 1937.941 |
| 3987 | 80.854 | 3924 | 6.94E+07 | 0.68746 | 1915.703 |
| 4020 | 80.8463 | 3956 | 6.61E+07 | 0.64309 | 1916.182 |
| 4052 | 80.8456 | 3988 | 6.40E+07 | 0.68208 | 1896.752 |
| 4084 | 80.8248 | 4020 | 6.91E+07 | 0.63536 | 1928.984 |
| 4116 | 80.7899 | 4052 | 7.20E+07 | 0.624 | 1942.102 |
| 4148 | 80.7895 | 4084 | 7.24E+07 | 0.68712 | 1926.075 |
| 4180 | 80.7811 | 4116 | 7.10E+07 | 0.62161 | 1940.858 |
| 4212 | 80.7678 | 4149 | 7.02E+07 | 0.68254 | 1920.768 |
| 4245 | 80.7743 | 4181 | 6.96E+07 | 0.68263 | 1918.609 |
| 4277 | 80.7387 | 4213 | 6.91E+07 | 0.6266 | 1932.059 |
| 4309 | 80.7424 | 4245 | 6.76E+07 | 0.63487 | 1923.804 |
| 4341 | 80.7173 | 4277 | 6.67E+07 | 0.68213 | 1907.906 |
| 4373 | 80.7098 | 4309 | 6.59E+07 | 0.69484 | 1899.894 |
| 4405 | | 4341 | 6.53E+07 | 0.69743 | 1896.737 |
| 4438 | 80.714 | 4374 | 6.39E+07 | 0.70053 | 1893.333 |
| 4469 | 80.6881 | 4406 | 6.31E+07 | 0.64786 | 1905.57 |
| 4502 | 80.6719 | 4438 | 6.24E+07 | 0.65687 | 1898.866 |
| 4534 | 80.6683 | 4470 | 6.46E+07 | 0.66913 | 1902.803 |
| 4566 | 80.6642 | 4502 | 6.48E+07 | 0.70808 | 1894.659 |
| 4598 | 80.6371 | 4534 | 6.40E+07 | 0.64726 | 1907.949 |
| 4630 | 80.6351 | 4566 | 6.34E+07 | 0.65967 | 1901.877 |
| 4662 | 80.6312 | 4598 | 6.23E+07 | 0.70161 | 1887.212 |
| 4694 | 80.5994 | 4631 | 6.17E+07 | 0.64894 | 1898.989 |
| 4727 | 80.5973 | 4663 | 6.17E+07 | 0.64616 | 1899.852 |
| 4759 | 80.6115 | 4695 | 6.50E+07 | 0.62769 | 1917.413 |
| 4791 | 80.5855 | 4727 | 6.84E+07 | 0.68965 | 1913.684 |
| 4823 | 80.5452 | 4759 | 7.07E+07 | 0.69595 | 1919.449 |
| 4855 | 80.5196 | 4791 | 7.24E+07 | 0.6286 | 1943.857 |
| 4887 | 80.4921 | 4823 | 7.28E+07 | 0.69049 | 1928.53 |
| 4919 | 80.4636 | 4855 | 7.36E+07 | 0.63315 | 1945.988 |
| 4951 | 80.454 | 4888 | 7.32E+07 | 0.63419 | 1944.71 |
| 4984 | | 4920 | 7.27E+07 | 0.6256 | 1945.095 |
| 5012 | | 4952 | 7.27E+07 | 0.67453 | 1929.252 |
| 5049 | 80.393 | 4984 | 7.24E+07 | 0.68417 | 1925.313 |
| 5081 | 80.381 | 5016 | 7.42E+07 | 0.61876 | 1949.848 |
| 5112 | 80.3444 | 5048 | 7.10E+07 | 0.67982 | 1923.732 |
| 5144 | 80.3418 | 5080 | 7.11E+07 | 0.63347 | 1936.46 |
| 5176 | 80.3161 | 5113 | 7.01E+07 | 0.69106 | 1917.408 |
| 5209 | 80.3232 | 5145 | 6.88E+07 | 0.69453 | 1909.701 |
| 5240 | 80.2979 | 5177 | 6.77E+07 | 0.69671 | 1907.272 |
| 5273 | 80.3096 | 5209 | 6.81E+07 | 0.65424 | 1919.047 |
| 5305 | 80.3022 | 5241 | 6.87E+07 | 0.7077 | 1905.021 |
| 5337 | 80.2827 | 5273 | 6.80E+07 | 0.66124 | 1917.053 |
| 5369 | 80.2705 | 5305 | 6.81E+07 | 0.66102 | 1916.656 |
| 5401 | 80.2633 | 5337 | 6.82E+07 | 0.64237 | 1923.961 |
| 5433 | | 5369 | 6.77E+07 | 0.69858 | 1906.713 |
| 5466 | 80.2634 | 5402 | 6.70E+07 | 0.70789 | 1897.66 |
| 5497 | 80.2872 | 5434 | 6.66E+07 | 0.70793 | 1897.093 |
| 5529 | 80.2081 | 5466 | 6.70E+07 | 0.65954 | 1913.485 |
| 5562 | 80.2255 | 5498 | 6.59E+07 | 0.66142 | 1909.457 |
| 5594 | 80.1955 | 5530 | 6.56E+07 | 0.72143 | 1891.896 |
| 5626 | | 5562 | 6.63E+07 | 0.65706 | 1912.677 |
| 5658 | 80.1948 | 5594 | 6.64E+07 | 0.71555 | 1893.831 |
| 5690 | 80.1646 | 5626 | 6.50E+07 | 0.66718 | 1904.537 |
| 5723 | 80.1884 | 5658 | 6.53E+07 | 0.71875 | 1889.564 |
| 5754 | 80.1626 | 5691 | 6.33E+07 | 0.68644 | 1891.056 |
| 5787 | 80.1678 | 5723 | 6.37E+07 | 0.67016 | 1898.592 |
| 5819 | 80.1561 | 5755 | 6.58E+07 | 0.7262 | 1891.446 |
| 5851 | 80.1572 | 5787 | 6.57E+07 | 0.67201 | 1905.448 |
| 5883 | 80.1283 | 5819 | 6.56E+07 | 0.67649 | 1903.066 |
| 5916 | 80.1602 | 5851 | 6.58E+07 | 0.72057 | 1891.136 |
| 5947 | 80.1086 | 5883 | 6.55E+07 | 0.72774 | 1888.406 |
| 5980 | 80.128 | 5915 | 6.54E+07 | 0.68064 | 1900.328 |
| 6011 | 80.0923 | 5948 | 6.60E+07 | 0.73026 | 1889.218 |
| 6043 | 80.0627 | 5980 | 6.68E+07 | 0.67178 | 1909.012 |
| 6076 | 80.0618 | 6012 | 6.67E+07 | 0.74151 | 1889.722 |
| 6108 | 80.0493 | 6044 | 6.72E+07 | 0.67487 | 1909.461 |
| 6136 | | 6076 | 6.81E+07 | 0.7348 | 1897.439 |
| 6176 | 80.0175 | 6108 | 6.67E+07 | 0.67577 | 1908.762 |
| 6204 | 80.0021 | 6140 | 6.76E+07 | 0.73473 | 1894.907 |
| 6236 | 79.989 | 6172 | 6.71E+07 | 0.74795 | 1888.271 |
| 6268 | 80.0035 | 6204 | 6.63E+07 | 0.68155 | 1902.978 |
| 6300 | 79.9742 | 6237 | 6.77E+07 | 0.67921 | 1910.086 |
| 6333 | 79.9677 | 6269 | 6.73E+07 | 0.67491 | 1909.23 |
| 6365 | 79.9553 | 6301 | 6.64E+07 | 0.69028 | 1900.891 |
| 6397 | 79.9371 | 6333 | 6.63E+07 | 0.67615 | 1905.273 |
| 6429 | 79.9562 | 6365 | 6.65E+07 | 0.66088 | 1910.662 |
| 6461 | 79.8994 | 6397 | 6.57E+07 | 0.7101 | 1894.615 |
| 6494 | 79.909 | 6429 | 6.46E+07 | 0.71659 | 1889.6 |
| 6526 | 79.9291 | 6462 | 6.26E+07 | 0.72534 | 1876.838 |
| 6557 | 79.8916 | 6494 | 6.26E+07 | 0.7327 | 1877.918 |
| 6590 | 79.8938 | 6526 | 6.33E+07 | 0.68229 | 1893.029 |
| 6622 | 79.9043 | 6558 | 6.43E+07 | 0.74908 | 1878.197 |
| 6654 | 79.8852 | 6590 | 6.55E+07 | 0.75726 | 1879.269 |
| 6686 | | 6622 | 6.68E+07 | 0.69485 | 1900.119 |
| 6718 | 79.8853 | 6654 | 6.82E+07 | 0.68707 | 1908.297 |
| 6751 | 79.8783 | 6687 | 6.92E+07 | 0.6784 | 1914.355 |
| 6782 | 79.8506 | 6719 | 7.05E+07 | 0.67766 | 1919.321 |
| 6815 | 79.8417 | 6751 | 7.02E+07 | 0.734 | 1904.302 |
| 6847 | 79.8207 | 6783 | 6.83E+07 | 0.65968 | 1918.057 |
| 6879 | 79.8185 | 6815 | 6.48E+07 | 0.72173 | 1888.279 |
| 6911 | 79.8009 | 6847 | 6.61E+07 | 0.72901 | 1890.4 |
| 6943 | 79.7965 | 6879 | 6.90E+07 | 0.6692 | 1917.253 |
| 6975 | 79.7736 | 6911 | 7.21E+07 | 0.73786 | 1908.035 |
| 7008 | 79.7645 | 6944 | 7.37E+07 | 0.66832 | 1933.633 |
| 7039 | 79.7547 | 6976 | 7.50E+07 | 0.66996 | 1937.078 |
| 7072 | 79.7651 | 7008 | 7.24E+07 | 0.72962 | 1911.589 |
| 7104 | 79.7403 | 7040 | 6.63E+07 | 0.66024 | 1910.836 |
| 7136 | 79.7264 | 7072 | 6.06E+07 | 0.72786 | 1870.443 |
| 7168 | 79.7202 | 7104 | 6.32E+07 | 0.67706 | 1892.944 |
| 7200 | 79.6907 | 7136 | 6.66E+07 | 0.6583 | 1912.344 |
| 7232 | 79.6628 | 7168 | 6.91E+07 | 0.65608 | 1921.513 |
| 7265 | 79.686 | 7200 | 7.04E+07 | 0.64211 | 1929.92 |
| 7296 | 79.649 | 7232 | 6.94E+07 | 0.63867 | 1927.045 |
| 7329 | 79.6471 | 7265 | 6.56E+07 | 0.68717 | 1901.513 |
| 7361 | | 7297 | 5.98E+07 | 0.68929 | 1876.848 |
| 7392 | 79.6576 | 7329 | 5.93E+07 | 0.6958 | 1871.978 |
| 7425 | 79.6644 | 7361 | 6.10E+07 | 0.70414 | 1875.873 |
| 7457 | 79.6502 | 7393 | 6.26E+07 | 0.67936 | 1889.235 |
| 7489 | 79.6411 | 7425 | 6.42E+07 | 0.68315 | 1895.63 |
| 7521 | 79.6433 | 7457 | 6.48E+07 | 0.72486 | 1886.464 |
| 7553 | 79.6291 | 7489 | 6.30E+07 | 0.73233 | 1877.923 |
| 7586 | 79.6235 | 7522 | 6.04E+07 | 0.70008 | 1876.485 |
| 7618 | 79.6147 | 7554 | 6.22E+07 | 0.67573 | 1891.224 |
| 7650 | 79.6238 | 7586 | 6.38E+07 | 0.69361 | 1889.84 |
| 7682 | | 7618 | 6.55E+07 | 0.67623 | 1902.086 |
| 7714 | 79.6023 | 7650 | 6.66E+07 | 0.73038 | 1891.753 |
| 7746 | 79.6069 | 7682 | 6.84E+07 | 0.66044 | 1917.541 |
| 7778 | 79.595 | 7714 | 7.02E+07 | 0.65291 | 1925.974 |
| 7810 | 79.5533 | 7746 | 7.17E+07 | 0.71685 | 1913.214 |
| 7843 | 79.565 | 7779 | 7.19E+07 | 0.65299 | 1930.752 |
| 7875 | 79.554 | 7811 | 7.10E+07 | 0.71685 | 1912.299 |
| 7907 | 79.5304 | 7843 | 6.74E+07 | 0.71401 | 1898.023 |
| 7939 | 79.5316 | 7875 | 6.28E+07 | 0.67791 | 1892.312 |
| 7971 | 79.5268 | 7907 | 5.83E+07 | 0.73051 | 1861.067 |
| 8003 | 79.4988 | 7939 | 5.41E+07 | 0.73338 | 1840.602 |
| 8035 | 79.5073 | 7971 | 5.58E+07 | 0.71029 | 1854.576 |
| 8067 | 79.5128 | 8003 | 5.85E+07 | 0.70684 | 1865.854 |
| 8099 | 79.4776 | 8036 | 6.02E+07 | 0.69374 | 1876.617 |
| 8132 | 79.4951 | 8068 | 6.17E+07 | 0.7507 | 1867.361 |
| 8164 | 79.5144 | 8100 | 6.26E+07 | 0.70555 | 1881.455 |
| 8196 | 79.4831 | 8132 | 6.44E+07 | 0.74909 | 1878.387 |
| 8228 | 79.4988 | 8164 | 6.40E+07 | 0.75724 | 1873.704 |
| 8260 | 79.4674 | 8196 | 6.22E+07 | 0.76783 | 1863.958 |
| 8292 | 79.4839 | 8228 | 5.96E+07 | 0.70878 | 1870.811 |
| 8324 | 79.4838 | 8260 | 6.18E+07 | 0.70081 | 1882.328 |
| 8356 | 79.4727 | 8293 | 6.13E+07 | 0.69991 | 1881.311 |
| 8389 | 79.4813 | 8325 | 5.84E+07 | 0.76326 | 1851.692 |
| 8420 | 79.4267 | 8357 | 5.75E+07 | 0.69846 | 1865.656 |
| 8453 | 79.4707 | 8389 | 5.83E+07 | 0.71771 | 1863.867 |
| 8484 | 79.4136 | 8421 | 5.98E+07 | 0.71287 | 1870.735 |
| 8517 | 79.4285 | 8453 | 6.05E+07 | 0.68252 | 1882.633 |
| 8549 | 79.4317 | 8485 | 6.13E+07 | 0.75071 | 1866.972 |
| 8581 | 79.3909 | 8517 | 6.22E+07 | 0.68922 | 1886.187 |
| 8614 | 79.4106 | 8550 | 6.21E+07 | 0.74614 | 1870.542 |
| 8645 | 79.402 | 8582 | 6.18E+07 | 0.75211 | 1867.884 |
| 8678 | 79.4088 | 8614 | 6.11E+07 | 0.69865 | 1879.195 |
| 8710 | 79.4009 | 8646 | 6.01E+07 | 0.691 | 1877.789 |
| 8742 | 79.3545 | 8678 | 5.77E+07 | 0.70074 | 1864.942 |
| 8774 | 79.345 | 8710 | 5.72E+07 | 0.72472 | 1856.328 |
| 8806 | 79.3675 | 8742 | 6.04E+07 | 0.68403 | 1882.273 |
| 8838 | 79.3568 | 8774 | 6.16E+07 | 0.73482 | 1871.516 |
| 8870 | 79.3473 | 8806 | 6.31E+07 | 0.67171 | 1895.225 |
| 8902 | 79.3495 | 8839 | 6.37E+07 | 0.73837 | 1878.637 |
| 8934 | 79.349 | 8871 | 6.50E+07 | 0.67519 | 1899.948 |
| 8967 | | 8903 | 6.59E+07 | 0.73587 | 1888.303 |
| 8999 | 79.3264 | 8935 | 6.75E+07 | 0.67563 | 1910.384 |
| 9031 | 79.2836 | 8967 | 6.91E+07 | 0.73432 | 1899.916 |
| 9063 | 79.3023 | 8999 | 7.05E+07 | 0.67628 | 1919.882 |
| 9095 | 79.2863 | 9031 | 7.14E+07 | 0.66448 | 1927.743 |
| 9127 | 79.2555 | 9063 | 7.14E+07 | 0.66245 | 1926.959 |
| 9159 | 79.2714 | 9095 | 7.11E+07 | 0.71677 | 1912.089 |
| 9191 | 79.2719 | 9128 | 7.05E+07 | 0.7254 | 1905.668 |
| 9223 | 79.231 | 9160 | 6.96E+07 | 0.65462 | 1924.148 |
| 9256 | 79.246 | 9192 | 6.54E+07 | 0.71898 | 1891.228 |
| 9288 | 79.229 | 9224 | 6.40E+07 | 0.73043 | 1880.128 |
| 9320 | 79.2077 | 9256 | 6.44E+07 | 0.7447 | 1880.964 |
| 9352 | 79.1708 | 9288 | 6.50E+07 | 0.69807 | 1892.48 |
| 9384 | 79.1867 | 9320 | 6.45E+07 | 0.68019 | 1898.269 |
| 9416 | 79.1655 | 9352 | 6.35E+07 | 0.7502 | 1875.02 |
| 9449 | 79.1659 | 9385 | 6.29E+07 | 0.68468 | 1889.452 |
| 9481 | 79.1754 | 9417 | 6.37E+07 | 0.67599 | 1895.819 |
| 9513 | 79.175 | 9449 | 6.40E+07 | 0.67018 | 1897.636 |
| 9545 | 79.1585 | 9481 | 6.41E+07 | 0.67267 | 1898.061 |
| 9577 | 79.157 | 9513 | 6.44E+07 | 0.72528 | 1886.382 |
| 9609 | 79.1334 | 9545 | 6.48E+07 | 0.72005 | 1887.127 |
| 9641 | 79.1342 | 9577 | 6.51E+07 | 0.72747 | 1887.163 |
| 9673 | 79.1324 | 9610 | 6.57E+07 | 0.6733 | 1903.604 |
| 9706 | 79.1354 | 9642 | 6.57E+07 | 0.66857 | 1905.881 |
| 9738 | 79.1158 | 9674 | 6.59E+07 | 0.73808 | 1886.753 |
| 9770 | 79.1163 | 9706 | 6.62E+07 | 0.74077 | 1886.798 |
| 9802 | 79.1076 | 9738 | 6.63E+07 | 0.74512 | 1885.996 |
| 9834 | | 9770 | 6.60E+07 | 0.67751 | 1904.054 |
| 9867 | 79.1267 | 9802 | 6.48E+07 | 0.67831 | 1899.841 |
| 9898 | 79.109 | 9834 | 6.46E+07 | 0.68016 | 1897.809 |
| 9930 | 79.0817 | 9866 | 6.48E+07 | 0.67006 | 1901.565 |
| 9962 | 79.0589 | 9899 | 6.56E+07 | 0.72474 | 1887.277 |
| 9995 | 79.0624 | 9931 | 6.65E+07 | 0.72046 | 1893.168 |
| 10027 | 79.0574 | 9963 | 6.66E+07 | 0.66001 | 1910.4 |
| 10059 | 79.0462 | 9995 | 6.63E+07 | 0.72564 | 1892.13 |
| 10091 | 79.0255 | 10027 | 6.63E+07 | 0.66517 | 1908.451 |
| 10123 | 79.0027 | 10059 | 6.54E+07 | 0.73417 | 1887.458 |
| 10156 | 79.0412 | 10091 | 6.64E+07 | 0.73615 | 1889.924 |
| 10187 | 79.0121 | 10124 | 6.66E+07 | 0.66902 | 1908.213 |
| 10220 | 79.0196 | 10156 | 6.61E+07 | 0.67073 | 1905.755 |
| 10252 | 79.0076 | 10188 | 6.67E+07 | 0.7298 | 1892.021 |
| 10284 | 79.0111 | 10220 | 6.58E+07 | 0.67257 | 1903.575 |
| 10316 | 78.9751 | 10252 | 6.49E+07 | 0.67726 | 1898.853 |
| 10348 | 79.0145 | 10284 | 6.44E+07 | 0.71807 | 1885.881 |
| 10380 | 78.9709 | 10316 | 6.38E+07 | 0.72837 | 1880.281 |
| 10412 | 78.952 | 10348 | 6.40E+07 | 0.66995 | 1899.021 |
| 10445 | 78.993 | 10380 | 6.52E+07 | 0.72791 | 1888.34 |
| 10476 | 78.9897 | 10413 | 6.40E+07 | 0.6691 | 1899.38 |
| 10508 | 78.9476 | 10445 | 6.39E+07 | 0.67786 | 1896.3 |
| 10541 | 78.933 | 10477 | 6.41E+07 | 0.72937 | 1883.103 |
| 10573 | 78.9515 | 10509 | 6.42E+07 | 0.72958 | 1883.199 |
| 10605 | 78.9511 | 10541 | 6.41E+07 | 0.7421 | 1879.809 |
| 10637 | 78.9666 | 10573 | 6.46E+07 | 0.67412 | 1900.026 |
| 10669 | 78.9439 | 10605 | 6.48E+07 | 0.67508 | 1900.284 |
| 10701 | 78.9142 | 10637 | 6.44E+07 | 0.67455 | 1898.71 |
| 10733 | 78.9288 | 10669 | 6.47E+07 | 0.67593 | 1898.708 |
| 10765 | 78.9228 | 10701 | 6.57E+07 | 0.72277 | 1888.806 |
| 10797 | 78.9317 | 10734 | 6.65E+07 | 0.66337 | 1909.09 |
| 10829 | 78.8895 | 10766 | 6.70E+07 | 0.64876 | 1916.034 |
| 10862 | 78.8997 | 10798 | 6.69E+07 | 0.65717 | 1912.504 |
| 10894 | 78.8932 | 10830 | 6.72E+07 | 0.64869 | 1916.592 |
| 10926 | 78.8615 | 10862 | 6.71E+07 | 0.64992 | 1915.776 |
| 10958 | 78.8617 | 10894 | 6.68E+07 | 0.64636 | 1916.752 |
| 10990 | 78.8739 | 10926 | 6.67E+07 | 0.68269 | 1904.363 |
| 11022 | 78.8611 | 10958 | 6.61E+07 | 0.68981 | 1902.617 |
| 11054 | 78.87 | 10991 | 6.58E+07 | 0.69667 | 1897.106 |
| 11086 | 78.8262 | 11023 | 6.46E+07 | 0.70674 | 1891.875 |
| 11119 | 78.8489 | 11055 | 6.53E+07 | 0.67051 | 1903.565 |
| 11150 | 78.8121 | 11087 | 6.77E+07 | 0.66544 | 1912.462 |
| 11183 | 78.8037 | 11119 | 7.00E+07 | 0.72013 | 1906.172 |
| 11215 | 78.7664 | 11151 | 7.02E+07 | 0.72649 | 1904.249 |
| 11247 | 78.7759 | 11183 | 7.18E+07 | 0.66297 | 1927.791 |
| 11279 | 78.7582 | 11215 | 7.29E+07 | 0.72356 | 1915.32 |
| 11311 | 78.7389 | 11248 | 7.30E+07 | 0.66473 | 1931.565 |
| 11343 | 78.7227 | 11280 | 7.32E+07 | 0.73456 | 1911.869 |
| 11376 | 78.7404 | 11312 | 7.33E+07 | 0.73759 | 1911.478 |
| 11408 | 78.7132 | 11344 | 7.30E+07 | 0.66945 | 1930.777 |
| 11440 | 78.6639 | 11376 | 7.25E+07 | 0.67068 | 1928.871 |
| 11472 | 78.6982 | 11408 | 7.24E+07 | 0.66063 | 1931.602 |
| 11504 | 78.6914 | 11440 | 7.16E+07 | 0.65985 | 1928.997 |
| 11536 | 78.6284 | 11472 | 7.19E+07 | 0.71362 | 1914.639 |
| 11569 | 78.6228 | 11504 | 7.15E+07 | 0.70859 | 1912.591 |
| 11600 | 78.6091 | 11537 | 7.10E+07 | 0.64854 | 1930.355 |
| 11633 | 78.5863 | 11569 | 7.06E+07 | 0.71216 | 1909.783 |
| 11665 | 78.6193 | 11601 | 7.05E+07 | 0.65412 | 1926.189 |
| 11696 | 78.6043 | 11633 | 7.00E+07 | 0.65304 | 1926.822 |
| 11729 | 78.5966 | 11665 | 6.92E+07 | 0.70905 | 1905.099 |
| 11761 | 78.5899 | 11697 | 6.93E+07 | 0.64211 | 1926.194 |
| 11793 | 78.5917 | 11729 | 6.87E+07 | 0.64836 | 1922.455 |
| 11825 | 78.572 | 11762 | 6.86E+07 | 0.6944 | 1908.914 |
| 11858 | | 11794 | 6.84E+07 | 0.64872 | 1920.636 |
| 11889 | 78.5501 | 11826 | 6.74E+07 | 0.70196 | 1904.24 |
| 11922 | 78.5485 | 11858 | 6.76E+07 | 0.70047 | 1903.275 |
| 11954 | 78.5563 | 11890 | 6.72E+07 | 0.653 | 1914.774 |
| 11986 | 78.5461 | 11922 | 6.67E+07 | 0.71369 | 1894.466 |
| 12018 | 78.5394 | 11954 | 6.61E+07 | 0.64939 | 1911.871 |
| 12050 | 78.4948 | 11986 | 6.58E+07 | 0.71789 | 1890.453 |
| 12078 | 78.4569 | 12018 | 6.54E+07 | 0.6683 | 1902.466 |
| 12119 | 78.4469 | 12051 | 6.46E+07 | 0.66248 | 1903.586 |
| 12147 | 78.4767 | 12083 | 6.39E+07 | 0.71647 | 1884.259 |
| 12178 | 78.4568 | 12115 | 6.38E+07 | 0.71705 | 1885.695 |
| 12211 | 78.4336 | 12147 | 6.39E+07 | 0.73126 | 1879.168 |
| 12243 | 78.4312 | 12179 | 6.42E+07 | 0.69069 | 1891.949 |
| 12275 | 78.4369 | 12211 | 6.40E+07 | 0.68219 | 1894.13 |
| 12307 | 78.3993 | 12243 | 6.33E+07 | 0.74291 | 1873.541 |
| 12340 | 78.4114 | 12275 | 6.34E+07 | 0.67718 | 1893.96 |
| 12371 | 78.4055 | 12307 | 6.43E+07 | 0.6719 | 1897.464 |
| 12404 | 78.3672 | 12340 | 6.48E+07 | 0.72762 | 1884.946 |
| 12436 | 78.3644 | 12372 | 6.41E+07 | 0.67649 | 1896.149 |
| 12468 | 78.3447 | 12404 | 6.48E+07 | 0.73418 | 1883.217 |
| 12500 | 78.3573 | 12436 | 6.50E+07 | 0.73235 | 1881.573 |
| 12532 | 78.3385 | 12468 | 6.56E+07 | 0.67309 | 1901.908 |
| 12564 | 78.3312 | 12500 | 6.64E+07 | 0.67331 | 1904.437 |
| 12596 | 78.2845 | 12533 | 6.67E+07 | 0.73278 | 1888.043 |
| 12629 | 78.325 | 12565 | 6.76E+07 | 0.7446 | 1887.968 |
| 12660 | 78.2881 | 12597 | 6.76E+07 | 0.74679 | 1887.817 |
| 12693 | 78.2866 | 12629 | 6.75E+07 | 0.74982 | 1887.053 |
| 12725 | 78.2607 | 12661 | 6.86E+07 | 0.69298 | 1905.42 |
| 12757 | 78.2251 | 12693 | 6.94E+07 | 0.77006 | 1886.065 |
| 12789 | 78.249 | 12725 | 7.04E+07 | 0.70381 | 1909.384 |
| 12821 | 78.1877 | 12757 | 7.09E+07 | 0.69606 | 1912.497 |
| 12854 | 78.1788 | 12790 | 7.19E+07 | 0.68642 | 1919.212 |
| 12886 | 78.1593 | 12822 | 7.26E+07 | 0.67437 | 1926.192 |
| 12918 | 78.1852 | 12854 | 7.22E+07 | 0.67084 | 1925.593 |
| 12950 | 78.1641 | 12886 | 7.20E+07 | 0.72061 | 1909.679 |
| 12982 | 78.1124 | 12918 | 7.17E+07 | 0.72414 | 1909.526 |
| 13014 | 78.1152 | 12950 | 7.15E+07 | 0.73004 | 1907.37 |
| 13046 | 78.0723 | 12982 | 7.09E+07 | 0.6688 | 1920.445 |
| 13079 | 78.0813 | 13015 | 7.16E+07 | 0.66774 | 1924.138 |
| 13110 | 78.0849 | 13047 | 7.09E+07 | 0.73738 | 1901.542 |
| 13142 | 78.0436 | 13079 | 6.99E+07 | 0.7409 | 1896.778 |
| 13175 | 78.008 | 13111 | 6.85E+07 | 0.74442 | 1892.379 |
| 13207 | | 13143 | 6.66E+07 | 0.74744 | 1885.338 |
| 13239 | 78.0175 | 13175 | 6.31E+07 | 0.76396 | 1867.61 |
| 13271 | 78.0007 | 13207 | 6.26E+07 | 0.70948 | 1880.251 |
| 13303 | 77.984 | 13239 | 6.35E+07 | 0.70418 | 1884.624 |
| 13335 | 77.9745 | 13272 | 6.44E+07 | 0.6972 | 1889.898 |
| 13367 | 77.9679 | 13304 | 6.32E+07 | 0.68823 | 1888.716 |
| 13400 | 77.9522 | 13336 | 6.20E+07 | 0.69025 | 1882.576 |
| 13432 | 77.9401 | 13368 | 6.17E+07 | 0.68225 | 1884.913 |
| 13464 | 77.9518 | 13400 | 6.23E+07 | 0.68499 | 1886.327 |
| 13496 | 77.9158 | 13432 | 6.28E+07 | 0.72301 | 1878.477 |
| 13528 | 77.873 | 13464 | 6.30E+07 | 0.72844 | 1876.958 |
| 13561 | 77.9002 | 13496 | 6.28E+07 | 0.73688 | 1874.6 |
| 13592 | 77.8818 | 13529 | 6.10E+07 | 0.69737 | 1877.481 |
| 13624 | 77.8802 | 13561 | 6.18E+07 | 0.69139 | 1882.43 |
| 13656 | 77.86 | 13593 | 6.33E+07 | 0.69198 | 1887.515 |
| 13689 | 77.8263 | 13625 | 6.43E+07 | 0.68857 | 1891.927 |
| 13721 | 77.8369 | 13657 | 6.60E+07 | 0.66618 | 1905.854 |
| 13753 | 77.8124 | 13689 | 6.78E+07 | 0.65274 | 1916.421 |
| 13785 | 77.8075 | 13721 | 6.84E+07 | 0.64427 | 1921.047 |
| 13817 | 77.7998 | 13753 | 6.90E+07 | 0.64508 | 1923.571 |
| 13849 | 77.7985 | 13785 | 6.93E+07 | 0.69234 | 1909.223 |
| 13881 | 77.7905 | 13818 | 6.89E+07 | 0.69313 | 1909.471 |
| 13913 | 77.7512 | 13850 | 6.90E+07 | 0.69878 | 1904.649 |
| 13946 | 77.7329 | 13882 | 6.90E+07 | 0.70845 | 1904.343 |
| 13978 | 77.7118 | 13914 | 6.92E+07 | 0.65318 | 1919.978 |
| 14010 | 77.7195 | 13946 | 6.90E+07 | 0.65472 | 1918.881 |
| 14042 | 77.7045 | 13978 | 6.85E+07 | 0.64835 | 1919.581 |
| 14074 | | 14010 | 6.78E+07 | 0.70633 | 1901.768 |
| 14106 | 77.679 | 14042 | 6.63E+07 | 0.64167 | 1914.001 |
| 14138 | 77.6615 | 14075 | 6.40E+07 | 0.69389 | 1889.164 |
| 14170 | 77.6449 | 14107 | 6.17E+07 | 0.70244 | 1879.827 |
| 14203 | 77.6332 | 14139 | 6.17E+07 | 0.71246 | 1877.308 |
| 14235 | 77.6427 | 14171 | 6.12E+07 | 0.72593 | 1869.501 |
| 14267 | 77.6222 | 14203 | 6.16E+07 | 0.68263 | 1883.515 |
| 14299 | 77.6049 | 14235 | 6.20E+07 | 0.73864 | 1871.258 |
| 14331 | 77.5936 | 14267 | 6.25E+07 | 0.74804 | 1869.107 |
| 14363 | 77.5731 | 14299 | 6.26E+07 | 0.68546 | 1887.965 |
| 14395 | 77.5663 | 14332 | 6.26E+07 | 0.75195 | 1867.222 |
| 14428 | 77.5413 | 14364 | 6.46E+07 | 0.68528 | 1894.741 |
| 14460 | 77.5199 | 14396 | 6.58E+07 | 0.68578 | 1899.169 |
| 14492 | 77.5191 | 14428 | 6.68E+07 | 0.74687 | 1884.889 |
| 14524 | 77.5278 | 14460 | 6.67E+07 | 0.74662 | 1884.977 |
| 14556 | 77.5174 | 14492 | 6.52E+07 | 0.68714 | 1896.226 |
| 14588 | 77.4897 | 14524 | 6.63E+07 | 0.67482 | 1903.435 |
| 14620 | 77.4863 | 14557 | 6.76E+07 | 0.67283 | 1909.747 |
| 14652 | 77.4592 | 14589 | 6.85E+07 | 0.67037 | 1913.298 |
| 14685 | 77.4444 | 14621 | 6.94E+07 | 0.65584 | 1920.325 |
| 14717 | 77.42 | 14653 | 6.94E+07 | 0.70581 | 1907.74 |
| 14749 | 77.4168 | 14685 | 6.94E+07 | 0.70973 | 1903.691 |
| 14781 | 77.3942 | 14717 | 6.79E+07 | 0.71706 | 1897.06 |
| 14813 | 77.3717 | 14749 | 6.69E+07 | 0.72635 | 1893.163 |
| 14845 | 77.3802 | 14781 | 6.77E+07 | 0.73811 | 1890.802 |
| 14877 | 77.372 | 14814 | 6.85E+07 | 0.68344 | 1909.163 |
| 14909 | 77.3553 | 14846 | 6.87E+07 | 0.67919 | 1912.206 |
| 14942 | 77.3333 | 14878 | 6.84E+07 | 0.67339 | 1912.228 |
| 14974 | 77.3183 | 14910 | 6.86E+07 | 0.7351 | 1894.814 |
| 15006 | 77.3005 | 14942 | 6.83E+07 | 0.73694 | 1894.114 |
| 15038 | 77.2723 | 14974 | 6.94E+07 | 0.73669 | 1898.061 |
| 15070 | 77.2806 | 15006 | 6.95E+07 | 0.74788 | 1895.034 |
| 15102 | 77.2722 | 15038 | 7.01E+07 | 0.6897 | 1913.431 |
| 15134 | 77.268 | 15071 | 7.14E+07 | 0.6928 | 1916.172 |
| 15166 | 77.222 | 15103 | 7.10E+07 | 0.68476 | 1916.683 |
| 15199 | 77.1959 | 15135 | 7.10E+07 | 0.74588 | 1900.723 |
| 15231 | 77.1644 | 15167 | 7.16E+07 | 0.67659 | 1923.369 |
| 15263 | 77.1353 | 15199 | 7.19E+07 | 0.66305 | 1926.057 |
| 15295 | 77.1161 | 15231 | 7.22E+07 | 0.65799 | 1930.096 |
| 15327 | 77.1301 | 15263 | 7.26E+07 | 0.65137 | 1933.391 |
| 15359 | 77.0954 | 15295 | 7.24E+07 | 0.71293 | 1915.287 |
| 15391 | 77.0868 | 15327 | 7.19E+07 | 0.63877 | 1934.918 |
| 15423 | 77.0646 | 15360 | 7.21E+07 | 0.63189 | 1936.604 |
| 15456 | 77.0812 | 15392 | 7.09E+07 | 0.63009 | 1934.048 |
| 15487 | 77.0398 | 15424 | 7.08E+07 | 0.68566 | 1919.48 |
| 15520 | 77.0348 | 15456 | 7.03E+07 | 0.69286 | 1912.524 |
| 15552 | 77.0059 | 15488 | 6.96E+07 | 0.68865 | 1910.692 |
| 15584 | 77.0009 | 15520 | 6.97E+07 | 0.63438 | 1928.54 |
| 15616 | 76.991 | 15552 | 6.98E+07 | 0.63982 | 1926.492 |
| 15643 | 76.9746 | 15584 | 6.98E+07 | 0.70169 | 1907.663 |
| 15685 | | 15616 | 6.97E+07 | 0.63791 | 1928.306 |
| 15712 | 76.9415 | 15648 | 6.92E+07 | 0.69351 | 1907.132 |
| 15744 | 76.9062 | 15681 | 6.86E+07 | 0.63648 | 1923.52 |
| 15777 | 76.9274 | 15713 | 6.80E+07 | 0.69498 | 1904.316 |
| 15808 | 76.9008 | 15745 | 6.74E+07 | 0.64724 | 1915.93 |
| 15841 | 76.8969 | 15777 | 6.72E+07 | 0.69513 | 1902.189 |
| 15873 | 76.8417 | 15809 | 6.72E+07 | 0.70714 | 1899.49 |
| 15906 | | 15841 | 6.66E+07 | 0.71001 | 1893.368 |
| 15937 | 76.8332 | 15873 | 6.60E+07 | 0.72544 | 1886.689 |
| 15969 | 76.8474 | 15906 | 6.57E+07 | 0.66214 | 1905.157 |
| 16001 | 76.833 | 15938 | 6.45E+07 | 0.66497 | 1899.758 |
| 16033 | 76.8019 | 15970 | 6.38E+07 | 0.72864 | 1878.257 |
| 16066 | 76.793 | 16002 | 6.38E+07 | 0.66333 | 1898.722 |
| 16098 | 76.7865 | 16034 | 6.52E+07 | 0.65446 | 1906.13 |
| 16125 | 76.746 | 16066 | 6.58E+07 | 0.66136 | 1906.252 |
| 16167 | 76.746 | 16098 | 6.67E+07 | 0.64724 | 1914.788 |
| 16194 | 76.7578 | 16130 | 6.61E+07 | 0.70456 | 1893.14 |
| 16226 | 76.7312 | 16162 | 6.56E+07 | 0.6995 | 1895.033 |
| 16259 | 76.7365 | 16195 | 6.58E+07 | 0.66021 | 1905.223 |
| 16290 | 76.7121 | 16227 | 6.54E+07 | 0.7083 | 1889.945 |
| 16323 | 76.7225 | 16259 | 6.42E+07 | 0.71816 | 1885.892 |
| 16354 | 76.6803 | 16291 | 6.41E+07 | 0.66799 | 1896.435 |
| 16387 | 76.6811 | 16323 | 6.35E+07 | 0.67891 | 1890.837 |
| 16419 | 76.6811 | 16355 | 6.26E+07 | 0.72105 | 1877.342 |
| 16451 | 76.6669 | 16387 | 6.27E+07 | 0.66504 | 1892.568 |
| 16483 | 76.687 | 16419 | 6.21E+07 | 0.68264 | 1885.5 |
| 16515 | 76.6676 | 16452 | 6.13E+07 | 0.71728 | 1871.528 |
| 16548 | 76.6619 | 16484 | 6.20E+07 | 0.72763 | 1873.406 |
| 16580 | 76.6415 | 16516 | 6.18E+07 | 0.67234 | 1886.51 |
| 16612 | 76.6556 | 16548 | 6.24E+07 | 0.68444 | 1884.786 |
| 16644 | 76.6423 | 16580 | 6.25E+07 | 0.73171 | 1871.909 |
| 16676 | 76.641 | 16612 | 6.23E+07 | 0.74202 | 1869.308 |
| 16708 | 76.6409 | 16644 | 6.27E+07 | 0.68573 | 1886.29 |
| 16740 | 76.6034 | 16676 | 6.24E+07 | 0.68377 | 1886.326 |
| 16773 | 76.5997 | 16709 | 6.25E+07 | 0.68319 | 1886.138 |
| 16805 | 76.5919 | 16741 | 6.21E+07 | 0.74055 | 1868.234 |
| 16837 | 76.6057 | 16773 | 6.18E+07 | 0.67463 | 1886.251 |
| 16869 | 76.5815 | 16805 | 6.14E+07 | 0.72825 | 1871.087 |
| 16901 | 76.602 | 16837 | 6.16E+07 | 0.73718 | 1869.211 |
| 16933 | | 16869 | 6.17E+07 | 0.6849 | 1882.721 |
| 16965 | 76.5512 | 16902 | 6.15E+07 | 0.74876 | 1865.357 |
| 16998 | 76.5526 | 16934 | 6.16E+07 | 0.75048 | 1864.771 |
| 17029 | 76.5303 | 16966 | 6.16E+07 | 0.69342 | 1880.986 |
| 17062 | 76.5554 | 16998 | 6.15E+07 | 0.70671 | 1874.973 |
| 17094 | 76.5325 | 17030 | 6.18E+07 | 0.68475 | 1884.018 |
| 17126 | 76.5148 | 17062 | 6.11E+07 | 0.69117 | 1878.252 |
| 17158 | 76.5113 | 17094 | 6.10E+07 | 0.74057 | 1866.23 |
| 17190 | 76.5034 | 17127 | 6.13E+07 | 0.73849 | 1867.648 |
| 17223 | 76.4834 | 17159 | 6.14E+07 | 0.74796 | 1864.889 |
| 17255 | 76.4981 | 17191 | 6.12E+07 | 0.69284 | 1878.886 |
| 17287 | 76.5185 | 17223 | 6.17E+07 | 0.75832 | 1863.211 |
| 17319 | 76.489 | 17255 | 6.12E+07 | 0.77242 | 1856.911 |
| 17351 | 76.5101 | 17287 | 6.19E+07 | 0.7049 | 1878.643 |
| 17383 | 76.4713 | 17319 | 6.33E+07 | 0.69852 | 1885.806 |
| 17415 | 76.4624 | 17352 | 6.25E+07 | 0.69608 | 1882.541 |
| 17447 | 76.4671 | 17384 | 6.26E+07 | 0.68975 | 1885.289 |
| 17479 | 76.4255 | 17416 | 6.25E+07 | 0.74795 | 1869.695 |
| 17512 | 76.4184 | 17448 | 6.46E+07 | 0.67207 | 1898.827 |
| 17544 | 76.4065 | 17480 | 6.29E+07 | 0.73591 | 1873.865 |
| 17576 | 76.4133 | 17512 | 6.30E+07 | 0.68122 | 1888.51 |
| 17608 | 76.4056 | 17544 | 6.28E+07 | 0.67568 | 1890.14 |
| 17640 | 76.3661 | 17576 | 6.24E+07 | 0.67975 | 1887.64 |
| 17673 | 76.3745 | 17608 | 6.28E+07 | 0.72337 | 1876.44 |
| 17704 | 76.3737 | 17641 | 6.17E+07 | 0.73415 | 1869.392 |
| 17736 | 76.3366 | 17673 | 6.21E+07 | 0.68821 | 1884.474 |
| 17769 | 76.353 | 17705 | 6.20E+07 | 0.73524 | 1871.271 |
| 17800 | 76.3185 | 17737 | 6.23E+07 | 0.68814 | 1884.918 |
| 17833 | 76.306 | 17769 | 6.18E+07 | 0.68167 | 1885.051 |
| 17865 | 76.3095 | 17801 | 6.13E+07 | 0.69172 | 1879.395 |
| 17897 | 76.2728 | 17833 | 6.14E+07 | 0.67972 | 1883.928 |
| 17929 | 76.2779 | 17865 | 6.13E+07 | 0.68797 | 1881.246 |
| 17961 | 76.2983 | 17897 | 6.12E+07 | 0.72014 | 1871.359 |
| 17993 | 76.2706 | 17929 | 6.19E+07 | 0.72686 | 1874.217 |
| 18026 | 76.27 | 17962 | 6.19E+07 | 0.73634 | 1871.225 |
| 18057 | 76.2428 | 17994 | 6.25E+07 | 0.74571 | 1870.903 |
| 18090 | 76.2686 | 18026 | 6.19E+07 | 0.69165 | 1881.951 |
| 18122 | 76.2477 | 18058 | 6.27E+07 | 0.68793 | 1888.178 |
| 18154 | 76.2394 | 18090 | 6.18E+07 | 0.68854 | 1883.24 |
| 18186 | | 18122 | 6.28E+07 | 0.74554 | 1871.934 |
| 18219 | 76.2251 | 18154 | 6.28E+07 | 0.67628 | 1890.906 |
| 18250 | 76.1905 | 18187 | 6.25E+07 | 0.73289 | 1873.323 |
| 18283 | 76.1975 | 18219 | 6.20E+07 | 0.69439 | 1882.297 |
| 18314 | 76.1955 | 18251 | 6.20E+07 | 0.74291 | 1870.35 |
| 18347 | 76.2096 | 18283 | 6.22E+07 | 0.68248 | 1887.074 |
| 18379 | 76.2229 | 18315 | 6.21E+07 | 0.6788 | 1887.382 |
| 18411 | 76.1993 | 18347 | 6.22E+07 | 0.74161 | 1871.586 |
| 18443 | 76.1949 | 18379 | 6.22E+07 | 0.74344 | 1870.709 |
| 18475 | 76.1785 | 18412 | 6.18E+07 | 0.68195 | 1886.097 |
| 18508 | 76.1603 | 18444 | 6.15E+07 | 0.74094 | 1869.054 |
| 18540 | 76.1625 | 18476 | 6.15E+07 | 0.68815 | 1883.059 |
| 18572 | 76.1706 | 18508 | 6.17E+07 | 0.68326 | 1885.792 |
| 18604 | 76.1413 | 18540 | 6.13E+07 | 0.69744 | 1879.277 |
| 18636 | 76.1481 | 18572 | 6.18E+07 | 0.67849 | 1887.808 |
| 18668 | 76.1583 | 18604 | 6.17E+07 | 0.68679 | 1884.448 |
| 18700 | 76.1476 | 18636 | 6.11E+07 | 0.72117 | 1871.601 |
| 18732 | 76.161 | 18668 | 6.14E+07 | 0.7281 | 1872.708 |
| 18764 | 76.1488 | 18701 | 6.15E+07 | 0.73752 | 1869.858 |
| 18796 | 76.1255 | 18733 | 6.15E+07 | 0.74593 | 1868.52 |
| 18829 | 76.1128 | 18765 | 6.14E+07 | 0.69297 | 1881.526 |
| 18861 | 76.0987 | 18797 | 6.18E+07 | 0.76139 | 1864.646 |
| 18893 | 76.0878 | 18829 | 6.20E+07 | 0.70528 | 1879.933 |
| 18925 | 76.0915 | 18861 | 6.16E+07 | 0.76505 | 1860.532 |
| 18957 | 76.0895 | 18893 | 6.14E+07 | 0.69742 | 1879.009 |
| 18989 | 76.11 | 18926 | 6.15E+07 | 0.6882 | 1882.739 |
| 19021 | 76.0986 | 18958 | 6.15E+07 | 0.70352 | 1877.615 |
| 19054 | 76.1124 | 18990 | 6.14E+07 | 0.69072 | 1881.67 |
| 19085 | 76.0979 | 19022 | 6.18E+07 | 0.74443 | 1869.001 |
| 19118 | 76.066 | 19054 | 6.17E+07 | 0.7401 | 1870.443 |
| 19150 | 76.0769 | 19086 | 6.17E+07 | 0.74906 | 1865.603 |
| 19182 | 76.0764 | 19118 | 6.17E+07 | 0.6936 | 1881.535 |
| 19214 | 76.0634 | 19150 | 6.20E+07 | 0.68741 | 1884.59 |
| 19246 | 76.0776 | 19182 | 6.21E+07 | 0.75907 | 1865.778 |
| 19278 | 76.0295 | 19215 | 6.29E+07 | 0.68956 | 1887.691 |
| 19311 | 76.0464 | 19247 | 6.37E+07 | 0.67995 | 1893.634 |
| 19343 | 76.0367 | 19279 | 6.39E+07 | 0.73701 | 1877.468 |
| 19375 | 76.0059 | 19311 | 6.51E+07 | 0.68614 | 1896.584 |
| 19407 | 76.0077 | 19343 | 6.51E+07 | 0.74586 | 1881.449 |
| 19439 | 75.9908 | 19375 | 6.50E+07 | 0.68183 | 1897.337 |
| 19472 | 75.9975 | 19408 | 6.58E+07 | 0.74147 | 1883.318 |
| 19503 | 75.9704 | 19440 | 6.61E+07 | 0.68254 | 1901.629 |
| 19536 | 75.9357 | 19472 | 6.60E+07 | 0.67219 | 1904.05 |
| 19568 | 75.9517 | 19504 | 6.59E+07 | 0.66954 | 1905.103 |
| 19600 | 75.9485 | 19536 | 6.63E+07 | 0.72399 | 1890.35 |
| 19632 | 75.9368 | 19568 | 6.65E+07 | 0.66305 | 1908.467 |
| 19664 | 75.9115 | 19601 | 6.63E+07 | 0.71637 | 1892.087 |
| 19697 | 75.9255 | 19633 | 6.66E+07 | 0.72348 | 1890.332 |
| 19728 | 75.9052 | 19665 | 6.65E+07 | 0.66819 | 1906.814 |
| 19761 | 75.9111 | 19697 | 6.62E+07 | 0.73397 | 1888.416 |
| 19793 | 75.9096 | 19729 | 6.47E+07 | 0.73606 | 1882.021 |
| 19825 | 75.8882 | 19761 | 6.39E+07 | 0.67983 | 1894.409 |
| 19857 | 75.8968 | 19793 | 6.34E+07 | 0.68758 | 1890.279 |
| 19889 | 75.8579 | 19825 | 6.34E+07 | 0.73618 | 1877.5 |
| 19922 | 75.8827 | 19858 | 6.36E+07 | 0.75046 | 1873.625 |
| 19954 | 75.8791 | 19890 | 6.42E+07 | 0.68212 | 1894.924 |
| 19986 | 75.8698 | 19922 | 6.52E+07 | 0.7426 | 1882.196 |
| 20018 | 75.8762 | 19954 | 6.53E+07 | 0.6838 | 1898.387 |
| 20050 | 75.8549 | 19986 | 6.51E+07 | 0.75499 | 1877.48 |
| 20082 | 75.835 | 20018 | 6.48E+07 | 0.68665 | 1895.016 |
| 20114 | 75.8359 | 20050 | 6.46E+07 | 0.67598 | 1898.317 |
| 20146 | 75.8282 | 20083 | 6.47E+07 | 0.67336 | 1898.567 |
| 20179 | 75.8384 | 20115 | 6.52E+07 | 0.67562 | 1898.857 |
| 20211 | 75.7988 | 20147 | 6.55E+07 | 0.65995 | 1905.473 |
| 20243 | 75.7996 | 20179 | 6.65E+07 | 0.70803 | 1894.884 |
| 20275 | 75.7833 | 20211 | 6.67E+07 | 0.64756 | 1915.44 |
| 20307 | 75.773 | 20243 | 6.71E+07 | 0.64939 | 1914.748 |
| 20340 | 75.7257 | 20276 | 6.74E+07 | 0.70152 | 1899.971 |
| 20373 | 75.7422 | 20308 | 6.71E+07 | 0.64647 | 1915.602 |
| 20404 | 75.7543 | 20340 | 6.77E+07 | 0.70412 | 1899.978 |
| 20436 | 75.7379 | 20373 | 6.80E+07 | 0.7033 | 1902.834 |
| 20469 | 75.7142 | 20405 | 6.78E+07 | 0.71501 | 1899.715 |
| 20501 | 75.7142 | 20437 | 6.90E+07 | 0.72879 | 1899.202 |
| 20533 | 75.7133 | 20469 | 6.95E+07 | 0.66456 | 1919.522 |
| 20566 | 75.7343 | 20501 | 6.96E+07 | 0.668 | 1917.115 |
| 20597 | 75.6994 | 20534 | 6.90E+07 | 0.66114 | 1917.879 |
| 20630 | 75.6999 | 20566 | 6.89E+07 | 0.72187 | 1902.097 |
| 20662 | 75.6975 | 20598 | 6.83E+07 | 0.65906 | 1916.269 |
| 20694 | 75.6739 | 20630 | 6.60E+07 | 0.70867 | 1893.84 |
| 20726 | 75.6556 | 20662 | 6.58E+07 | 0.71793 | 1890.415 |
| 20758 | 75.6463 | 20694 | 6.63E+07 | 0.65905 | 1908.947 |
| 20791 | 75.6535 | 20727 | 6.69E+07 | 0.65865 | 1911.826 |
| 20823 | 75.6113 | 20759 | 6.67E+07 | 0.65393 | 1912.052 |
| 20855 | 75.6305 | 20791 | 6.65E+07 | 0.65023 | 1912.293 |
| 20887 | 75.5826 | 20823 | 6.63E+07 | 0.70136 | 1898.888 |
| 20920 | 75.6313 | 20856 | 6.56E+07 | 0.69787 | 1894.977 |
| 20951 | 75.6049 | 20888 | 6.66E+07 | 0.70603 | 1897.084 |
| 20984 | 75.5692 | 20920 | 6.74E+07 | 0.71598 | 1898.17 |
| 21016 | 75.5949 | 20952 | 6.67E+07 | 0.66358 | 1908.717 |
| 21048 | 75.5686 | 20984 | 6.59E+07 | 0.73181 | 1886.353 |
| 21080 | 75.5567 | 21016 | 6.57E+07 | 0.6825 | 1900.485 |
| 21112 | 75.5512 | 21049 | 6.68E+07 | 0.67091 | 1906.79 |
| 21145 | 75.5619 | 21081 | 6.59E+07 | 0.67271 | 1903.441 |
| 21176 | 75.5272 | 21113 | 6.62E+07 | 0.71852 | 1893.017 |
| 21209 | 75.5582 | 21145 | 6.67E+07 | 0.66053 | 1910.323 |
| 21241 | 75.5205 | 21177 | 6.86E+07 | 0.64698 | 1921.132 |
| 21273 | 75.5102 | 21209 | 6.93E+07 | 0.70763 | 1904.033 |
| 21305 | 75.514 | 21241 | 6.89E+07 | 0.65248 | 1920.642 |
| 21337 | 75.5026 | 21274 | 6.92E+07 | 0.71109 | 1906.011 |
| 21370 | 75.5396 | 21306 | 6.92E+07 | 0.70953 | 1906.666 |
| 21401 | 75.5192 | 21338 | 6.92E+07 | 0.65563 | 1921.148 |
| 21434 | 75.5056 | 21370 | 6.90E+07 | 0.65291 | 1921.324 |
| 21466 | | 21402 | 6.87E+07 | 0.71014 | 1902.665 |
| 21498 | | 21434 | 6.80E+07 | 0.65927 | 1914.797 |
| 21535 | 75.4665 | 21466 | 6.79E+07 | 0.65347 | 1916.358 |
| 21558 | 75.456 | 21499 | 6.79E+07 | 0.6385 | 1921.347 |
| 21595 | 75.4588 | 21531 | 6.73E+07 | 0.696 | 1903.089 |
| 21627 | 75.4148 | 21563 | 6.63E+07 | 0.70505 | 1898.807 |
| 21660 | 75.4442 | 21595 | 6.56E+07 | 0.70398 | 1896.681 |
| 21691 | 75.4263 | 21627 | 6.57E+07 | 0.71337 | 1893.49 |
| 21723 | 75.3766 | 21660 | 6.54E+07 | 0.65681 | 1908.684 |
| 21756 | 75.37 | 21692 | 6.49E+07 | 0.72979 | 1884.149 |
| 21788 | | 21724 | 6.43E+07 | 0.67946 | 1897.009 |
| 21819 | 75.3327 | 21756 | 6.44E+07 | 0.66185 | 1902.856 |
| 21853 | 75.3589 | 21788 | 6.47E+07 | 0.72267 | 1887.553 |
| 21884 | 75.3237 | 21820 | 6.54E+07 | 0.72462 | 1889.703 |
| 21917 | 75.32 | 21853 | 6.61E+07 | 0.73886 | 1885.437 |
| 21949 | 75.3262 | 21885 | 6.80E+07 | 0.74077 | 1893.295 |
| 21981 | 75.29 | 21917 | 6.82E+07 | 0.75364 | 1890.378 |
| 22013 | 75.2966 | 21949 | 6.86E+07 | 0.77292 | 1885.09 |
| 22045 | | 21981 | 6.87E+07 | 0.70885 | 1902.69 |
| 22078 | 75.2772 | 22013 | 6.86E+07 | 0.78016 | 1882.909 |
| 22110 | 75.2686 | 22046 | 6.89E+07 | 0.71371 | 1903.705 |
| 22142 | 75.2533 | 22078 | 6.90E+07 | 0.70856 | 1905.16 |
| 22174 | 75.2143 | 22110 | 6.89E+07 | 0.77368 | 1885.309 |
| 22206 | 75.1948 | 22142 | 6.86E+07 | 0.77384 | 1884.403 |
| 22239 | 75.2146 | 22174 | 6.87E+07 | 0.71249 | 1902.099 |
| 22271 | 75.2212 | 22207 | 6.81E+07 | 0.77328 | 1884.621 |
| 22303 | 75.1869 | 22239 | 6.57E+07 | 0.71095 | 1893.064 |
| 22335 | 75.1971 | 22271 | 6.34E+07 | 0.71141 | 1884.336 |
| 22367 | 75.2045 | 22303 | 6.30E+07 | 0.77321 | 1865.59 |
| 22399 | 75.2065 | 22336 | 6.56E+07 | 0.69946 | 1895.617 |
| 22432 | 75.204 | 22368 | 6.45E+07 | 0.69698 | 1892.692 |
| 22464 | 75.2035 | 22400 | 6.33E+07 | 0.75432 | 1873.124 |
| 22496 | 75.1946 | 22432 | 6.25E+07 | 0.69137 | 1885.898 |
| 22529 | 75.1697 | 22464 | 6.17E+07 | 0.70858 | 1878.636 |
| 22561 | 75.158 | 22496 | 6.19E+07 | 0.67794 | 1888.662 |
| 22593 | 75.146 | 22529 | 6.19E+07 | 0.67262 | 1889.846 |
| 22625 | 75.1621 | 22561 | 6.13E+07 | 0.73601 | 1870.614 |
| 22657 | 75.1525 | 22593 | 6.14E+07 | 0.73217 | 1872.483 |
| 22690 | | 22626 | 6.16E+07 | 0.68753 | 1884.168 |
| 22722 | 75.1267 | 22658 | 6.17E+07 | 0.73632 | 1872.464 |
| 22754 | 75.1486 | 22690 | 6.18E+07 | 0.74858 | 1869.675 |
| 22786 | 75.122 | 22722 | 6.11E+07 | 0.69988 | 1879.253 |
| 22819 | 75.1176 | 22755 | 6.09E+07 | 0.75694 | 1861.352 |
| 22851 | 75.1149 | 22787 | 5.98E+07 | 0.71727 | 1869.095 |
| 22883 | 75.1051 | 22819 | 5.88E+07 | 0.76839 | 1852.702 |
| 22915 | 75.0842 | 22851 | 5.83E+07 | 0.77512 | 1848.066 |
| 22948 | 75.1182 | 22883 | 5.76E+07 | 0.70903 | 1862.672 |
| 22979 | 75.1022 | 22915 | 5.81E+07 | 0.73826 | 1855.851 |
| 23012 | 75.0431 | 22948 | 5.86E+07 | 0.72722 | 1860.656 |
| 23046 | 75.0634 | 22980 | 5.92E+07 | 0.70642 | 1870.647 |
| 23077 | 75.1025 | 23013 | 5.99E+07 | 0.69465 | 1876.306 |
| 23109 | 75.0868 | 23046 | 6.10E+07 | 0.69958 | 1880.051 |
| 23142 | 75.1029 | 23078 | 6.11E+07 | 0.67862 | 1885.591 |
| 23174 | 75.0765 | 23110 | 6.10E+07 | 0.73636 | 1867.498 |
| 23207 | 75.0898 | 23142 | 6.09E+07 | 0.73828 | 1869.309 |
| 23238 | 75.0771 | 23174 | 6.11E+07 | 0.67909 | 1886.467 |
| 23271 | 75.0632 | 23207 | 6.07E+07 | 0.69863 | 1878.301 |
| 23303 | 75.0565 | 23239 | 6.05E+07 | 0.74151 | 1866.981 |
| 23335 | 75.0625 | 23271 | 6.07E+07 | 0.69577 | 1878.936 |
| 23367 | 75.055 | 23303 | 6.00E+07 | 0.69407 | 1877.532 |
| 23399 | 75.0402 | 23335 | 6.03E+07 | 0.7369 | 1866.834 |
| 23432 | 75.0444 | 23368 | 6.22E+07 | 0.69436 | 1884.539 |
| 23465 | | 23400 | 6.44E+07 | 0.74815 | 1878.512 |
| 23496 | 75.0498 | 23433 | 6.71E+07 | 0.74954 | 1888.571 |
| 23530 | 75.0786 | 23465 | 6.78E+07 | 0.76083 | 1887.593 |
| 23561 | 75.0538 | 23497 | 7.15E+07 | 0.70106 | 1916.869 |
| 23594 | 75.0325 | 23530 | 7.33E+07 | 0.76733 | 1902.587 |
| 23626 | 75.0028 | 23562 | 7.43E+07 | 0.70811 | 1923.634 |
| 23659 | 74.9845 | 23594 | 7.45E+07 | 0.77424 | 1905.076 |
| 23691 | 74.9648 | 23627 | 7.45E+07 | 0.7052 | 1925.833 |
| 23723 | | 23659 | 7.46E+07 | 0.69463 | 1928.463 |
| 23757 | 74.9496 | 23691 | 7.39E+07 | 0.69341 | 1926.499 |
| 23788 | 74.929 | 23724 | 7.36E+07 | 0.67936 | 1930.204 |
| 23821 | 74.9145 | 23756 | 7.33E+07 | 0.6739 | 1931.188 |
| 23853 | 74.8988 | 23789 | 7.33E+07 | 0.72376 | 1917.024 |
| 23886 | 74.9012 | 23821 | 7.29E+07 | 0.65836 | 1934.255 |
| 23918 | 74.8767 | 23854 | 7.28E+07 | 0.72005 | 1914.332 |
| 23950 | 74.8517 | 23886 | 7.29E+07 | 0.72564 | 1915.102 |
| 23983 | 74.8415 | 23918 | 7.20E+07 | 0.66533 | 1928.76 |
| 24015 | 74.8231 | 23950 | 7.19E+07 | 0.6637 | 1928.941 |
| 24047 | 74.8104 | 23983 | 7.05E+07 | 0.65235 | 1927.93 |
| 24079 | 74.8044 | 24015 | 7.06E+07 | 0.71968 | 1906.985 |
| 24112 | | 24047 | 7.02E+07 | 0.65107 | 1927.426 |
| 24143 | 74.7974 | 24080 | 6.93E+07 | 0.6369 | 1928.819 |
| 24176 | 74.7903 | 24112 | 6.92E+07 | 0.6315 | 1931.229 |
| 24208 | 74.7707 | 24144 | 6.85E+07 | 0.69122 | 1908.572 |
| 24241 | | 24177 | 6.75E+07 | 0.69938 | 1905.567 |
| 24273 | 74.7711 | 24209 | 6.83E+07 | 0.69512 | 1909.76 |
| 24305 | 74.7609 | 24241 | 6.82E+07 | 0.70437 | 1904.034 |
| 24337 | 74.7169 | 24273 | 6.79E+07 | 0.71853 | 1898.745 |
| 24370 | 74.7161 | 24305 | 6.85E+07 | 0.66664 | 1916.473 |
| 24402 | 74.714 | 24338 | 6.82E+07 | 0.66197 | 1917.2 |
| 24434 | 74.6854 | 24370 | 6.81E+07 | 0.65486 | 1920.215 |
| 24464 | | 24402 | 6.78E+07 | 0.65591 | 1917.545 |
| 24499 | | 24435 | 6.69E+07 | 0.70516 | 1899.646 |
| 24533 | 74.6599 | 24467 | 6.87E+07 | 0.70235 | 1906.04 |
| 24563 | 74.6716 | 24499 | 6.60E+07 | 0.70985 | 1894.09 |
| 24595 | 74.6675 | 24531 | 6.48E+07 | 0.71991 | 1888.924 |
| 24628 | 74.6501 | 24563 | 6.55E+07 | 0.73357 | 1886.748 |
| 24660 | 74.6304 | 24596 | 6.51E+07 | 0.69189 | 1895.314 |
| 24692 | 74.6389 | 24628 | 6.48E+07 | 0.67788 | 1899.816 |
| 24724 | 74.6342 | 24660 | 6.42E+07 | 0.67823 | 1896.74 |
| 24757 | 74.6133 | 24693 | 6.39E+07 | 0.73531 | 1878.608 |
| 24789 | 74.6056 | 24725 | 6.38E+07 | 0.67955 | 1894.577 |
| 24821 | 74.5885 | 24757 | 6.38E+07 | 0.7272 | 1882.634 |
| 24853 | 74.5822 | 24789 | 6.34E+07 | 0.73462 | 1879.45 |
| 24886 | 74.5599 | 24821 | 6.29E+07 | 0.74653 | 1872.743 |
| 24919 | 74.5745 | 24854 | 6.34E+07 | 0.69267 | 1889.005 |
| 24951 | 74.5465 | 24886 | 6.37E+07 | 0.68031 | 1894.043 |
| 24984 | | 24919 | 6.36E+07 | 0.75208 | 1875.151 |
| 25017 | 74.5549 | 24952 | 6.28E+07 | 0.70415 | 1884.248 |
| 25048 | 74.5329 | 24984 | 6.36E+07 | 0.69532 | 1889.485 |
| 25081 | | 25017 | 6.24E+07 | 0.74015 | 1873.121 |
| 25113 | 74.508 | 25049 | 6.41E+07 | 0.67868 | 1898.375 |
| 25146 | 74.4811 | 25081 | 6.89E+07 | 0.67837 | 1915.714 |
| 25178 | 74.4748 | 25114 | 7.25E+07 | 0.7343 | 1912.721 |
| 25205 | 74.4272 | 25146 | 7.54E+07 | 0.6701 | 1939.59 |
| 25247 | 74.3465 | 25178 | 7.83E+07 | 0.66531 | 1951.163 |
| 25276 | 74.3745 | 25211 | 8.06E+07 | 0.65109 | 1961.984 |
| 25307 | 74.3208 | 25243 | 8.20E+07 | 0.64584 | 1967.548 |
| 25340 | 74.3597 | 25275 | 8.26E+07 | 0.63966 | 1971.885 |
| 25371 | 74.3286 | 25308 | 8.27E+07 | 0.68712 | 1959.552 |
| 25404 | 74.3098 | 25340 | 8.23E+07 | 0.62491 | 1975.593 |
| 25436 | 74.2965 | 25372 | 8.17E+07 | 0.61724 | 1976.092 |
| 25469 | 74.299 | 25404 | 8.14E+07 | 0.6751 | 1956.586 |
| 25501 | 74.2572 | 25437 | 8.10E+07 | 0.61637 | 1973.577 |
| 25533 | 74.2473 | 25469 | 7.98E+07 | 0.67635 | 1951.464 |
| 25565 | 74.2192 | 25501 | 7.93E+07 | 0.67409 | 1950.156 |
| 25598 | 74.1841 | 25533 | 7.84E+07 | 0.61924 | 1964.655 |
| 25630 | 74.1593 | 25566 | 7.78E+07 | 0.61962 | 1963.193 |
| 25663 | 74.191 | 25598 | 7.64E+07 | 0.67319 | 1944.039 |
| 25694 | 74.184 | 25630 | 7.60E+07 | 0.68512 | 1937.338 |
| 25727 | 74.1877 | 25663 | 7.45E+07 | 0.68697 | 1933.537 |
| 25759 | 74.152 | 25695 | 7.48E+07 | 0.62399 | 1951.268 |
| 25792 | 74.1587 | 25728 | 7.44E+07 | 0.69067 | 1932.43 |
| 25824 | 74.1297 | 25760 | 7.40E+07 | 0.70559 | 1924.86 |
| 25856 | 74.1222 | 25792 | 7.39E+07 | 0.64454 | 1941.509 |
| 25889 | 74.133 | 25824 | 7.37E+07 | 0.70387 | 1926.018 |
| 25921 | 74.1379 | 25856 | 7.33E+07 | 0.70742 | 1922.941 |
| 25953 | 74.0946 | 25889 | 7.32E+07 | 0.6559 | 1936.722 |
| 25986 | 74.0746 | 25921 | 7.24E+07 | 0.71763 | 1914.917 |
| 26018 | 74.0659 | 25953 | 7.21E+07 | 0.65327 | 1932.468 |
| 26050 | 74.0537 | 25986 | 7.14E+07 | 0.65324 | 1931.443 |
| 26082 | 74.0619 | 26018 | 7.02E+07 | 0.71136 | 1909.659 |
| 26115 | 74.0486 | 26050 | 7.02E+07 | 0.71219 | 1908.853 |
| 26147 | | 26083 | 6.96E+07 | 0.65506 | 1924.05 |
| 26180 | 74.0511 | 26115 | 6.76E+07 | 0.71088 | 1903.093 |
| 26212 | 74.0138 | 26147 | 6.70E+07 | 0.72297 | 1894.349 |
| 26245 | 74.0076 | 26180 | 6.75E+07 | 0.73762 | 1892.283 |
| 26277 | 74.0254 | 26212 | 6.82E+07 | 0.67236 | 1914.519 |
| 26309 | 74.0049 | 26245 | 6.90E+07 | 0.73477 | 1899.298 |
| 26342 | 73.9886 | 26277 | 6.81E+07 | 0.67993 | 1911.411 |
| 26374 | 73.9918 | 26310 | 6.74E+07 | 0.74501 | 1890.853 |
| 26407 | 73.9975 | 26342 | 6.66E+07 | 0.67824 | 1907.623 |
| 26439 | 73.9704 | 26374 | 6.56E+07 | 0.67716 | 1903.237 |
| 26471 | 73.9676 | 26407 | 6.53E+07 | 0.73817 | 1884.599 |
| 26504 | 73.9733 | 26439 | 6.26E+07 | 0.73889 | 1872.861 |
| 26536 | 73.9559 | 26472 | 6.02E+07 | 0.75127 | 1862.999 |
| 26569 | 73.9779 | 26504 | 6.06E+07 | 0.75258 | 1861.01 |
| 26600 | 73.957 | 26536 | 6.01E+07 | 0.71628 | 1869.526 |
| 26633 | 73.946 | 26569 | 5.83E+07 | 0.73322 | 1858.163 |
| 26666 | 73.989 | 26601 | 5.69E+07 | 0.77537 | 1842.845 |
| 26698 | 73.9988 | 26633 | 5.62E+07 | 0.78303 | 1837.977 |
| 26730 | 73.9869 | 26666 | 5.30E+07 | 0.76511 | 1828.592 |
| 26763 | 73.9762 | 26698 | 5.36E+07 | 0.79233 | 1822.597 |
| 26795 | | 26731 | 5.29E+07 | 0.75519 | 1830.077 |
| 26828 | 73.9581 | 26763 | 5.48E+07 | 0.77294 | 1833.388 |
| 26860 | 73.9819 | 26796 | 5.38E+07 | 0.74074 | 1838.475 |
| 26892 | 73.9522 | 26828 | 5.41E+07 | 0.75646 | 1834.113 |
| 26925 | 73.9571 | 26860 | 5.52E+07 | 0.80171 | 1828.248 |
| 26957 | 73.9533 | 26893 | 5.57E+07 | 0.73342 | 1847.86 |
| 26989 | 73.9476 | 26925 | 5.71E+07 | 0.7991 | 1834.801 |
| 27022 | 73.9673 | 26957 | 5.91E+07 | 0.72797 | 1864.066 |
| 27054 | 73.9491 | 26990 | 6.10E+07 | 0.80192 | 1850.489 |
| 27087 | 73.9232 | 27022 | 6.24E+07 | 0.73418 | 1874.444 |
| 27120 | 73.9542 | 27055 | 6.34E+07 | 0.72135 | 1881.486 |
| 27151 | 73.9184 | 27087 | 6.46E+07 | 0.71906 | 1887.349 |
| 27184 | 73.9016 | 27119 | 6.48E+07 | 0.70385 | 1892.104 |
| 27217 | 73.9557 | 27152 | 6.55E+07 | 0.69927 | 1895.991 |
| 27248 | 73.9238 | 27184 | 6.61E+07 | 0.75132 | 1885.205 |
| 27281 | 73.9175 | 27217 | 6.65E+07 | 0.68302 | 1904.795 |
| 27313 | 73.8917 | 27249 | 6.69E+07 | 0.74687 | 1887.996 |
| 27346 | 73.9122 | 27281 | 6.69E+07 | 0.68625 | 1905.608 |
| 27378 | 73.8963 | 27314 | 6.55E+07 | 0.74901 | 1882.613 |
| 27411 | 73.8868 | 27346 | 6.51E+07 | 0.69737 | 1896.11 |
| 27443 | 73.8625 | 27379 | 6.47E+07 | 0.74813 | 1881.864 |
| 27476 | 73.8904 | 27411 | 6.51E+07 | 0.69651 | 1895.502 |
| 27508 | 73.8695 | 27443 | 6.61E+07 | 0.67968 | 1905.63 |
| 27540 | 73.8506 | 27476 | 6.57E+07 | 0.68792 | 1900.56 |
| 27573 | 73.8475 | 27508 | 6.63E+07 | 0.73702 | 1889.924 |
| 27605 | 73.8357 | 27541 | 6.65E+07 | 0.74535 | 1889.654 |
| 27638 | 73.8458 | 27573 | 6.72E+07 | 0.74476 | 1892.383 |
| 27670 | 73.827 | 27605 | 6.77E+07 | 0.75489 | 1889.749 |
| 27702 | 73.8191 | 27638 | 6.79E+07 | 0.69729 | 1906.513 |
| 27735 | 73.8367 | 27670 | 6.84E+07 | 0.69907 | 1907.422 |
| 27767 | 73.8502 | 27702 | 6.88E+07 | 0.76468 | 1891.717 |
| 27799 | 73.8497 | 27735 | 6.83E+07 | 0.6963 | 1906.872 |
| 27832 | 73.8113 | 27767 | 6.88E+07 | 0.68632 | 1911.966 |
| 27865 | 73.8127 | 27800 | 6.89E+07 | 0.74417 | 1895.582 |
| 27897 | 73.8182 | 27832 | 6.90E+07 | 0.6822 | 1913.563 |
| 27929 | 73.8089 | 27865 | 6.95E+07 | 0.75048 | 1895.455 |
| 27962 | 73.7981 | 27897 | 6.96E+07 | 0.6763 | 1918.421 |
| 27994 | 73.8128 | 27929 | 6.99E+07 | 0.6725 | 1920.255 |
| 28026 | 73.828 | 27962 | 7.00E+07 | 0.66987 | 1920.944 |
| 28059 | 73.8067 | 27994 | 7.00E+07 | 0.72479 | 1906.846 |
| 28091 | 73.7582 | 28027 | 7.05E+07 | 0.66385 | 1926.097 |
| 28124 | 73.795 | 28059 | 7.06E+07 | 0.71554 | 1910.408 |
| 28156 | 73.7523 | 28092 | 7.11E+07 | 0.65484 | 1929.703 |
| 28188 | 73.7668 | 28124 | 7.07E+07 | 0.72007 | 1908.099 |
| 28220 | 73.7319 | 28156 | 7.10E+07 | 0.73061 | 1909.196 |
| 28253 | 73.7336 | 28188 | 7.14E+07 | 0.66213 | 1929.185 |
| 28285 | 73.7266 | 28221 | 7.18E+07 | 0.73147 | 1908.818 |
| 28318 | 73.6975 | 28253 | 7.25E+07 | 0.66483 | 1932.771 |
| 28351 | 73.7271 | 28286 | 7.21E+07 | 0.66673 | 1930.068 |
| 28383 | 73.7089 | 28318 | 7.23E+07 | 0.6566 | 1933.955 |
| 28415 | 73.7126 | 28351 | 7.10E+07 | 0.71017 | 1913.908 |
| 28448 | 73.6722 | 28383 | 7.07E+07 | 0.72026 | 1909.227 |
| 28481 | 73.7114 | 28416 | 7.11E+07 | 0.7179 | 1911.441 |
| 28512 | | 28448 | 7.07E+07 | 0.65963 | 1927.193 |
| 28545 | 73.6556 | 28481 | 6.98E+07 | 0.66303 | 1922.036 |
| 28578 | 73.654 | 28513 | 6.90E+07 | 0.65949 | 1921.803 |
| 28611 | 73.6866 | 28545 | 6.90E+07 | 0.71827 | 1903.012 |
| 28642 | 73.6442 | 28578 | 6.87E+07 | 0.71785 | 1902.332 |
| 28675 | 73.6254 | 28610 | 6.91E+07 | 0.71784 | 1904.066 |
| 28708 | 73.6391 | 28643 | 6.89E+07 | 0.73006 | 1902.182 |
| 28740 | 73.6348 | 28675 | 6.79E+07 | 0.74413 | 1893.118 |
| 28772 | 73.6117 | 28708 | 6.76E+07 | 0.68935 | 1906.862 |
| 28805 | 73.6102 | 28740 | 6.77E+07 | 0.75584 | 1889.554 |
| 28837 | 73.6026 | 28772 | 6.75E+07 | 0.69107 | 1907.156 |
| 28870 | 73.6088 | 28805 | 6.82E+07 | 0.75819 | 1890.491 |
| 28902 | 73.604 | 28837 | 6.82E+07 | 0.7641 | 1889.252 |
| 28934 | 73.5692 | 28870 | 6.75E+07 | 0.69512 | 1905.181 |
| 28967 | 73.5536 | 28902 | 6.78E+07 | 0.76951 | 1886.275 |
| 29000 | 73.5701 | 28935 | 6.74E+07 | 0.77238 | 1883.167 |
| 29032 | 73.5892 | 28967 | 6.71E+07 | 0.71427 | 1899.302 |
| 29064 | 73.5638 | 29000 | 6.65E+07 | 0.70592 | 1899.066 |
| 29092 | 73.5275 | 29032 | 6.67E+07 | 0.69487 | 1903.743 |
| 29134 | | 29065 | 6.61E+07 | 0.69482 | 1900.199 |
| 29162 | 73.525 | 29097 | 6.64E+07 | 0.75228 | 1885.608 |
| 29194 | 73.5118 | 29129 | 6.58E+07 | 0.74805 | 1885.692 |
| 29226 | 73.4942 | 29162 | 6.63E+07 | 0.75515 | 1884.284 |
| 29259 | 73.4845 | 29194 | 6.68E+07 | 0.69258 | 1902.328 |
| 29291 | 73.4748 | 29227 | 6.72E+07 | 0.765 | 1884.616 |
| 29324 | 73.473 | 29259 | 6.77E+07 | 0.70548 | 1903.92 |
| 29356 | 73.4538 | 29292 | 6.73E+07 | 0.69659 | 1904.425 |
| 29389 | 73.4439 | 29324 | 6.69E+07 | 0.68681 | 1904.245 |
| 29421 | 73.4266 | 29357 | 6.72E+07 | 0.67385 | 1910.316 |
| 29454 | 73.4177 | 29389 | 6.65E+07 | 0.74082 | 1890.177 |
| 29486 | 73.4123 | 29421 | 6.61E+07 | 0.73575 | 1887.971 |
| 29519 | 73.4334 | 29454 | 6.57E+07 | 0.68576 | 1901.671 |
| 29551 | 73.4238 | 29486 | 6.48E+07 | 0.73983 | 1881.944 |
| 29584 | 73.4044 | 29519 | 6.52E+07 | 0.69293 | 1897.559 |
| 29616 | 73.4251 | 29551 | 6.51E+07 | 0.68406 | 1899.496 |
| 29648 | 73.3959 | 29584 | 6.51E+07 | 0.74305 | 1883.123 |
| 29681 | 73.3769 | 29616 | 6.44E+07 | 0.74458 | 1878.857 |
| 29713 | 73.3904 | 29649 | 6.47E+07 | 0.68763 | 1896.965 |
| 29746 | 73.4111 | 29681 | 6.47E+07 | 0.74597 | 1879.522 |
| 29778 | 73.4031 | 29713 | 6.41E+07 | 0.68556 | 1895.675 |
| 29811 | 73.4064 | 29746 | 6.43E+07 | 0.68776 | 1894.953 |
| 29843 | 73.3824 | 29778 | 6.36E+07 | 0.69658 | 1890.934 |
| 29876 | 73.3644 | 29811 | 6.38E+07 | 0.66846 | 1900.457 |
| 29908 | 73.3709 | 29844 | 6.37E+07 | 0.73171 | 1881.801 |
| 29941 | 73.407 | 29876 | 6.37E+07 | 0.72872 | 1883.764 |
| 29973 | 73.3664 | 29908 | 6.26E+07 | 0.73778 | 1874.917 |
| 30006 | 73.3575 | 29941 | 6.29E+07 | 0.69441 | 1889.367 |
| 30038 | 73.3356 | 29973 | 6.35E+07 | 0.74782 | 1877.234 |
| 30071 | 73.3586 | 30006 | 6.40E+07 | 0.76449 | 1873.759 |
| 30102 | 73.3275 | 30038 | 6.42E+07 | 0.77138 | 1871.936 |
| 30136 | 73.3283 | 30071 | 6.47E+07 | 0.78034 | 1871.827 |
| 30168 | 73.3204 | 30103 | 6.52E+07 | 0.78927 | 1871.819 |
| 30200 | 73.3184 | 30136 | 6.55E+07 | 0.73213 | 1887.222 |
| 30232 | 73.29 | 30168 | 6.58E+07 | 0.8059 | 1867.802 |
| 30265 | 73.2889 | 30200 | 6.59E+07 | 0.73701 | 1887.712 |
| 30298 | 73.3053 | 30232 | 6.56E+07 | 0.80737 | 1866.181 |
| 30330 | 73.2913 | 30265 | 6.54E+07 | 0.73562 | 1885.997 |
| 30362 | | 30298 | 6.54E+07 | 0.72543 | 1888.601 |
| 30395 | 73.2815 | 30330 | 6.60E+07 | 0.72415 | 1891.17 |
| 30427 | 73.2713 | 30363 | 6.60E+07 | 0.70862 | 1896.142 |
| 30460 | 73.2785 | 30395 | 6.61E+07 | 0.77725 | 1878.358 |
| 30492 | 73.2627 | 30427 | 6.61E+07 | 0.6975 | 1899.508 |
| 30525 | 73.2468 | 30460 | 6.61E+07 | 0.69026 | 1902.064 |
| 30557 | 73.2248 | 30492 | 6.64E+07 | 0.68771 | 1903.049 |
| 30591 | 73.2351 | 30525 | 6.65E+07 | 0.74788 | 1887.175 |
| 30623 | 73.2311 | 30558 | 6.67E+07 | 0.68977 | 1903.114 |
| 30655 | 73.1958 | 30590 | 6.63E+07 | 0.67539 | 1906.918 |
| 30689 | 73.2002 | 30623 | 6.63E+07 | 0.73311 | 1889.651 |
| 30721 | | 30656 | 6.62E+07 | 0.74211 | 1887.005 |
| 30753 | 73.2186 | 30688 | 6.59E+07 | 0.68177 | 1903.232 |
| 30785 | 73.1859 | 30721 | 6.62E+07 | 0.73965 | 1889.017 |
| 30819 | 73.2359 | 30753 | 6.63E+07 | 0.67868 | 1905.795 |
| 30850 | 73.1867 | 30786 | 6.65E+07 | 0.67931 | 1906.907 |
| 30884 | 73.2065 | 30819 | 6.67E+07 | 0.73952 | 1891.716 |
| 30915 | 73.1641 | 30851 | 6.67E+07 | 0.73898 | 1890.46 |
| 30948 | 73.1713 | 30883 | 6.64E+07 | 0.68186 | 1905.838 |
| 30981 | 73.1632 | 30916 | 6.66E+07 | 0.73886 | 1889.602 |
| 31013 | 73.1393 | 30949 | 6.69E+07 | 0.75149 | 1888.15 |
| 31046 | 73.1433 | 30981 | 6.67E+07 | 0.76665 | 1883.069 |
| 31078 | 73.1492 | 31013 | 6.67E+07 | 0.69879 | 1900.933 |
| 31111 | 73.1337 | 31046 | 6.65E+07 | 0.69115 | 1903.002 |
| 31144 | 73.1316 | 31078 | 6.70E+07 | 0.69292 | 1906.559 |
| 31176 | 73.1068 | 31111 | 6.67E+07 | 0.68841 | 1903.807 |
| 31208 | 73.1289 | 31144 | 6.64E+07 | 0.74296 | 1886.498 |
| 31240 | 73.0808 | 31176 | 6.66E+07 | 0.73799 | 1891.086 |
| 31274 | 73.0925 | 31209 | 6.73E+07 | 0.74622 | 1890.305 |
| 31305 | 73.0768 | 31241 | 6.72E+07 | 0.68562 | 1906.391 |
| 31338 | 73.0469 | 31273 | 6.74E+07 | 0.68423 | 1909.127 |
| 31372 | 73.0698 | 31306 | 6.77E+07 | 0.68343 | 1908.187 |
| 31404 | 73.0426 | 31338 | 6.75E+07 | 0.67212 | 1911.197 |
| 31437 | 73.0552 | 31371 | 6.75E+07 | 0.72904 | 1895.185 |
| 31469 | 73.047 | 31404 | 6.72E+07 | 0.72543 | 1896.646 |
| 31502 | 73.0289 | 31437 | 6.75E+07 | 0.66391 | 1915.118 |
| 31535 | 73.0217 | 31469 | 6.67E+07 | 0.72953 | 1891.768 |
| 31567 | 73.0076 | 31502 | 6.63E+07 | 0.66881 | 1908.232 |
| 31600 | 73.0098 | 31534 | 6.64E+07 | 0.66926 | 1908.887 |
| 31632 | 72.9876 | 31567 | 6.66E+07 | 0.66346 | 1911.135 |
| 31664 | 72.9697 | 31599 | 6.60E+07 | 0.71294 | 1893.782 |
| 31697 | 72.9804 | 31632 | 6.64E+07 | 0.72324 | 1894.97 |
| 31729 | | 31665 | 6.64E+07 | 0.72192 | 1895.843 |
| 31761 | | 31697 | 6.68E+07 | 0.73307 | 1894.685 |
| 31794 | 72.968 | 31730 | 6.60E+07 | 0.74647 | 1884.176 |
| 31827 | 72.9621 | 31762 | 6.69E+07 | 0.69115 | 1904.026 |
| 31859 | 72.9322 | 31794 | 6.61E+07 | 0.68701 | 1903.157 |
| 31892 | 72.9369 | 31827 | 6.58E+07 | 0.75172 | 1885.141 |
| 31925 | 72.9398 | 31859 | 6.59E+07 | 0.75798 | 1883.681 |
| 31957 | 72.9232 | 31892 | 6.58E+07 | 0.69052 | 1901.514 |
| 31990 | 72.9191 | 31925 | 6.58E+07 | 0.67989 | 1903.61 |
| 32022 | 72.9186 | 31958 | 6.56E+07 | 0.68647 | 1900.465 |
| 32055 | 72.9357 | 31990 | 6.58E+07 | 0.74898 | 1884.673 |
| 32088 | 72.9259 | 32023 | 6.74E+07 | 0.74841 | 1891.075 |
| 32121 | 72.8927 | 32055 | 6.79E+07 | 0.67531 | 1913.506 |
| 32153 | 72.8907 | 32088 | 6.79E+07 | 0.67242 | 1913.29 |
| 32185 | 72.8892 | 32121 | 6.81E+07 | 0.671 | 1914.443 |
| 32218 | 72.8705 | 32153 | 6.86E+07 | 0.73839 | 1898.036 |
| 32251 | 72.8701 | 32186 | 6.92E+07 | 0.73616 | 1898.72 |
| 32283 | 72.8643 | 32218 | 6.89E+07 | 0.66373 | 1919.145 |
| 32316 | 72.8556 | 32251 | 6.93E+07 | 0.73116 | 1899.633 |
| 32349 | 72.8288 | 32284 | 6.95E+07 | 0.67268 | 1917.718 |
| 32381 | | 32316 | 7.01E+07 | 0.673 | 1920.412 |
| 32414 | 72.8116 | 32349 | 7.02E+07 | 0.66058 | 1924.192 |
| 32446 | 72.7982 | 32382 | 6.95E+07 | 0.71424 | 1907.791 |
| 32479 | | 32414 | 6.83E+07 | 0.65616 | 1918.975 |
| 32511 | 72.7835 | 32447 | 6.64E+07 | 0.7091 | 1896.331 |
| 32545 | 72.8315 | 32479 | 6.61E+07 | 0.71675 | 1894.605 |
| 32576 | 72.8137 | 32512 | 6.66E+07 | 0.6568 | 1912.026 |
| 32609 | 72.7886 | 32544 | 6.66E+07 | 0.72514 | 1891.866 |
| 32642 | | 32577 | 6.64E+07 | 0.72743 | 1892.982 |
| 32674 | 72.7474 | 32610 | 6.70E+07 | 0.67169 | 1909.82 |
| 32708 | 72.7603 | 32642 | 6.79E+07 | 0.73454 | 1894.238 |
| 32740 | 72.756 | 32675 | 6.86E+07 | 0.66796 | 1916.697 |
| 32772 | 72.7259 | 32707 | 6.91E+07 | 0.74062 | 1898.303 |
| 32805 | 72.7436 | 32740 | 7.00E+07 | 0.67302 | 1920.724 |
| 32837 | 72.7123 | 32772 | 7.10E+07 | 0.73245 | 1907.409 |
| 32870 | 72.7174 | 32805 | 7.24E+07 | 0.67423 | 1927.301 |
| 32902 | 72.685 | 32838 | 7.36E+07 | 0.66277 | 1935.295 |
| 32936 | 72.6762 | 32870 | 7.47E+07 | 0.66179 | 1939.149 |
| 32968 | 72.6529 | 32903 | 7.61E+07 | 0.7149 | 1930.325 |
| 33001 | | 32935 | 7.80E+07 | 0.65477 | 1953.392 |
| 33032 | 72.6414 | 32968 | 7.91E+07 | 0.70738 | 1939.439 |
| 33066 | 72.6286 | 33001 | 8.00E+07 | 0.71434 | 1941.864 |
| 33098 | 72.6119 | 33033 | 8.06E+07 | 0.6561 | 1961.214 |
| 33131 | | 33066 | 8.02E+07 | 0.72397 | 1939.123 |
| 33163 | 72.5816 | 33098 | 7.98E+07 | 0.65674 | 1959.165 |
| 33197 | 72.5771 | 33131 | 7.96E+07 | 0.65712 | 1957.085 |
| 33229 | 72.5819 | 33164 | 7.90E+07 | 0.64757 | 1957.315 |
| 33261 | 72.591 | 33196 | 7.76E+07 | 0.63489 | 1957.86 |
| 33294 | 72.5754 | 33229 | 7.57E+07 | 0.69726 | 1933.6 |
| 33327 | 72.5772 | 33261 | 7.12E+07 | 0.69165 | 1921.945 |
| 33360 | 72.5665 | 33294 | 7.09E+07 | 0.69898 | 1918.77 |
| 33392 | 72.5233 | 33327 | 6.94E+07 | 0.64037 | 1928.509 |
| 33426 | 72.5553 | 33360 | 6.50E+07 | 0.70626 | 1895.336 |
| 33458 | 72.5841 | 33393 | 6.18E+07 | 0.72125 | 1879.519 |
| 33490 | 72.5622 | 33425 | 6.05E+07 | 0.72629 | 1873.047 |
| 33523 | 72.6191 | 33458 | 5.74E+07 | 0.73356 | 1857.748 |
| 33554 | 72.5659 | 33490 | 5.63E+07 | 0.74082 | 1851.443 |
| 33588 | 72.5664 | 33523 | 5.66E+07 | 0.76192 | 1844.521 |
| 33620 | 72.5628 | 33555 | 5.69E+07 | 0.77361 | 1844.283 |
| 33653 | 72.5675 | 33588 | 5.80E+07 | 0.74245 | 1855.06 |
| 33686 | 72.5499 | 33620 | 5.87E+07 | 0.71295 | 1866.451 |
| 33718 | 72.546 | 33653 | 5.88E+07 | 0.71717 | 1865.081 |
| 33751 | 72.5792 | 33686 | 5.92E+07 | 0.7787 | 1851.488 |
| 33783 | 72.5526 | 33718 | 5.89E+07 | 0.7118 | 1867.834 |
| 33817 | | 33751 | 5.89E+07 | 0.76924 | 1853.729 |
| 33849 | 72.5427 | 33784 | 5.85E+07 | 0.77487 | 1849.959 |
| 33882 | | 33817 | 5.98E+07 | 0.71284 | 1872.487 |
| 33913 | 72.5069 | 33849 | 6.01E+07 | 0.71202 | 1872.747 |
| 33948 | 72.5314 | 33882 | 6.47E+07 | 0.69935 | 1894.544 |
| 33980 | 72.4959 | 33914 | 6.92E+07 | 0.69698 | 1910.576 |
| 34013 | 72.512 | 33947 | 7.14E+07 | 0.68325 | 1922.827 |
| 34045 | 72.4933 | 33980 | 7.19E+07 | 0.73593 | 1909.56 |
| 34078 | 72.4729 | 34013 | 7.01E+07 | 0.74078 | 1901.873 |
| 34111 | | 34046 | 6.94E+07 | 0.67569 | 1917.55 |
| 34143 | 72.4697 | 34078 | 7.09E+07 | 0.6712 | 1923.93 |
| 34176 | 72.4673 | 34111 | 7.04E+07 | 0.6663 | 1923.977 |
| 34209 | 72.4845 | 34144 | 7.01E+07 | 0.66308 | 1924.711 |
| 34241 | 72.4654 | 34176 | 6.73E+07 | 0.71758 | 1901.155 |
| 34274 | 72.4819 | 34209 | 6.28E+07 | 0.71271 | 1882.803 |
| 34306 | 72.4737 | 34241 | 6.24E+07 | 0.72049 | 1879.784 |
| 34339 | 72.4624 | 34274 | 6.04E+07 | 0.73083 | 1868.787 |
| 34372 | 72.4247 | 34307 | 6.27E+07 | 0.6861 | 1891.319 |
| 34405 | 72.4238 | 34339 | 6.48E+07 | 0.74922 | 1882.022 |
| 34438 | | 34372 | 6.75E+07 | 0.68388 | 1909.458 |
| 34469 | 72.4333 | 34405 | 6.81E+07 | 0.6784 | 1912.858 |
| 34502 | 72.3997 | 34438 | 6.76E+07 | 0.6704 | 1913.471 |
| 34535 | 72.4807 | 34470 | 6.57E+07 | 0.65892 | 1909.841 |
| 34566 | 72.3627 | 34502 | 6.56E+07 | 0.65375 | 1912.356 |
| 34600 | 72.3217 | 34535 | 6.35E+07 | 0.70252 | 1889.457 |
| 34634 | 72.4596 | 34568 | 6.18E+07 | 0.70644 | 1880.635 |
| 34664 | 72.4279 | 34600 | 6.02E+07 | 0.71233 | 1875.799 |
| 34698 | 72.3778 | 34633 | 5.82E+07 | 0.72113 | 1862.669 |
| 34731 | 72.3823 | 34665 | 5.97E+07 | 0.7342 | 1867.362 |
| 34763 | 72.3788 | 34698 | 5.88E+07 | 0.75109 | 1857.99 |
| 34796 | 72.3819 | 34731 | 5.77E+07 | 0.76129 | 1852.072 |
| 34828 | 72.3553 | 34763 | 5.55E+07 | 0.77281 | 1838.658 |
| 34862 | 72.3685 | 34796 | 5.60E+07 | 0.7213 | 1853.246 |
| 34894 | 72.3454 | 34829 | 5.65E+07 | 0.73196 | 1852.456 |
| 34927 | | 34862 | 5.75E+07 | 0.72577 | 1858.309 |
| 34958 | 72.3525 | 34894 | 5.70E+07 | 0.72177 | 1857.016 |
| 34992 | 72.3412 | 34927 | 5.71E+07 | 0.72818 | 1856.022 |
| 35024 | 72.3591 | 34959 | 5.73E+07 | 0.76209 | 1849.938 |
| 35057 | 72.3578 | 34992 | 5.76E+07 | 0.71139 | 1863.288 |
| 35090 | 72.3651 | 35024 | 5.75E+07 | 0.76633 | 1846.532 |
| 35122 | 72.3493 | 35057 | 5.72E+07 | 0.77485 | 1844.638 |
| 35155 | 72.3574 | 35090 | 5.65E+07 | 0.786 | 1839.087 |
| 35187 | 72.3617 | 35123 | 5.77E+07 | 0.72971 | 1857.729 |
| 35220 | 72.3331 | 35155 | 5.77E+07 | 0.71597 | 1861.353 |
| 35253 | 72.3449 | 35188 | 5.69E+07 | 0.72649 | 1855.576 |
| 35285 | | 35220 | 5.77E+07 | 0.71611 | 1862.638 |
| 35319 | 72.3667 | 35253 | 5.76E+07 | 0.7716 | 1848.449 |
| 35350 | 72.3347 | 35286 | 5.96E+07 | 0.76857 | 1856.515 |
| 35384 | 72.3392 | 35318 | 5.95E+07 | 0.7054 | 1872.939 |
| 35416 | 72.3062 | 35351 | 6.03E+07 | 0.70075 | 1877.862 |
| 35449 | 72.3063 | 35384 | 5.91E+07 | 0.76911 | 1855.631 |
| 35481 | | 35417 | 5.97E+07 | 0.70814 | 1873.031 |
| 35514 | 72.3245 | 35449 | 5.97E+07 | 0.76692 | 1857.98 |
| 35546 | 72.3076 | 35482 | 6.00E+07 | 0.76657 | 1857.291 |
| 35579 | 72.3124 | 35514 | 6.19E+07 | 0.77941 | 1862.243 |
| 35611 | 72.3046 | 35547 | 6.01E+07 | 0.72013 | 1871.17 |
| 35644 | 72.3004 | 35579 | 6.01E+07 | 0.71082 | 1874.683 |
| 35677 | 72.3132 | 35612 | 6.30E+07 | 0.71133 | 1884.513 |
| 35709 | 72.2722 | 35644 | 6.59E+07 | 0.70077 | 1898.421 |
| 35742 | 72.2902 | 35677 | 6.97E+07 | 0.68827 | 1914.849 |
| 35774 | 72.2942 | 35710 | 7.27E+07 | 0.67095 | 1930.865 |
| 35807 | 72.2416 | 35742 | 7.15E+07 | 0.66161 | 1928.864 |
| 35841 | 72.2513 | 35775 | 7.36E+07 | 0.65225 | 1939.638 |
| 35873 | 72.2615 | 35808 | 7.51E+07 | 0.64103 | 1949.384 |
| 35906 | 72.2311 | 35841 | 7.70E+07 | 0.69596 | 1940.784 |
| 35939 | 72.2591 | 35873 | 7.78E+07 | 0.63114 | 1960.823 |
| 35970 | 72.1992 | 35906 | 7.70E+07 | 0.68922 | 1943.65 |
| 36004 | 72.2042 | 35939 | 7.58E+07 | 0.62882 | 1955.98 |
| 36036 | 72.188 | 35971 | 7.60E+07 | 0.6901 | 1939.805 |
| 36069 | 72.1817 | 36004 | 7.63E+07 | 0.6337 | 1956.162 |
| 36101 | | 36037 | 7.62E+07 | 0.68733 | 1940.78 |
| 36133 | 72.1277 | 36069 | 7.64E+07 | 0.63254 | 1956.838 |
| 36167 | 72.1363 | 36102 | 7.54E+07 | 0.68722 | 1934.881 |
| 36199 | 72.1428 | 36134 | 7.23E+07 | 0.63287 | 1942.435 |
| 36232 | 72.1461 | 36167 | 7.10E+07 | 0.68747 | 1920.781 |
| 36265 | 72.1277 | 36200 | 6.95E+07 | 0.63221 | 1932.451 |
| 36298 | 72.1148 | 36233 | 6.96E+07 | 0.68719 | 1915.323 |
| 36330 | 72.1015 | 36265 | 7.24E+07 | 0.69866 | 1923.72 |
| 36363 | 72.0986 | 36298 | 7.32E+07 | 0.70106 | 1927.564 |
| 36395 | 72.0937 | 36330 | 7.51E+07 | 0.64725 | 1947.383 |
| 36428 | 72.0623 | 36363 | 7.62E+07 | 0.63961 | 1954.665 |
| 36461 | 72.075 | 36396 | 7.65E+07 | 0.70849 | 1936.559 |
| 36493 | 72.0465 | 36428 | 7.65E+07 | 0.64506 | 1953.05 |
| 36527 | 72.0479 | 36461 | 7.57E+07 | 0.63488 | 1954.558 |
| 36559 | 72.0245 | 36494 | 7.49E+07 | 0.68781 | 1938.301 |
| 36592 | 72.0363 | 36526 | 7.27E+07 | 0.69633 | 1925.733 |
| 36624 | 72.032 | 36559 | 6.92E+07 | 0.70824 | 1911.595 |
| 36657 | 72.0574 | 36592 | 7.00E+07 | 0.64074 | 1932.54 |
| 36689 | 72.0234 | 36624 | 6.85E+07 | 0.70728 | 1909.306 |
| 36723 | 72.0341 | 36657 | 6.63E+07 | 0.66393 | 1912.353 |
| 36755 | | 36690 | 6.66E+07 | 0.71747 | 1897.957 |
| 36788 | 71.9994 | 36722 | 6.61E+07 | 0.66142 | 1913.245 |
| 36821 | 72.025 | 36755 | 6.56E+07 | 0.71208 | 1898.365 |
| 36853 | 71.9989 | 36788 | 6.58E+07 | 0.65782 | 1914.031 |
| 36886 | 72.0133 | 36821 | 6.12E+07 | 0.71667 | 1880.178 |
| 36919 | 71.9975 | 36854 | 6.27E+07 | 0.72775 | 1880.853 |
| 36952 | 72.0149 | 36886 | 6.40E+07 | 0.66102 | 1905.375 |
| 36984 | 72.0031 | 36919 | 6.29E+07 | 0.6722 | 1898.281 |
| 37017 | 71.9833 | 36952 | 6.09E+07 | 0.72145 | 1876.313 |
| 37050 | 71.9636 | 36985 | 5.91E+07 | 0.73481 | 1864.857 |
| 37082 | 71.9464 | 37017 | 6.00E+07 | 0.73862 | 1869.443 |
| 37116 | | 37050 | 5.80E+07 | 0.7409 | 1859.272 |
| 37147 | 71.9928 | 37083 | 6.40E+07 | 0.68551 | 1898.27 |
| 37181 | 72.0063 | 37115 | 6.35E+07 | 0.76047 | 1875.146 |
| 37213 | 72.0015 | 37148 | 6.28E+07 | 0.69546 | 1890.149 |
| 37246 | 71.9857 | 37181 | 5.91E+07 | 0.7604 | 1856.886 |
| 37279 | 71.9867 | 37214 | 6.12E+07 | 0.76497 | 1863.153 |
| 37311 | 71.9738 | 37246 | 6.19E+07 | 0.77003 | 1867.122 |
| 37344 | 71.9878 | 37279 | 6.46E+07 | 0.71275 | 1890.86 |
| 37376 | 71.9635 | 37311 | 6.05E+07 | 0.70369 | 1879.135 |
| 37409 | | 37344 | 6.20E+07 | 0.70525 | 1883.239 |
| 37440 | 71.9506 | 37377 | 6.16E+07 | 0.69477 | 1885.431 |
| 37475 | 71.958 | 37409 | 6.21E+07 | 0.75316 | 1870.783 |
| 37507 | 71.9149 | 37442 | 5.76E+07 | 0.74962 | 1855.125 |
| 37540 | 71.9528 | 37474 | 5.66E+07 | 0.75736 | 1847.175 |
| 37572 | 71.9522 | 37507 | 5.87E+07 | 0.69543 | 1875.899 |
| 37604 | 71.9152 | 37540 | 6.27E+07 | 0.69357 | 1890.619 |
| 37638 | 71.9271 | 37572 | 6.20E+07 | 0.69334 | 1888.946 |
| 37671 | 71.9853 | 37605 | 6.36E+07 | 0.68837 | 1895.327 |
| 37702 | 71.9346 | 37638 | 6.24E+07 | 0.7409 | 1876.013 |
| 37736 | 71.8904 | 37671 | 6.37E+07 | 0.67435 | 1900.786 |
| 37769 | 71.8965 | 37703 | 6.21E+07 | 0.73301 | 1876.219 |
| 37802 | 71.8994 | 37736 | 6.13E+07 | 0.74129 | 1872.109 |
| 37835 | 71.9278 | 37769 | 5.97E+07 | 0.75136 | 1862.876 |
| 37867 | 71.8885 | 37802 | 6.09E+07 | 0.76459 | 1865.524 |
| 37901 | 71.8971 | 37835 | 6.16E+07 | 0.76957 | 1866.344 |
| 37933 | 71.8754 | 37867 | 6.32E+07 | 0.71437 | 1887.24 |
| 37966 | 71.8577 | 37900 | 6.40E+07 | 0.71083 | 1890.056 |
| 37999 | 71.8421 | 37934 | 6.93E+07 | 0.70504 | 1911.947 |
| 38032 | | 37966 | 7.31E+07 | 0.76952 | 1906.998 |
| 38063 | 71.8739 | 37999 | 7.39E+07 | 0.69717 | 1929.357 |
| 38097 | 71.8219 | 38032 | 7.59E+07 | 0.68341 | 1941.609 |
| 38130 | 71.8152 | 38064 | 7.65E+07 | 0.67753 | 1943.683 |
| 38163 | 71.8041 | 38097 | 7.56E+07 | 0.67121 | 1944.004 |
| 38196 | | 38130 | 7.57E+07 | 0.66478 | 1945.993 |
| 38227 | 71.7904 | 38163 | 7.62E+07 | 0.64488 | 1953.831 |
| 38262 | 71.7861 | 38196 | 7.69E+07 | 0.63495 | 1958.019 |
| 38294 | 71.7956 | 38229 | 7.86E+07 | 0.6916 | 1946.846 |
| 38326 | 71.77 | 38261 | 7.91E+07 | 0.62841 | 1967.919 |
| 38360 | 71.7715 | 38294 | 7.94E+07 | 0.68557 | 1950.505 |
| 38392 | 71.7502 | 38327 | 7.88E+07 | 0.62423 | 1967.264 |
| 38425 | 71.7504 | 38360 | 7.82E+07 | 0.6856 | 1948.593 |
| 38458 | | 38392 | 7.76E+07 | 0.62993 | 1962.677 |
| 38490 | | 38425 | 7.73E+07 | 0.61821 | 1966.271 |
| 38524 | 71.7446 | 38458 | 7.58E+07 | 0.68114 | 1941.077 |
| 38557 | 71.7557 | 38491 | 7.48E+07 | 0.68021 | 1940.889 |
| 38589 | 71.7361 | 38524 | 7.39E+07 | 0.69197 | 1935.184 |
| 38622 | 71.7078 | 38557 | 7.42E+07 | 0.69331 | 1934.82 |
| 38655 | 71.698 | 38590 | 7.33E+07 | 0.70662 | 1927.516 |
| 38688 | 71.6785 | 38622 | 7.32E+07 | 0.65612 | 1940.546 |
| 38720 | 71.6673 | 38655 | 7.16E+07 | 0.65136 | 1936.195 |
| 38754 | 71.6756 | 38688 | 7.05E+07 | 0.71387 | 1913.376 |

Graphite Data

Figure 4A | Figure 4C

| T avg | Iavg/Mi | T avg | Emissivity Parameter, n |
|---|---|---|---|
| 1812.94787 | 0.52104 | 1812.94787 | 0.92197 |
| 1902.57648 | 0.71115 | 1902.57648 | 0.77167 |
| 2006.89379 | 0.91229 | 2006.89379 | 0.62154 |
| 2080.73537 | 1.08538 | 2080.73537 | 0.52346 |

| T avg | Iavg/Mi | T avg | n |
|---|---|---|---|
| 1793.90168 | 0.59474 | 1793.90168 | 0.83404 |
| 1893.8928 | 0.75276 | 1893.8928 | 0.71028 |
| 2015.90787 | 0.97367 | 2015.90787 | 0.56332 |
| 2232.13757 | 1.46208 | 2232.13757 | 0.33077 |

| T avg | Iavg/Mi | T avg | n |
|---|---|---|---|
| 1910.57947 | 0.70945 | 1910.57947 | 0.59607 |
| 2019.54073 | 0.98146 | 2019.54073 | 0.43751 |
| 2095.5591 | 1.20947 | 2095.5591 | 0.32353 |

| T avg | Iavg/Mi | T avg | n |
|---|---|---|---|
| 1904.73075 | 1.42324 | 1904.73075 | 0.36469 |
| 2004.29508 | 1.7483 | 2004.29508 | 0.23764 |
| 2104.20153 | 2.19528 | 2104.20153 | 0.12677 |
| 2181.60988 | 2.59535 | 2181.60988 | 0.09192 |

| T avg | Iavg/Mi | T avg | n |
|---|---|---|---|
| 1807.21116 | 0.59927 | 1807.21116 | 0.70571 |
| 1913.09339 | 0.82796 | 1913.09339 | 0.54238 |
| 2011.0964 | 1.03358 | 2011.0964 | 0.38377 |
| 2091.13587 | 1.244 | 2091.13587 | 0.2808 |

| T avg | Iavg/Mi | T avg | n |
|---|---|---|---|
| 2091.13587 | 1.244 | 2091.13587 | 0.2808 |
| 1998.86577 | 1.04945 | 1998.86577 | 0.35315 |
| 1888.48525 | 0.81605 | 1888.48525 | 0.44452 |
| 1804.2565 | 0.69331 | 1804.2565 | 0.57764 |

| T avg | Iavg/Mi | T avg | n |
|---|---|---|---|
| 1816.6796 | 0.45408 | 1816.6796 | 0.64803 |
| 1850.41986 | 0.49726 | 1850.41986 | 0.61784 |
| 1907.03962 | 0.56625 | 1907.03962 | 0.53133 |
| 2011.65633 | 0.76362 | 2011.65633 | 0.41492 |

| T avg | Iavg/Mi | T avg | n |
|---|---|---|---|
| 1914.95414 | 0.22332 | 1914.95414 | 0.49566 |
| 1950.33472 | 0.30857 | 1950.33472 | 0.45233 |
| 2007.86418 | 0.47301 | 2007.86418 | 0.38736 |
| 2066.33719 | 0.57567 | 2066.33719 | 0.32015 |

| T avg | Iavg/Mi | T avg | n |
|---|---|---|---|
| 1834.00836 | 0.27712 | 1834.00836 | 0.97192 |
| 1957.78381 | 0.44802 | 1957.78381 | 0.75699 |
| 2077.98058 | 0.58617 | 2077.98058 | 0.57543 |

| T avg | Iavg/Mi | T avg | n |
|---|---|---|---|
| 1791.51129 | 0.35457 | 1791.51129 | 0.76023 |
| 1899.948 | 0.47144 | 1899.948 | 0.63051 |
| 1950.10993 | 0.52911 | 1950.10993 | 0.58901 |
| 2012.1849 | 0.6197 | 2012.1849 | 0.54152 |

| T avg | Iavg/Mi | T avg | n |
|---|---|---|---|
| 1911.16464 | 0.70863 | 1911.16464 | 0.68727 |
| 2031.59671 | 1.04463 | 2031.59671 | 0.53749 |
| 2112.79093 | 1.2285 | 2112.79093 | 0.42348 |
| 2173.09917 | 1.41818 | 2173.09917 | 0.35341 |

| T avg | Iavg/Mi | T avg | n |
|---|---|---|---|
| 1837.75104 | 0.33243 | 1837.75104 | 0.75333 |
| 1891.62185 | 0.35987 | 1891.62185 | 0.73126 |

Graphene Data

Figure 4B | Figure 4D

| T avg | Iavg/Mi | T avg | Emissivity Parameter, n |
|---|---|---|---|
| 1925.80261 | 1.06172 | 1925.80261 | 0.58017 |
| 1954.81185 | 1.23212 | 1954.81185 | 0.5892 |
| 2003.01204 | 1.3617 | 2003.01204 | 0.5303 |
| 2140.85915 | 1.76694 | 2140.85915 | 0.34939 |

| T avg | Iavg/Mi | T avg | n |
|---|---|---|---|
| 2140.85915 | 1.76694 | 2140.85915 | 0.34939 |
| 1969.90275 | 1.40486 | 1969.90275 | 0.53877 |
| 1729.9279 | 0.78726 | 1729.9279 | 0.81734 |
| 1637.87023 | 0.5955 | 1637.87023 | 0.92972 |

| T avg | Iavg/Mi | T avg | n |
|---|---|---|---|
| 1869.4201 | 1.87254 | 1869.4201 | 0.32537 |
| 1896.3167 | 2.05556 | 1896.3167 | 0.29487 |
| 1933.05436 | 2.41554 | 1933.05436 | 0.28314 |
| 1957.30791 | 2.67071 | 1957.30791 | 0.27541 |
| 1924.04544 | 2.48714 | 1924.04544 | 0.30441 |
| 2199.78351 | 4.63578 | 2199.78351 | 0.13901 |

| T avg | Iavg/Mi | T avg | n |
|---|---|---|---|
| 1815.31168 | 1.76364 | 1815.31168 | 0.45089 |
| 2040.25966 | 2.91101 | 2040.25966 | 0.20872 |
| 1962.12547 | 2.74473 | 1962.12547 | 0.32549 |
| 2215.22794 | 4.33478 | 2215.22794 | 0.00139 |

| T avg | Iavg/Mi | T avg | n |
|---|---|---|---|
| 1881.15027 | 0.31964 | 1881.15027 | 0.72559 |
| 2017.6305 | 0.74081 | 2017.6305 | 0.52295 |
| 2130.57911 | 0.91831 | 2130.57911 | 0.37578 |

| T avg | Iavg/Mi | T avg | n |
|---|---|---|---|
| 2130.57911 | 0.91831 | 2130.57911 | 0.37578 |
| 2054.09987 | 0.77959 | 2054.09987 | 0.38724 |
| 1881.44809 | 0.69686 | 1881.44809 | 0.81789 |

| T avg | Iavg/Mi | T avg | n |
|---|---|---|---|
| 1881.44809 | 0.69686 | 1881.44809 | 0.81789 |
| 1952.5489 | 0.76392 | 1952.5489 | 0.67247 |
| 2197.05722 | 1.08966 | 2197.05722 | 0.29344 |

| T avg | Iavg/Mi | T avg | n |
|---|---|---|---|
| 1911.88077 | 0.67543 | 1911.88077 | 0.50718 |
| 2031.68143 | 0.88793 | 2031.68143 | 0.39642 |
| 2062.49722 | 0.93109 | 2062.49722 | 0.35024 |

| T avg | Iavg/Mi | T avg | n |
|---|---|---|---|
| 2062.49722 | 0.93109 | 2062.49722 | 0.35024 |
| 1932.42258 | 0.69659 | 1932.42258 | 0.47208 |
| 1786.59117 | 0.45111 | 1786.59117 | 0.52945 |

| T avg | Iavg/Mi | T avg | n |
|---|---|---|---|
| 1786.59117 | 0.45111 | 1786.59117 | 0.52945 |
| 1930.36183 | 0.70054 | 1930.36183 | 0.48213 |
| 2033.686 | 0.89621 | 2033.686 | 0.39405 |
| 2108.19491 | 1.04937 | 2108.19491 | 0.33287 |

| T avg | Iavg/Mi | T avg | n |
|---|---|---|---|
| 1806.11444 | 0.27981 | 1806.11444 | 0.65754 |
| 1898.53523 | 0.45427 | 1898.53523 | 0.56332 |
| 2004.415 | 0.65489 | 2004.415 | 0.45161 |
| 2062.86757 | 0.78344 | 2062.86757 | 0.38243 |

| T avg | Iavg/Mi | T avg | n |
|---|---|---|---|
| 2062.86757 | 0.78344 | 2062.86757 | 0.38243 |
| 1976.97512 | 0.6651 | 1976.97512 | 0.42787 |
| 1930.796 | 0.62477 | 1930.796 | 0.51201 |
| 1887.02754 | 0.56931 | 1887.02754 | 0.55841 |
| 1998.72679 | 0.51533 | 1998.72679 | 0.68952 |
| 2101.26115 | 0.65307 | 2101.26115 | 0.56224 |

| T avg | Iavg/Mi | T avg | n |
|---|---|---|---|
| 1812.80194 | 0.16609 | 1812.80194 | 0.8584 |
| 1908.7598 | 0.356 | 1908.7598 | 0.73431 |
| 2021.79725 | 0.55746 | 2021.79725 | 0.64617 |
| 2031.07336 | 0.60644 | 2031.07336 | 0.54986 |

| T avg | Iavg/Mi | T avg | n |
|---|---|---|---|
| 1804.341 | 0.3824 | 1804.341 | 0.66985 |
| 1912.95433 | 0.50069 | 1912.95433 | 0.55964 |
| 2005.92659 | 0.62354 | 2005.92659 | 0.46987 |
| 2062.062 | 0.7003 | 2062.062 | 0.40713 |
| 2102.03614 | 0.76546 | 2102.03614 | 0.36244 |

| T avg | Iavg/Mi | T avg | n |
|---|---|---|---|
| 1928.25571 | 0.58538 | 1928.25571 | 0.52555 |
| 2016.55175 | 0.70879 | 2016.55175 | 0.43708 |
| 2128.52844 | 0.91083 | 2128.52844 | 0.35948 |

| T avg | Iavg/Mi | T avg | n |
|---|---|---|---|
| 1770.84544 | 0.39601 | 1770.84544 | 0.81446 |
| 1905.542 | 0.61669 | 1905.542 | 0.68495 |
| 1980.20358 | 0.79062 | 1980.20358 | 0.56876 |
| 2051.76829 | 0.95684 | 2051.76829 | 0.48259 |

| T avg | Iavg/Mi | T avg | n |
|---|---|---|---|
| 1768.71524 | 0.32821 | 1768.71524 | 0.7962 |
| 1895.10273 | 0.53047 | 1895.10273 | 0.79984 |
| 1971.94773 | 0.6388 | 1971.94773 | 0.66358 |
| 2117.1979 | 1.00741 | 2117.1979 | 0.51305 |

| T avg | Iavg/Mi | T avg | n |
|---|---|---|---|
| 1783.17246 | 0.46709 | 1783.17246 | 0.68869 |
| 1885.10191 | 0.69557 | 1885.10191 | 0.72845 |
| 2009.20633 | 0.94568 | 2009.20633 | 0.63787 |
| 2107.40065 | 1.15692 | 2107.40065 | 0.48295 |

| T avg | Iavg/Mi | T avg | n |
|---|---|---|---|
| 1810.97731 | 0.39756 | 1810.97731 | 0.81428 |
| 1973.53413 | 0.69406 | 1973.53413 | 0.64432 |
| 2067.19067 | 0.95196 | 2067.19067 | 0.50022 |
| 2200.12836 | 1.47119 | 2200.12836 | 0.34695 |

| T avg | Iavg/Mi | T avg | n |
|---|---|---|---|
| 1813.75131 | 0.52275 | 1813.75131 | 0.40586 |
| 1868.78471 | 0.70569 | 1868.78471 | 0.50946 |
| 1940.2738 | 0.91651 | 1940.2738 | 0.52113 |
| 2113.57813 | 1.37227 | 2113.57813 | 0.39287 |
| 2222.59346 | 1.74156 | 2222.59346 | 0.30926 |

| T avg | Iavg/Mi | T avg | n |
|---|---|---|---|
| 1813.13689 | 0.58416 | 1813.13689 | 0.39858 |
| 1907.16446 | 0.89011 | 1907.16446 | 0.4932 |
| 2008.60925 | 1.20887 | 2008.60925 | 0.47286 |
| 2199.87307 | 1.85581 | 2199.87307 | 0.36372 |

| T avg | Iavg/Mi | T avg | n |
|---|---|---|---|
| 1809.30126 | 0.50094 | 1809.30126 | 0.91347 |
| 1884.95574 | 0.63884 | 1884.95574 | 0.79164 |
| 1836.59233 | 0.57928 | 1836.59233 | 0.82418 |
| 1940.065 | 0.76291 | 1940.065 | 0.73034 |
| 2120.87767 | 1.13411 | 2120.87767 | 0.51507 |

| T avg | Iavg/Mi | T avg | n |
|---|---|---|---|
| 1988.22033 | 0.88386 | 1988.22033 | 0.63022 |
| 2061.57845 | 1.07028 | 2061.57845 | 0.54358 |
| 1990.78448 | 0.93173 | 1990.78448 | 0.6059 |
| 2067.5995 | 1.1117 | 2067.5995 | 0.54032 |
| 1810.39881 | 0.48182 | 1810.39881 | 0.66126 |

| T avg | Iavg/Mi | T avg | n |
|---|---|---|---|
| 1810.39881 | 0.48182 | 1810.39881 | 0.66126 |
| 1911.73767 | 0.60937 | 1911.73767 | 0.54402 |
| 2101.37029 | 1.02982 | 2101.37029 | 0.46304 |

| T avg | Iavg/Mi | T avg | n |
|---|---|---|---|
| 1808.94671 | 0.28177 | 1808.94671 | 0.8245 |
| 1911.17276 | 0.46236 | 1911.17276 | 0.682 |
| 2006.72732 | 0.68556 | 2006.72732 | 0.51922 |
| 2099.85628 | 0.87808 | 2099.85628 | 0.39264 |

| T avg | Iavg/Mi | T avg | n |
|---|---|---|---|
| 2099.85628 | 0.87808 | 2099.85628 | 0.39264 |
| 2000.7878 | 0.7601 | 2000.7878 | 0.46961 |
| 1898.88905 | 0.6152 | 1898.88905 | 0.58107 |
| 1791.37136 | 0.47266 | 1791.37136 | 0.69871 |

| T avg | Iavg/Mi | T avg | n |
|---|---|---|---|
| 1791.37136 | 0.47266 | 1791.37136 | 0.69871 |
| 1897.58431 | 0.61588 | 1897.58431 | 0.58077 |
| 1992.0879 | 0.75949 | 1992.0879 | 0.47783 |
| 2097.37678 | 0.96054 | 2097.37678 | 0.35732 |
| 2116.3208 | 1.19737 | 2116.3208 | 0.46835 |

| T avg | Iavg/Mi | T avg | n |
|---|---|---|---|
| 1803.23565 | 0.31884 | 1803.23565 | 0.84765 |
| 1863.37417 | 0.38093 | 1863.37417 | 0.78571 |
| 1955.46471 | 0.44777 | 1955.46471 | 0.63939 |
| 2112.47811 | 0.68197 | 2112.47811 | 0.52761 |

| T avg | Iavg/Mi | T avg | n |
|---|---|---|---|
| 1812.07972 | 0.22429 | 1812.07972 | 0.92217 |
| 1885.40783 | 0.2994 | 1885.40783 | 0.85852 |
| 1955.85308 | 0.37606 | 1955.85308 | 0.80997 |
| 2037.44133 | 0.5303 | 2037.44133 | 0.67069 |
| 2134.44329 | 0.70163 | 2134.44329 | 0.50502 |

| T avg | Iavg/Mi | T avg | n |
|---|---|---|---|
| 2134.44329 | 0.70163 | 2134.44329 | 0.50502 |
| 2024.95862 | 0.59447 | 2024.95862 | 0.53679 |
| 1883.99942 | 0.4348 | 1883.99942 | 0.65144 |

| T avg | Iavg/Mi | T avg | n |
|---|---|---|---|
| 1796.7533 | 0.36182 | 1796.7533 | 0.7755 |
| 1911.14691 | 0.48268 | 1911.14691 | 0.66274 |
| 2032.23155 | 0.65014 | 2032.23155 | 0.5677 |
| 2144.65911 | 0.85719 | 2144.65911 | 0.45681 |

| T avg | Iavg/Mi | T avg | n |
|---|---|---|---|
| 1805.57536 | 0.34218 | 1805.57536 | 1.35814 |
| 1851.6055 | 0.35246 | 1851.6055 | 1.06385 |
| 1904.37085 | 0.36029 | 1904.37085 | 0.88522 |
| 2006.86817 | 0.48171 | 2006.86817 | 0.77354 |
| 2177.47533 | 0.91703 | 2177.47533 | 0.62753 |

| T avg | Iavg/Mi | T avg | n |
|---|---|---|---|
| 2177.47533 | 0.91703 | 2177.47533 | 0.62753 |
| 2116.5406 | 0.77301 | 2116.5406 | 0.54759 |
| 2040.0975 | 0.66488 | 2040.0975 | 0.60887 |

| T avg | Iavg/Mi | T avg | n |
|---|---|---|---|
| 2253.35911 | 1.05627 | 2253.35911 | 0.4283 |
| 1858.02567 | 0.55259 | 1858.02567 | 0.81197 |
| 1792.11705 | 0.47525 | 1792.11705 | 0.90636 |

This data in all figures

| Temp | 1000/T | Theo ln(C/s/nm^2) |
|---|---|---|
| 3410 | 0.29326 | 14.69311 |
| 3405 | 0.29369 | 14.64847 |
| 3255 | 0.30722 | 13.24568 |
| 3150 | 0.31746 | 12.18424 |
| 3145 | 0.31797 | 12.13193 |
| 3065 | 0.32626 | 11.27171 |
| 3010 | 0.33223 | 10.65378 |
| 2955 | 0.33841 | 10.01286 |
| 2875 | 0.34783 | 9.03683 |
| 2825 | 0.35398 | 8.39873 |
| 2825 | 0.35398 | 8.39873 |
| 2725 | 0.36697 | 7.0523 |
| 2595 | 0.38536 | 5.1468 |
| 2590 | 0.3861 | 5.06969 |
| 2465 | 0.40568 | 3.04032 |
| 2425 | 0.41237 | 2.34673 |
| 2415 | 0.41408 | 2.16975 |
| 2405 | 0.4158 | 1.99129 |
| 2395 | 0.41754 | 1.81134 |
| 2385 | 0.41929 | 1.62988 |
| 2375 | 0.42105 | 1.44689 |
| 2365 | 0.42283 | 1.26236 |
| 2355 | 0.42463 | 1.07626 |
| 2345 | 0.42644 | 0.88857 |
| 2335 | 0.42827 | 0.69928 |
| 2325 | 0.43011 | 0.50835 |
| 2315 | 0.43197 | 0.31578 |
| 2305 | 0.43384 | 0.12154 |
| 2295 | 0.43573 | -0.0744 |
| 2285 | 0.43764 | -0.27205 |
| 2275 | 0.43956 | -0.47144 |
| 2265 | 0.4415 | -0.67259 |
| 2255 | 0.44346 | -0.87552 |
| 2245 | 0.44543 | -1.08027 |
| 2235 | 0.44743 | -1.28684 |
| 2225 | 0.44944 | -1.49527 |
| 2215 | 0.45147 | -1.70558 |
| 2205 | 0.45351 | -1.9178 |
| 2195 | 0.45558 | -2.13196 |
| 2185 | 0.45767 | -2.34807 |
| 2175 | 0.45977 | -2.56617 |
| 2165 | 0.46189 | -2.78629 |
| 2155 | 0.46404 | -3.00845 |
| 2145 | 0.4662 | -3.23268 |
| 2135 | 0.46838 | -3.45901 |
| 2125 | 0.47059 | -3.68747 |
| 2115 | 0.47281 | -3.91809 |
| 2105 | 0.47506 | -4.1509 |
| 2095 | 0.47733 | -4.38594 |
| 2085 | 0.47962 | -4.62323 |
| 2075 | 0.48193 | -4.86281 |
| 2065 | 0.48426 | -5.1047 |
| 2055 | 0.48662 | -5.34896 |
| 2045 | 0.489 | -5.5956 |
| 2035 | 0.4914 | -5.84466 |
| 2025 | 0.49383 | -6.09619 |
| 2015 | 0.49628 | -6.35021 |
| 2005 | 0.49875 | -6.60676 |
| 1995 | 0.50125 | -6.86589 |
| 1985 | 0.50378 | -7.12762 |
| 1975 | 0.50633 | -7.39201 |
| 1965 | 0.50891 | -7.65909 |
| 1955 | 0.51151 | -7.9289 |
| 1945 | 0.51414 | -8.20149 |
| 1935 | 0.5168 | -8.47689 |
| 1925 | 0.51948 | -8.75516 |
| 1915 | 0.52219 | -9.03633 |
| 1905 | 0.52493 | -9.32045 |
| 1895 | 0.5277 | -9.60757 |
| 1885 | 0.5305 | -9.89774 |
| 1875 | 0.53333 | -10.191 |
| 1865 | 0.53619 | -10.48741 |
| 1855 | 0.53908 | -10.78701 |
| 1845 | 0.54201 | -11.08986 |
| 1835 | 0.54496 | -11.39601 |
| 1825 | 0.54795 | -11.70552 |
| 1815 | 0.55096 | -12.01844 |
| 1805 | 0.55402 | -12.33482 |
| 1795 | 0.5571 | -12.65473 |
| 1785 | 0.56022 | -12.97823 |
| 1775 | 0.56338 | -13.30537 |
| 1765 | 0.56657 | -13.63621 |
| 1755 | 0.5698 | -13.97083 |
| 1745 | 0.57307 | -14.30928 |
| 1735 | 0.57637 | -14.65163 |
| 1725 | 0.57971 | -14.99796 |
| 1715 | 0.58309 | -15.34832 |
| 1705 | 0.58651 | -15.70279 |
| 1695 | 0.58997 | -16.06144 |
| 1685 | 0.59347 | -16.42435 |
| 1675 | 0.59701 | -16.7916 |
| 1665 | 0.6006 | -17.16325 |
| 1655 | 0.60423 | -17.5394 |
| 1645 | 0.6079 | -17.92012 |
| 1635 | 0.61162 | -18.3055 |
| 1625 | 0.61538 | -18.69562 |
| 1615 | 0.6192 | -19.09057 |
| 1605 | 0.62305 | -19.49044 |
| 1595 | 0.62696 | -19.89533 |
| 1585 | 0.63091 | -20.30532 |
| 1575 | 0.63492 | -20.72052 |
| 1565 | 0.63898 | -21.14103 |
| 1555 | 0.64309 | -21.56695 |
| 1545 | 0.64725 | -21.99838 |
| 1535 | 0.65147 | -22.43543 |
| 1525 | 0.65574 | -22.87821 |
| 1515 | 0.66007 | -23.32684 |

Figure 5A Graphite

| 1000/T | Avg Rate 1st Ramp | 1000/T | Extrapolated Data 1st Ramp |
|---|---|---|---|
| 0.55556 | -6.34733 | 0.625 | -9.07355 |
| 0.52632 | -5.30061 | 0.58824 | -7.65141 |
| 0.5 | -4.27797 | 0.55556 | -6.38728 |
| 0.47619 | -3.2731 | 0.52632 | -5.25621 |
| | | 0.5 | -4.23826 |
| | | 0.47619 | -3.31725 |
| | | 0.45455 | -2.47997 |
| | | 0.43478 | -1.7155 |
| | | 0.41667 | -1.01473 |

| 1000/T | Exp Data | Fit Data |
|---|---|---|
| 0.55159 | -6.08665 | -6.17875 |
| 0.5256 | -5.47885 | -5.34386 |
| 0.49828 | -4.49221 | -4.46605 |
| 0.4806 | -3.82882 | -3.89789 |

| 1000/T | Exp Data | Fit Data |
|---|---|---|
| 0.5234 | -5.32349 | -5.32862 |
| 0.49516 | -4.13407 | -4.12089 |
| 0.4772 | -3.34462 | -3.35268 |

| 1000/T | Exp Data | Fit Data |
|---|---|---|
| 0.52501 | -8.71276 | -7.08224 |
| 0.49893 | -5.6092 | -5.71035 |
| 0.47524 | -4.70749 | -4.46424 |
| 0.45838 | -3.43512 | -3.57722 |

| 1000/T | Exp Data | Fit Data |
|---|---|---|
| 0.55334 | -6.09254 | -6.043 |
| 0.52271 | -5.01761 | -5.14457 |
| 0.49724 | -4.49864 | -4.3973 |
| 0.47821 | -3.81505 | -3.83896 |

| 1000/T | Exp Data | Fit Data |
|---|---|---|
| 0.55045 | -6.59245 | -7.01882 |
| 0.54042 | -6.47018 | -6.45243 |
| 0.52437 | -5.5188 | -5.54699 |
| 0.4971 | -4.01856 | -4.00811 |

| 1000/T | Exp Data | Fit Data |
|---|---|---|
| 0.52221 | -5.66367 | -5.6479 |
| 0.51273 | -5.90026 | -5.41791 |
| 0.49804 | -5.01844 | -5.06125 |
| 0.48395 | -4.74613 | -4.71909 |

| 1000/T | Exp Data | Fit Data |
|---|---|---|
| 0.54525 | -6.63474 | -6.56653 |
| 0.51078 | -5.40125 | -5.54906 |
| 0.48124 | -4.7566 | -4.67702 |

| 1000/T | Exp Data | Fit Data |
|---|---|---|
| 0.55819 | -9.25146 | -9.34889 |
| 0.52633 | -7.26314 | -7.06328 |
| 0.51279 | -6.09809 | -6.09196 |
| 0.49697 | -4.84845 | -4.95701 |

| 1000/T | Exp Data | Fit Data |
|---|---|---|
| 0.52324 | -4.75897 | -4.75415 |
| 0.49222 | -3.64941 | -3.69483 |
| 0.47331 | -3.1365 | -3.0488 |
| 0.46017 | -2.55311 | -2.6002 |

| 1000/T | Exp Data | Fit Data |
|---|---|---|
| 0.55163 | -8.03496 | -5.68218 |
| 0.5239 | -4.85757 | -4.85617 |
| 0.49461 | -2.55202 | -3.98372 |
| 0.49235 | -3.91256 | -3.91643 |
| 0.47445 | -3.38565 | -3.38319 |

| 1000/T | Exp Data | Fit Data |
|---|---|---|
| 0.52479 | -5.4425 | -5.42116 |
| 0.505 | -4.29319 | -4.33852 |
| 0.48738 | -3.39872 | -3.37474 |

| 1000/T | Exp Data | Fit Data |
|---|---|---|
| 0.55219 | -6.19434 | -6.0353 |
| 0.50671 | -3.67797 | -3.99136 |
| 0.48375 | -2.98779 | -2.95971 |
| 0.45452 | -1.77244 | -1.64617 |

| 1000/T | Exp Data | Fit Data |
|---|---|---|
| 0.53052 | -5.14511 | -5.18606 |
| 0.51545 | -4.53878 | -4.48379 |
| 0.4715 | -2.42191 | -2.43596 |

| 1000/T | Exp Data | Fit Data1 |
|---|---|---|
| 0.55185 | -6.5756 | -5.1629 |
| 0.53039 | -5.39258 | -4.58874 |
| 0.51129 | -4.05306 | -4.07769 |
| 0.49081 | -3.57721 | -3.52998 |
| 0.46851 | -2.91066 | -2.93327 |

| 1000/T | Exp Data | Fit Data |
|---|---|---|
| 0.52511 | -5.8351 | -5.8288 |
| 0.49829 | -4.44216 | -4.45279 |
| 0.45925 | -2.45401 | -2.44968 |

Figure 5C Graphite

| 1000/T | Avg Rate 2nd Ramp | 1000/T | Extrapolated Data 2nd Ramp |
|---|---|---|---|
| 0.55556 | -8.24182 | 0.625 | -12.33554 |
| 0.52632 | -6.585 | 0.58824 | -10.18223 |
| 0.5 | -5.04033 | 0.55556 | -8.26819 |
| 0.47619 | -3.59075 | 0.52632 | -6.55562 |
| | | 0.5 | -5.01431 |
| | | 0.47619 | -3.61979 |
| | | 0.45455 | -2.35204 |
| | | 0.43478 | -1.19453 |
| | | 0.41667 | -0.13349 |

| 1000/T | Exp Data | Fit Data |
|---|---|---|
| 0.55744 | -7.86741 | -7.93618 |
| 0.52801 | -6.40861 | -6.44403 |
| 0.49605 | -5.03933 | -4.82373 |
| 0.448 | -2.27603 | -2.38744 |

| 1000/T | Exp Data | Fit Data |
|---|---|---|
| 0.47821 | -3.81505 | -3.9402 |
| 0.50028 | -5.74074 | -5.5211 |
| 0.52952 | -7.52075 | -7.61523 |
| 0.55424 | -8.39706 | -9.38557 |

| 1000/T | Exp Data | Fit Data |
|---|---|---|
| 0.54414 | -8.26299 | -8.4897 |
| 0.52865 | -7.74353 | -7.42933 |
| 0.50032 | -5.15827 | -5.49092 |
| 0.4759 | -3.86695 | -3.82037 |

| 1000/T | Exp Data | Fit Data |
|---|---|---|
| 0.55422 | -7.65253 | -7.83902 |
| 0.52275 | -6.28627 | -6.34701 |
| 0.49852 | -5.32361 | -5.19822 |
| 0.48495 | -4.55455 | -4.55474 |

| 1000/T | Exp Data | Fit Data |
|---|---|---|
| 0.52768 | -5.47895 | -5.52515 |
| 0.50711 | -4.425 | -4.3515 |
| 0.47232 | -2.33849 | -2.3658 |

| 1000/T | Exp Data | Fit Data |
|---|---|---|
| 0.51539 | -6.77462 | -6.7512 |
| 0.47313 | -3.62359 | -3.68966 |
| 0.44992 | -2.05111 | -2.00846 |

| 1000/T | Exp Data | Fit Data |
|---|---|---|
| 0.50231 | -5.58364 | -5.56008 |
| 0.48365 | -4.51973 | -4.58274 |
| 0.47252 | -4.0391 | -3.99962 |

| 1000/T | Exp Data | Fit Data |
|---|---|---|
| 0.46851 | -2.91066 | -2.97007 |
| 0.49384 | -5.02934 | -4.92921 |
| 0.53079 | -7.74614 | -7.78686 |

| 1000/T | Exp Data | Fit Data |
|---|---|---|
| 0.45925 | -2.45401 | -2.54323 |
| 0.47247 | -3.9598 | -3.80394 |
| 0.49017 | -5.42538 | -5.49202 |

Figure 5E Graphite

| 1000/T | Average Rate 3rd Ramp | 1000/T | Extrapolated Data 3rd Ramp |
|---|---|---|---|
| 0.55556 | -8.07586 | 0.625 | -11.87026 |
| 0.52632 | -6.57588 | 0.58824 | -9.88257 |
| 0.5 | -5.15864 | 0.55556 | -8.11573 |
| 0.47619 | -3.77713 | 0.52632 | -6.53487 |
| | | 0.5 | -5.11209 |
| | | 0.47619 | -3.82482 |
| | | 0.45455 | -2.65458 |
| | | 0.43478 | -1.58609 |
| | | 0.41667 | -0.60665 |

| 1000/T | Exp Data | Fit Data |
|---|---|---|
| 0.5186 | -7.74233 | -7.59575 |
| 0.4959 | -5.5191 | -5.79328 |
| 0.46981 | -3.85006 | -3.72247 |

| 1000/T | Exp Data | Fit Data |
|---|---|---|
| 0.53048 | -5.69919 | -5.67346 |
| 0.49771 | -4.07229 | -4.13439 |
| 0.47452 | -3.08145 | -3.04509 |

| 1000/T | Exp Data | Fit Data |
|---|---|---|
| 0.55153 | -8.56271 | -10.79624 |
| 0.52434 | -8.51051 | -8.35494 |
| 0.49786 | -5.72661 | -5.97738 |
| 0.45457 | -2.18637 | -2.09119 |

| 1000/T | Exp Data | Fit Data |
|---|---|---|
| 0.55456 | -8.26027 | -8.56939 |
| 0.53666 | -7.72475 | -7.52705 |
| 0.51139 | -6.38624 | -6.05516 |
| 0.47338 | -3.62189 | -3.84154 |

| 1000/T | Exp Data | Fit Data |
|---|---|---|
| 0.55656 | -8.20293 | -8.24638 |
| 0.52325 | -6.54204 | -6.45132 |
| 0.49207 | -4.7232 | -4.77143 |
| 0.46627 | -3.38246 | -3.38147 |

| 1000/T | Exp Data | Fit Data |
|---|---|---|
| 0.44378 | -2.37675 | -2.37676 |
| 0.53821 | -8.25929 | -8.25929 |
| 0.558 | -8.57507 | -9.49242 |

Figure 5B Graphene Data

| 1000/T | Avg Rate 1st Ramp | 1000/T | Extrapolated Data 1st Ramp |
|---|---|---|---|
| 0.555555556 | -6.445195477 | 0.625 | -9.44662 |
| 0.526315789 | -5.330970131 | 0.58824 | -7.88839 |
| 0.5 | -4.201344388 | 0.55556 | -6.5033 |
| 0.476190476 | -3.077937104 | 0.52632 | -5.26401 |
| | | 0.5 | -4.14864 |
| | | 0.47619 | -3.1395 |
| | | 0.45455 | -2.2221 |
| | | 0.43478 | -1.38448 |
| | | 0.41667 | -0.61665 |

| 1000/T | Exp Data | Fit Data |
|---|---|---|
| 0.51926 | -6.13427 | -6.1837 |
| 0.51156 | -5.71971 | -5.81764 |
| 0.49925 | -5.44851 | -5.23286 |
| 0.4671 | -3.63752 | -3.7058 |

| 1000/T | Exp Data | Fit Data |
|---|---|---|
| 0.53493 | -5.75279 | -5.6154 |
| 0.52734 | -5.19768 | -5.27458 |
| 0.51732 | -4.73417 | -4.82438 |
| 0.51091 | -4.48419 | -4.53642 |
| 0.51974 | -4.98166 | -4.93319 |
| 0.45459 | -2.04015 | -2.00666 |

| 1000/T | Exp Data | Fit Data |
|---|---|---|
| 0.55087 | -6.03004 | -6.04253 |
| 0.49013 | -2.49424 | -2.52061 |
| 0.50965 | -3.69124 | -3.65239 |
| 0.45142 | -1.78411 | -0.27575 |

| 1000/T | Exp Data | Fit Data |
|---|---|---|
| 0.53159 | -5.29168 | -5.20771 |
| 0.49563 | -4.01641 | -4.21528 |
| 0.46936 | -3.60501 | -3.4901 |

| 1000/T | Exp Data | Fit Data |
|---|---|---|
| 0.52305 | -6.83621 | -6.82075 |
| 0.4922 | -5.25154 | -5.33187 |
| 0.48485 | -5.04172 | -4.97686 |

| 1000/T | Exp Data | Fit Data |
|---|---|---|
| 0.55367 | -6.27214 | -7.1979 |
| 0.52672 | -6.29428 | -6.40128 |
| 0.4989 | -5.89651 | -5.57893 |
| 0.48476 | -4.95053 | -5.16111 |

| 1000/T | Exp Data | Fit Data |
|---|---|---|
| 0.55281 | -5.51999 | -5.5152 |
| 0.52324 | -4.66442 | -4.71678 |
| 0.49832 | -4.13881 | -4.04401 |
| 0.47622 | -3.40003 | -3.44725 |

Figure 5D Graphene Data

| 1000/T | Avg Rate 2nd Ramp | 1000/T | Extrapolated Data 2nd Ramp |
|---|---|---|---|
| 0.55556 | -7.69967 | 0.625 | -10.82961 |
| 0.52632 | -6.60657 | 0.58824 | -9.21969 |
| 0.5 | -5.44542 | 0.55556 | -7.78864 |
| 0.47619 | -4.21434 | 0.52632 | -6.50823 |
| | | 0.5 | -5.35587 |
| | | 0.47619 | -4.31325 |
| | | 0.45455 | -3.36542 |
| | | 0.43478 | -2.5 |
| | | 0.41667 | -1.70671 |

| 1000/T | Exp Data | Fit Data |
|---|---|---|
| 0.4671 | -3.63752 | -3.98858 |
| 0.50764 | -6.60717 | -6.05404 |
| 0.57806 | -9.44 | -9.64209 |
| 0.61055 | -9.8567 | -11.29754 |

| 1000/T | Exp Data | Fit Data |
|---|---|---|
| 0.46936 | -3.60501 | -4.08724 |
| 0.48683 | -5.32216 | -4.6513 |
| 0.53151 | -5.90464 | -6.09327 |

| 1000/T | Exp Data | Fit Data |
|---|---|---|
| 0.48485 | -5.04172 | -5.32791 |
| 0.51749 | -6.66652 | -6.15921 |
| 0.55973 | -7.01402 | -7.23515 |

| 1000/T | Exp Data | Fit Data |
|---|---|---|
| 0.48476 | -4.95053 | -4.96819 |
| 0.50582 | -6.83096 | -6.61867 |
| 0.51792 | -7.20711 | -7.56673 |
| 0.52993 | -8.67316 | -8.50814 |
| 0.55236 | -9.94601 | -10.26593 |

| 1000/T | Exp Data | Fit Data |
|---|---|---|
| 0.47622 | -3.40003 | -3.53839 |
| 0.4998 | -5.15759 | -5.02902 |
| 0.52662 | -6.84583 | -6.7245 |
| 0.55823 | -8.61109 | -8.72262 |

Figure 5F Graphene Data

| 1000/T | Average Rate 3rd Ramp | 1000/T | Extrapolated Data 3rd Ramp |
|---|---|---|---|
| 0.55556 | -8.03932 | 0.625 | -11.71169 |
| 0.52632 | -6.66305 | 0.58824 | -9.80359 |
| 0.5 | -5.29762 | 0.55556 | -8.10751 |
| 0.47619 | -3.91011 | 0.52632 | -6.58996 |
| | | 0.5 | -5.22417 |
| | | 0.47619 | -3.98845 |
| | | 0.45455 | -2.86508 |
| | | 0.43478 | -1.83938 |
| | | 0.41667 | -0.89916 |

| 1000/T | Exp Data | Fit Data |
|---|---|---|
| 0.53151 | -5.90464 | -6.04075 |
| 0.51215 | -5.45852 | -5.27618 |
| 0.45515 | -2.97839 | -3.02461 |

| 1000/T | Exp Data | Fit Data |
|---|---|---|
| 0.55973 | -7.01402 | -9.58472 |
| 0.51804 | -7.17105 | -7.16339 |
| 0.49172 | -5.61543 | -5.63468 |
| 0.47434 | -4.63689 | -4.62529 |

| 1000/T | Exp Data | Fit Data |
|---|---|---|
| 0.55236 | -9.94601 | -9.92595 |
| 0.52308 | -7.44584 | -7.47833 |
| 0.47588 | -3.54483 | -3.53239 |

| 1000/T | Exp Data | Fit Data |
|---|---|---|
| 0.55823 | -8.61109 | -8.46645 |
| 0.52699 | -6.47127 | -6.70072 |
| 0.50199 | -5.27756 | -5.28795 |
| 0.47679 | -3.9591 | -3.86389 |

| Figure 6A | Graphite Data |
| --- | --- |
| M_I | Ea (kJ/mol) |
| 22.16349 | 267.14402 |
| 20.65761 | 421.54101 |
| 14.26879 | 355.58903 |
| 21.83925 | 437.36601 |
| 38.93047 | 243.9166506 |
| 36.91668 | 595.4463931 |
| 46.71086 | 469.19336 |
| 10.02041 | 201.85678 |
| 5.55622 | 245.40743 |
| 75.96802 | 596.51555 |
| 39.55572 | 283.95802 |
| 35.86965 | 568.92825 |
| 120.82357 | 247.65209 |
| 114.41283 | 394.22492 |
| 109.06675 | 659.98583 |
| 15.14995 | 454.93622 |
| 13.78304 | 474.55585 |
| 12.22539 | 390.53929 |
| 58.11882 | 373.64122 |
| 50.80775 | 602.34573 |
| 46.66514 | 746.4811 |
| 17.83133 | 387.46435 |
| 16.65207 | 435.43481 |
| 16.24699 | 484.2215 |
| 17.663499 | 222.4248456 |
| 16.897197 | 643.0542578 |
| 15.436965 | 448.0165869 |
| 19.35606 | 426.59304 |
| 18.80892 | 792.80297 |
| 17.45237 | 517.98512 |

| Figure 6B | Graphene Data |
| --- | --- |
| M_I | Ea (kJ/mol) |
| 11.26089 | 394.9702 |
| 11.08014 | 423.6423 |
| 39.37613 | 373.4958 |
| 70.86106 | 482.1347 |
| 41.79014 | 229.4739 |
| 38.94829 | 268.3702 |
| 37.87616 | 328.4503 |
| 41.06963 | 401.3761 |
| 40.87548 | 211.7857 |
| 40.70118 | 482.9261 |
| 42.84934 | 245.7435 |
| 42.53173 | 651.5733 |
| 42.12453 | 695.0297 |
| 64.03095 | 224.5084 |
| 61.79144 | 525.6055 |
| 59.67103 | 469.8583 |

| Figure 7 | Graphite Data | |
|---|---|---|
| S2-s1/S1 | I2-I1/I1 | Percent Change |
| -0.39501 | 0.06966 | 6.966 |
| -0.64032 | 0.02886 | 2.886 |
| -0.77881 | -0.4572 | -45.72 |
| -0.67502 | 0.19491 | 19.491 |
| 0.12692 | -0.15574 | -15.574 |
| -0.85718 | 0.28134 | 28.134 |
| -0.81267 | 0.0431 | 4.31 |
| -0.80408 | 0.17835 | 17.835 |

| Figure 7 Graphite Data | | |
|---|---|---|
| S3-S1/S1 | I3-I1/I1 | Percent Change |
| -0.86121 | 0.32832 | 32.832 |
| -0.16221 | 0.05557 | 5.557 |
| -0.91624 | 0.42786 | 42.786 |
| -0.80368 | -0.3883 | -38.83 |
| -0.759 | 0.30172 | 30.172 |

| Figure 7 | Graphene Data | |
|---|---|---|
| grapheneS2-s1/S1 | grapheneI2-I1/I1 | Percent Change |
| -0.32714 | 0.08361 | 8.361 |
| -0.52312 | 0.24147 | 24.147 |
| -0.00556 | -0.00953 | -0.953 |
| -0.42353 | 0.06164 | 6.164 |
| -0.61411 | 0.11622 | 11.622 |

| Figure 7 | Graphene Data | |
|---|---|---|
| grapheneS3-S1/S1 | grapheneI3-I1/I1 | Percent Change |
| -0.36892 | 0.34088 | 34.088 |
| -0.33467 | 0.01181 | 1.181 |
| 0.06485 | 0.25604 | 25.604 |
| -0.66259 | 0.16956 | 16.956 |